\documentclass[12pt,a4paper]{report}
\usepackage{titlesec}
\usepackage{longtable}

\usepackage{Packages/epsfig}
\usepackage{Packages/fancyheadings}
\usepackage{Packages/setspace}

\usepackage{color}
\usepackage{amssymb}
\usepackage{graphicx}
\usepackage{subfigure}
\usepackage{geometry}
\usepackage[lined,ruled,vlined,linesnumbered]{algorithm2e}
\usepackage{bmpsize}
\usepackage{capt-of}

\renewcommand{\theequation}{\mbox{\rm \thechapter.\arabic{equation}}}

\newcounter{proposition}

\newcounter{theorem}
\newtheorem{theorem}{Theorem}[chapter]

\newcounter{assumption}

\newcounter{lemma}

\newcounter{corollary}

\newcounter{definition}
\newtheorem{definition}{Definition}[chapter]

\renewcommand{\thefigure}{\thechapter.\arabic{figure}}

\def\LEQ.#1.#2.#3{#1\!\leqslant\!#2\!\leqslant\!#3}
\def\GEQ.#1.#2.#3{#1\!\geqslant\!#2\!\geqslant\!#3}

\renewcommand{\theenumi}{\rm \roman{enumi}}
\renewcommand{\labelenumi}{\rm (\theenumi)}
\renewcommand{\theenumii}{\rm \alph{enumii}}

\begin{document}

%-----------------------------------------------------------------------------
% File          PpCover.tex
% Author        Dinh Thai Hoang
% Description   The front cover page
%-----------------------------------------------------------------------------
%sc -- small caps typeface
\title{\sc
\vspace{-0.5in} Nanyang Technological University \\
\vspace*{0.3in} \centering %\vspace* is used even for the page breaking
\resizebox*{0.20\textwidth}{!}{\includegraphics{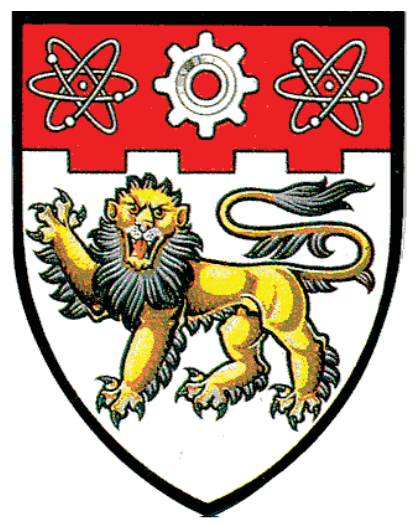}}\\[1em]
\vspace{0.2in}{\huge\bf GPU-based Commonsense Reasoning for Real-Time Query Answering and Multimodal Analysis}\\[1em]}

\author{
Ph.D Thesis\\[1em]
By\\[1em]
{\rm\bf Tran Ha Nguyen}\\[1em]
{\rm\bf Supervisor: Erik Cambria}\\[4em]
School of Computer Science and Engineering\\[1em]
A thesis submitted to the Nanyang Technological University \\in fulfillment of the requirement for the degree of
Doctor of Philosophy}

\date{October, 2016}
\maketitle
\thispagestyle{empty}        % no page number on this front cover page

%\clearpage

%-----------------------------------------------------------------------------
%==== The End Here ====%
%

\setlength{\baselineskip}{20pt plus 1pt minus 1pt} % plus 1pt minus 1pt are the elastic space
%\renewcommand{\baselinestretch}{1.24}?
% ~~~~~~~ contents, list of figures, list of tables ~~~~~~~~~
\pagenumbering{roman}       % page numbering with small roman digits for preface and contents

%----------------------------------------------------------------------------
% File          PpPref.tex
% Description   The abstract, acknowledgements, and vita
%-----------------------------------------------------------------------------

\chapter* {Acknowledgments}
\addcontentsline{toc}{section}{\numberline{}\hspace{-.35in}{\bf Acknowledgments}}

% Thesis Acknowledgements ------------------------------------------------
First and foremost, I would like to express my profound gratitude to my supervisor, Prof. Erik Cambria, for giving me the opportunity to pursue research works at Nanyang Technological University. I am truly privileged to have learned from his deep knowledge and research enthusiasm. Since the beginning of my research, he has inspired me and given me a passion for research. He has always shown me great research directions and unique ideas, along with encouragement and unwavering support regardless of his time. As well, I would greatly thank Prof. Kim Jung-Jae, Prof. He Bingsheng, Prof. James Cheng, and Prof. Foh Chuan Heng, for their helpful comments and advice.

In addition, I also want to thank all my co-authors, my colleagues in Nanyang Technological University and my friends for their numerous help. I gratefully acknowledge the scholarship towards my PhD from Nanyang Technological University. Last but not least, my deepest love and gratitude are devoted to all of my family members: my parents, my brothers, and especially my wife. Their loves will be endless motivation for me to strive and help me overcome difficult times in my life. % add to table of contents

%\chapter* {Abstract} % '*' is to avoid the number of Chapter
%\addcontentsline{toc}{section}{\numberline{}\hspace{-0.35in}{\bf Abstract}} % add Abstract into the table of contents
%\input{Abstract/abstract.tex}

%-----------------------------------------------------------------------------
%==== The End Here ====
%
%
          % preface: Acknowledgement 
\setlength{\baselineskip}{20pt plus 1pt minus 1pt}  % 1.5 spacing for whole document

\setcounter{secnumdepth}{3} % section numbering depth
\setcounter{tocdepth}{2}    % count to the depth of sections

\tableofcontents            % table of contents appears here

\newpage
%add the List of Figures section into the table of contents, toc is the file of table of contents, produced by \tablecontents
\addcontentsline{toc}{section}{\numberline{}\hspace{-.35in}{\bf List of Figures}}
\listoffigures              % list of figures appears here

\newpage
\addcontentsline{toc}{section}{\numberline{}\hspace{-.35in}{\bf List of Tables}}
\listoftables              % list of tables appears here

%list of notations
\def\listofnotations {\input symbols.tex \clearpage}
\def\addnotation #1: #2#3 {\parbox{6in}{$#1$ \hfill \parbox{5in}{#2 \dotfill  \pageref{#3}}\\}}
\def\newnot#1{\label{#1}}

\addtolength{\headheight}{3pt} \pagestyle{fancy} \setlength{\headrulewidth}{0.1pt}
\renewcommand{\chaptermark}[1]{\markboth{\chaptername\ \thechapter. #1}{}}
\lhead{\fancyplain{}{\bfseries\footnotesize\sc\leftmark}} \rhead{} \addtolength{\headsep}{-0.1in}
\cfoot{\fancyplain{}{\bfseries\rm\thepage}}

\newpage
\pagenumbering {arabic}     % page numbering with arabic

% ~~~~~~~ text part of the dissertation ~~~~~~~~~
\setlength{\baselineskip}{24pt plus 1pt minus 1pt}  % 1-1/2 spacing for whole document
%\setlength{\parindent}{0em} %indent of each paragraph
%\setlength{\parskip}{20pt} %space between two paragraph
%\setlength{\parskip}{0.5\baselineskip plus 1pt minus 1pt}
%\titlespacing{\subsectiong}{0.5\baselineskip}

\setlength{\belowcaptionskip}{5pt} %caption of the table
\renewcommand\arraystretch{1} %the row length of table

\chapter*{List of Abbreviations}
\pagenumbering{gobble}
% Thesis Abbreviations -----------------------------------------------------

\begin{longtable}{p{2.7cm}p{14.5cm}}

AI & Artificial Intelligence \\
DAG & Directed Acyclic Graph \\
DL & Description Logic \\
FOL & First Order Logic \\
GPU & Graphics Processing Unit \\
KB & Knowledge Base \\
KR & Knowledge representation \\
OWL & Web Ontology Language \\
RDF & Resource Description Framework \\
RDFS & Resource Description Framework schema\\
SIMD & Single Instruction Multiple Data \\
SIMT & Single Instruction, Multiple Thread \\
SM & Stream Multiprocessor \\
SP & Stream Processor

\end{longtable}

%\newpage
% !TEX spellcheck = en_US
\chapter* {Summary} % '*' is to avoid the number of Chapter
% Thesis Abstract -----------------------------------------------------

A commonsense knowledge base is a set of facts containing the information possessed by an ordinary person. A commonsense knowledge base is also called a fundamental ontology, as it consists of very general concepts across all domains. In order to represent such a database in practice, different approaches have been proposed in recent years. Most of them fall into either graph-based or rule-based knowledge representations. Reasoning and querying information on such kind of representations present two major implementation issues: performance and scalability, due to the fact that many new concepts (mined from the Web or learned through crowd-sourcing) are continuously integrated into the knowledge base. Some distributed computing based methods have recently been introduced to deal with those very large networks by utilizing parallelism, yet there remains the open problem of high communication costs between the participating machines.

In recent years, Graphics Processing Units (GPUs) have become popular computing devices owing to their massive parallel execution power. A typical GPU device consists of hundreds of cores running simultaneously. Modern General Purpose GPUs have been successfully adopted to accelerate heavy workload tasks such as relational database joining operations, fundamental large-scale graph algorithms, and big data analytics. Encouraged by those promising results, the dissertation investigates whether and how GPUs can be leveraged to accelerate the performance of commonsense reasoning and query answering systems on large-scale networks.

Firstly, to address the problem of reasoning and querying on large-scale graph-based commonsense knowledge bases, the thesis presents a GPU-friendly method, called GpSense, to solve the subgraph matching problem which is the core function of commonsense reasoning and query answering systems. Our approach is based on a novel filtering-and-joining strategy which is suitable to be implemented on massively parallel architectures. In order to optimize the performance in depth, we utilize a series of optimization techniques which contribute towards increasing GPU occupancy, reducing workload imbalances and in particular speeding up subgraph matching on commonsense graphs. To address the issue of large graphs which cannot fit into the GPU memory, we propose a multiple-level graph compression technique to reduce graph sizes while preserving all subgraph matching results. The graph compression method converts the data graph to a weighted graph which is small enough to be maintained in GPU memory. To highlight the efficiency of our solution, we perform an extensive evaluation of GpSense against state-of-the-art subgraph matching algorithms. Experiment results on both real and synthetic data show that our solution outperforms the existing methods on large graphs.

Secondly, in order to reason and retrieve information on rule-based knowledge bases, the thesis introduces gSparql, a fast and scalable inference and querying method on mass-storage RDF data with rule-based entailment regimes.  Our approach accepts different rulesets and executes the reasoning process at query time when the inferred triples are determined by the set of triple patterns defined in the query. To answer SPARQL queries in parallel, we first present a query rewriting algorithm to extend the queries and also eliminate redundant triple patterns based on the rulesets. Then, we convert the execution plan into a series of primitives such as sort, merge, prefix scan, and compaction which can be efficiently done on GPU devices. Extensive experimental evaluations show that our implementation scales in a linear way and outperforms current optimized CPU-based competitors.

Finally, we utilize commonsense knowledge bases to address the problem of real-time multimodal analysis. In particular, we focus on the problem of multimodal sentiment analysis, which consists in the simultaneous analysis of different modalities, e.g., speech and video, for emotion and polarity detection. Our approach takes advantages of the massively parallel processing power of modern GPUs to enhance the performance of feature extraction from different modalities. In addition, in order to extract important textual features from multimodal sources we generate domain-specific graphs based on commonsense knowledge and apply GPU-based graph traversal for fast feature detection. Then, powerful ELM classifiers are applied to build the sentiment analysis model based on the extracted features. We conduct our experiments on the YouTube dataset and achieve an accuracy of 78\% which outperforms all previous systems. In term of processing speed, our method shows improvements of several orders of magnitude for feature extraction compared to CPU-based counterparts.

          % Abstract
\addcontentsline{toc}{section}{\numberline{}\hspace{-0.35in}{\bf Abstract}} % add Abstract into the table of contents

% !TEX spellcheck = en_US
%\setcounter{equation}{0}
\chapter{Introduction}
\graphicspath{{Chapter1/fig/EPS/}{Chapter1/fig/}}
\label{tag:chap1}
\pagenumbering{arabic}

\section{Commonsense Reasoning}
\label{chap1:csreason}

This section gives a brief introduction to commonsense knowledge and popular strategies used to represent a commonsense knowledge base. Then, we discuss the challenging issues of for applications which require real-time reasoning and retrieving information on large-scale commonsense knowledge bases. Finally, we present GPU-based solutions to overcome those limitations on query answering and multimodal sentiment analysis applications.

\subsection{Commonsense Knowledge}
\label{chap1:cskb}

Commonsense knowledge is the most general understanding possessed by an ordinary person. From the AI perspective, commonsense knowledge consists of very basic concepts and their relationships which describe everyday life problems and allow humans to communicate with each other. In other words, commonsense is not the sort of knowledge that we can find in Wikipedia\footnote{wikipedia.org} but sounds obvious and natural to us. The commonsense knowledge problem is still an on-going project in the field of Knowledge Representation that aims to build a comprehensive knowledge base. Commonsense knowledge must be represented in a machine-readable manner such that different artificial intelligent programs can retrieve information and make inferences about the world \cite{camint}.

\begin{figure}[tp]
	\centering
	\includegraphics[width=0.7\textwidth]{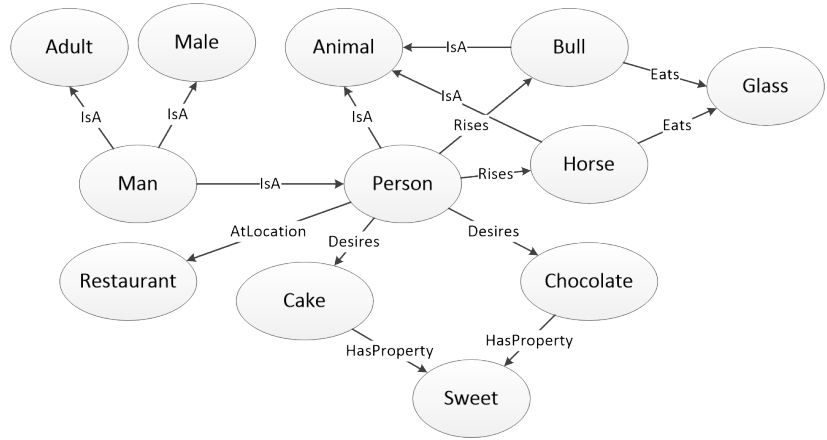}
	\caption{An example graph-based representation of commonsense knowledge}
	\label{fig:csgraph}
\end{figure}

One of the most popular strategies for representing knowledge bases is based on semantic graphs in which the collected pieces of knowledge are integrated as triples, using the format $<concept-relation-concept>$.  By considering triples as directed edges and concepts as nodes, the knowledge base naturally becomes a directed labeled graph. Figure~\ref{fig:csgraph} shows the semantic graph representation for a part of SenticNet, a commonsense knowledge base for sentiment analysis \cite{cambria2014senticnet}. Furthermore, commonsense knowledge graphs can be generally divided into six sub-types which are definition, assertion, implication, executable, learning, and hybrid semantic graphs.

In the thesis, we concentrate on the hybrid graph which is the combination of definition and assertion semantic graphs.
The formal focuses on building the hierarchical relationships between concepts which are called subsumption relations. In the resulting graph, the properties hold by a concept are inherited by its sub-concepts or its child nodes in the hierarchical graph. Due to this rule of inheritance, the definition graph is also called a generalization network.
The assertion graphs are meant to assert statements. The information is assumed to be contingently true which means that a proposition may be true in some worlds, but not in the others.
In general, semantic graphs are very expressive. This sort of knowledge representation is quite flexible and can be applied to express different models including relational and hierarchical. However, the vast amount of concepts and relations integrated into the graph causes a challenging problem at the implementation level.

Making use of logic-based systems such as  propositional logic (PL), description logic (DL), or  first-order logic (FOL) is another well-known strategy for representing commonsense knowledge. FOL and its fragment DL comprise axioms and rules of inferences and are widely used in artificial intelligence to describe and reason about the relevant concepts. Recently, the W3C has introduced the ontology web language (OWL)\footnote{w3.org/TR/owl-overview} which are built upon a decidable fragment of FOL. OWL is a semantic extension of resource description framework (RDF)\footnote{w3.org/TR/PR-rdf-syntax}, a subject-predicate-object model that makes statements about resources, to comprehensively represent ontologies.
OWL is a computational logic-based language such that knowledge expressed in OWL can be exploited by reasoners to verify the consistency of the knowledge or to make implicit facts explicit. OWL comprises different dialects corresponding to various levels of expressiveness and reasoning capabilities, namely OWL-Lite, OWL-DL, and OWL-Full. RDF and OWL have been widely used to represent knowledge bases including common knowledge bases such as DBPedia \cite{bizer2009dbpedia}, Freebase \cite{bollacker2008freebase}, or Yago \cite{suchanek2007yago} and commonsense knowledge bases such as OpenCyc \cite{lenat1989building}, SenticNet 3 \cite{cambria2014senticnet}.

\subsection{Commonsense Reasoning} Commonsense reasoning is a process that involves retrieving information about problems in daily life and making inferences based on our commonsense knowledge. Commonsense reasoning is a central part of intelligent behavior which has attached an increasing attention in recent years. In decades, there have been many research works on the representation of knowledge in formal logic and on inference algorithms to manipulate commonsense knowledge. They mainly focus on answering the following questions: 1) How to sufficiently develop a powerful and expressive formal language; 2) How to efficiently handle millions of commonsense facts and make inferences on them; 3) How to correctly encode the information as sentences in a logic; and 4) How to construct a query answering system that efficiently retrieves information from the commonsense knowledge bases.

To tackle the above problems, most state-of-the-art methods follow two main strategies. The first one tries to handle foundational problems methodically and painstakingly and then constructs small (``toy") formalizations to test the progress. Most of the works following the strategy focus on sufficiently developing powerful and expressive extensions of classical logics \cite{mccarthy1986applications,mcdermott1980non}. These methods ensure that the logical axioms precisely model the facts which they are attempting to formalize. Then, efficient reasoning systems are build make inferences. 
The second approach aims to encode a large volume of facts to enable broad commonsense reasoning. The works following this strategy focus on constructing very fast and large-scale knowledge bases for commonsense inference and query answering in different domains \cite{lenat1989building,lenat1995cyc}. Instead of producing comprehensive solutions and very powerful logics to fundamental representational problems, these methods are more concerned with the issues of performance and scalability of reasoning and querying on commonsense knowledge bases. After that, different expressiveness levels of logics/rules are built on top of these fast and large-scale systems.
The latter strategy is an interesting and practical direction which has gained an increasing attention from both academic and industrial researchers in recent years \cite{broekstra2002sesame, carroll2004jena, urbani2011querypie, peters2014scaling}. However, executing even a simple reasoning problem on commonsense knowledge is a huge task due to the massive number of commonsense facts. The search space produced during the semantic network exploration grows exponentially with the size of knowledge bases. As a consequence, our thesis will mainly focus on this emerging direction.

\begin{figure}[htp]
	\centering
	\includegraphics[width=0.6\textwidth]{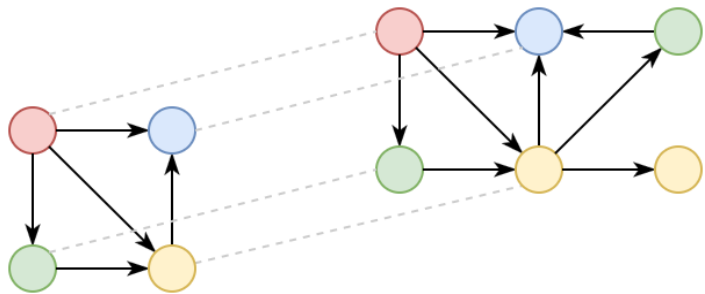}
	\caption{An example subgraph similarity search}
	\label{fig:graphiso}
\end{figure}

As described in the last subsection, there are two major strategies for representing commonsense knowledge, i.e. graph-based and logic-based representation. The recent approaches for reasoning and retrieving information on commonsense knowledge bases are, thus, proposed to deal with such kinds of data representation. Reasoning and query processing on semantic graph-based commonsense knowledge are closely related to the problem of subgraph similarity search (Figure~\ref{fig:graphiso}). The core function of the task is associated with the subgraph matching problem which is defined as finding all embeddings of a given small graph in a database graph. The subgraph matching problem is usually a bottleneck for the overall reasoning performance since it involves subgraph isomorphism which is known as an NP-complete problem \cite{cook1971complexity}. Most state-of-the-art solutions for subgraph isomorphism enumeration are generally based on a Filtering-and-Verification framework \cite{lee2012depth}. In those methods, all candidate nodes which cannot contribute to any subgraph isomorphism results are initially filtered out to reduce the search space for the network exploration. Then the verification phase follows, in which Backtracking-based algorithms are applied to find results in an incremental fashion. Those algorithms, however, are designed to work only in small-graph settings. The number of candidates grows significantly high in medium-to-large-scale graphs, resulting in an exorbitant number of costly verification operations.

\begin{figure}[htp]
	\centering
	\subfigure[Forward-chaining]{\includegraphics[width=0.45\textwidth]{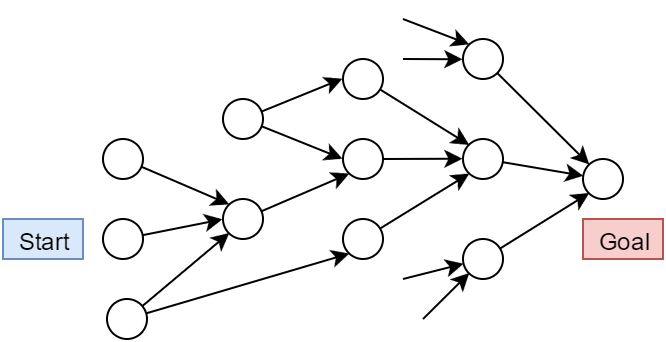}}
	\quad
	\subfigure[Backward-chaining]{\includegraphics[width=0.45\textwidth]{backward_chaining.png}}
	\caption{Rule-based reasoning and query processing approaches}
	\label{fig:rulereason}
\end{figure}

For commonsense knowledge bases represented on rule-based systems such as OWL or RDFS with different rule-sets, there are two popular schemes to perform the reasoning and query answering tasks, namely forward-chaining and backward-chaining reasoning (Figure~\ref{fig:rulereason}).
The former approach exhaustively generates all the facts from asserted ones through a certain collection of rules. In other words, this scheme makes explicit all implicit facts, called the \emph{materialization} process. The resulting facts are then explicitly written into the data storages of query engines including relational databases, RDF(S) triple stores, and graph-based query engines. The benefits of the forward-chaining inference scheme are 1) the time-consuming materialization is an off-line computation; 2) the inferred facts can be consumed as explicit ones without integrating the inference engine with the runtime query engine. However, the forward-chaining approaches are not suitable for frequently-changing knowledge bases since the costly materialization is required to re-run to catch up the modification. In addition, the unexpectedly large amount of inferred facts is also an issue. In contrast, the backward-chaining reasoning performs inference at query time in which the set of inferred facts is limited to the triple patterns defined in the query. This approach is more flexible than the former since different rules can directly be applied to the triple stores at runtime without performing the expensive off-line computation. The data storage used to maintain triples is also minimized. However, the response time of the backward-chaining approach is generally slow due to the real-time reasoning process.

\subsection{Issues for real-time commonsense-based applications}

Reasoning and retrieving information on commonsense knowledge bases have been applied in many different branches of artificial intelligence, such as healthcare \cite{camtow}, multimodality \cite{poria2016fusing}, and social data analysis \cite{cambria2014guest}. However, there remain challenging issues for such real-time applications based on commonsense knowledge and commonsense reasoning due to the following reasons:

First, the volumes of many knowledge bases including common KBs such as DBPedia, Yago, and Freebase and commonsense KBs like ConceptNet, OpenCyc, and SenticNet have grown beyond millions of concepts recently due to the fact that many new concepts learned through crowd-sourcing are continuously integrated into the knowledge graphs. As a consequence, there exists an increasing demand for building scalable systems to efficiently handle such large-scale knowledge networks.

Second, real-time reasoning on both common and commonsense knowledge bases faces serious problems in term of performance for current settings of single CPU-based solutions because of the time complexity. In particular, the subgraph similarity search is considered as an NP-complete problem. While the time complexity of inference within a family of recommendations by the W3C, ranging from simple entailment regime like RDFS to more complex ones like OWL DL or OWL Full, varies from P to EXPTIME\footnote{https://www.w3.org/Submission/owl11-tractable}. The OWL 2 Full language\footnote{http://www.w3.org/TR/owl2-primer} is even proven undecidable. As a result, the query-specific backward-chaining techniques adversely affect the response time due to real-time inference.
Some distributed computing based methods \cite{urbani2010owl,zeng2013distributed,mutharaju2015distributed} have recently been introduced to deal with those very large networks by utilizing parallelism, yet there remains the open problem of high communication costs between the participating machines.

\textbf{GPU-accelerated solutions:} In recent years, Graphics Processing Units (GPUs) have become popular computing devices owing to their massive parallel execution power. A typical GPU device consists of hundreds of cores running simultaneously. Modern General Purpose GPUs have been successfully adopted to accelerate heavy workload tasks such as relational database joining operations \cite{he2008relational}, fundamental large-scale graph algorithms \cite{zhong2014medusa}, and big data analytics \cite{he2008mars}. Encouraged by those promising results, the thesis investigates whether and how GPUs can be leveraged to accelerate the performance of commonsense reasoning on real-time query answering and multimodal sentiment analysis systems.

To deal with the problems of real-time reasoning and query processing on commonsense knowledge bases, the thesis presents two GPU-accelerated systems which correspond to two main strategies of commonsense knowledge representation. The first method focuses on large-scale subgraph similarity search in the context of commonsense reasoning while the second serves as a fast inference and querying system on mass-storage RDF data with rule-based entailment regimes. The traditional backtracking approaches cannot efficiently be adapted to GPUs due to irregular access patterns. Thus, in order to implement an effective solution for subgraph similarity search on GPUs, we introduce a novel Filtering-And-Joining strategy to prune out irrelevant candidate nodes and edges as well as combine partial results in parallel. Our algorithm is specially designed for concurrently executing on massively parallel architectures. In order to optimize the performance in depth, we utilize a series of optimization techniques which contribute towards increasing GPU occupancy, reducing workload imbalances and in particular speeding up subgraph matching on commonsense graphs. For fast inference and querying on large-scale RDF knowledge bases with customized rulesets, we first present a query rewriting algorithm to extend the queries and also eliminate redundant triple patterns based on the rulesets. After that, we convert the execution plan into a series of primitives such as sort, merge, prefix scan, and compaction which can be efficiently done on GPU devices.

After that, we focus on the problem of commonsense-based multimodal sentiment analysis, which consists in the simultaneous analysis of different modalities, e.g., speech and video, for emotion and polarity detection. Our approach takes advantages of the massively parallel processing power of modern GPUs to enhance the performance of feature extraction from different modalities. In addition, in order to extract important textual features from multimodal sources we generate domain-specific graphs based on commonsense knowledge and apply GPU-based graph traversal for fast feature detection.

% In order to understand the challenging issues for building parallel systems on GPUs, the next section gives a brief introduction to the modern GPU architecture and programming model as well as the implementation problems of GPU-based algorithms. 

\section{Graphics Processing Units}
\label{chap1:gpu}

Recently, the trend of utilizing Graphics Processing Units (GPUs) for high-performance parallel processing and analysis of vast data sets has emerged. With the support of programming environments such as Nvidia CUDA and OpenCL, developers can easily write their parallel programs on the GPUs. Consequently, many multi-threaded based graph processing methods have been introduced in the last decade. Although recent approaches have made remarkable progress in dealing with the large volume of data in parallel, many of them still fail to achieve high performance of their optimized sequential counterparts. The poor performance is mainly due to the parallel execution model supported by GPUs and the highly irregular characteristic of graph-based structures. In the section, we present an introduction to the architecture of a typical modern GPU, its memory hierarchy, and the programming model. After that, the following sub-sections present the most significant challenging issues for implementing parallel algorithms on GPUs.

\subsection{Architecture of Modern GPUs} 
\begin{figure}[htp]
	\centering
	\includegraphics[width=0.8\textwidth]{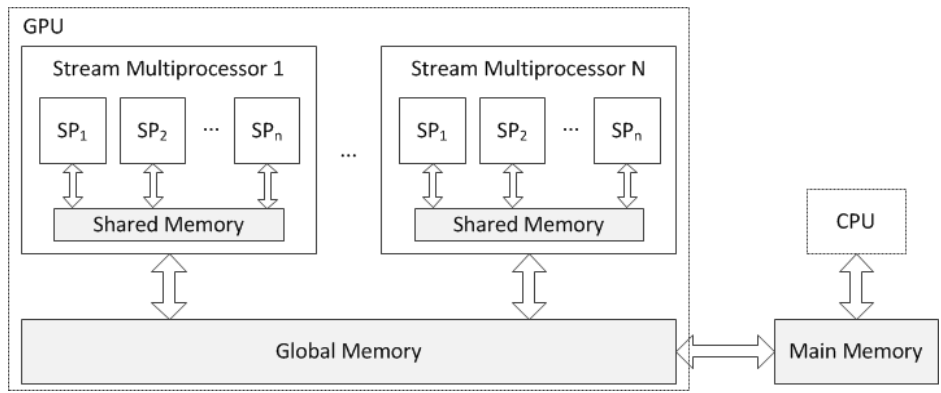}
	\caption{The Modern GPU Architecture}
	\label{fig:gpu}
\end{figure}

Figure~\ref{fig:gpu} shows the architecture of a typical programmable GPU. It is organized into an array of highly threaded \emph{stream multiprocessors} (SMs), each of which executes in parallel with the others. Each SM has multiple \emph{stream processors} (SPs) that share control logic and instruction cache. The stream processors in a multiprocessor execute in SIMT (Single Instruction, Multiple Thread) fashion in which SPs in the same SM execute the same instruction simultaneously. The massively parallel GeForce 8800 GTX GPU consists of 128 SPs (16 SMs, each of which has 8 SPs) with the total of over 500 gigaflops. With 448 SPs, the NVIDIA Tesla C2050 chip exceeds 1 teraflops. Since each SP is massively threaded, it allows thousands of threads running concurrently per application, called \emph{active threads}. The Tesla C2050 can support the maximum of 1526 active threads per SM, which sums up to 21,504 threads for this chip. In order to achieve high performance, it is very important to keep active threads busy as much as possible during parallel execution.

At runtime, groups of threads called \emph{thread blocks}, are assigned on a streaming multiprocessor for parallel execution. Each thread of a thread block is processed on an SP in the SM. In other words, an SM can run multiple thread blocks concurrently but a thread block cannot run on more than one SM. Since SPs are grouped to share a single instruction unit, threads mapped on these SPs execute the same instruction each cycle, but on different data (i.e., Single Instruction Multiple Data, or SIMD). 

Once a thread block is assigned to a streaming multiprocessor, it is further divided into 32-thread units called \emph{warps}. A warp is the unit of thread scheduling in SMs. When the threads in a warp issue a device memory operation, that instruction takes a very long time, in hundreds of clock cycles, due to the long memory latency. GPUs tolerate memory latency by using a high degree of multi-threading. When one warp stalls on a memory operation, the warp scheduler selects another active warp and switches to that one with little overhead.

Each GPU device has its own memory, called \emph{global memory}, which is an off-chip memory with high bandwidth and high access latency. The NVIDIA Tesla C2050 device memory, for example, has a size of 3 GB and  a bandwidth of 144 GBps. The global memory can be accessed by all thread blocks. Therefore, the data in the global memory can be used by all threads when the GPU executes kernels. 

There is also a small data cache attached to each multiprocessor, called \emph{shared memory}. The total size of shared memory is up to 64 KB. The data that reside in this on-chip memory can be accessed at much higher speed than accessing the data in the global memory. At runtime, shared memory is assigned to thread blocks. All threads within a thread block can access the corresponding shared memory, but cannot access to shared memory allocated to the other blocks. General speaking, shared memory is an efficient mean for threads within a block to communicate with each other by sharing the input/output data as well as the intermediate results during parallel execution.

\subsection{GPU Programming Model} 

To best support graphics processors for general purpose computation, some efficient GPGPU (General-Purpose computing on GPUs) programming models such as NVIDIA CUDA\footnote{https://developer.nvidia.com/what-cuda}, OpenCL \footnote{https://www.khronos.org/opencl}, and ADM CTM \footnote{http://ati.amd.com/products/streamprocessor} are introduced to developers to write parallel programs on GPUs easily. In these programming models, a typical program consists of multiple phases that are executed on either the GPU (host code) or the GPU (device code). The phases that exhibit little or no data parallelism are implemented in host code which runs as an ordinary CPU process. The phases that require a rich amount of data-parallelism are written in the device code. The device code contains predefined keywords for labeling parallel functions, called \emph{kernels}. The kernels often generate a large number of threads for efficiently performing GPU programs. However, there are some restrictions for kernel functions such as no-recursion is allowed within a kernel, no static variable declaration, and an unchangeable number of arguments.

A GPU program starts with the host execution. When a kernel is launched, the execution is moved to the device where a large number of threads are generated. These threads are organized into a three-level hierarchy. At the highest level, all threads generated by the kernel are collectively called a \emph{grid}. Each grid consists of thread blocks which are groups of threads. A grid can hold up to $2^{16} - 1$ blocks in either of two dimensions. When all threads in the kernel finish their jobs, the grid will terminate and the GPU will be ready for running the next kernel invoked by the host code.

\subsection{Challenges for GPU Implementation}
\label{sec:challenges}

Despite the massively parallel processing power of modern GPUs, many algorithms executed on GPUs still fail to exhibit acceptable performance and cannot make a significant improvement in comparison with its sequential counterpart running on CPUs. In this section, we first describe a list of challenges that prevents GPU-based algorithms from achieving high performance.

\textbf{Host-Device Data Transfer:}
A GPU is connected to a GPU through an IO bus slot, typically a PCI-Express in current high-performance systems. The PCI-Express bus, however, has a limited bandwidth which is up to 16 GBps in the current generation. Although GPUs provide a large amount of global memory, with 3 GB for Tesla C2050 chip, it still may not be big enough to maintain all data of large graphs whose sizes can be up to millions to billions of nodes and edges. In order to process such huge graphs on GPUs, we have to take the main memory (or even hard-disk if necessary) as the main data storage and then transfer the data to the device memory during the kernel execution. However, the bandwidth of the PCI-Express bus is very low. A large amount of time will be spent on copying the data from the main memory to the global memory and vice versa. This transferring process may stall other running kernels.
In addition, the performance overhead due to PCI calls is higher in the graph applications than the others. This can potentially become a bottleneck if the parallel graph algorithms are not well designed.

\textbf{Branch Divergence:}
As mentioned above, the kernel functions running on GPUs are based on warp execution. The group of threads in a warp performs as a Single Instruction Multiple Data (SIMD) unit. However, in order to provide the flexibility for GPU programming, the modern GPU architecture relaxes SIMD constraints by allowing threads within a warp to execute different instructions. Since threads in the warp are assigned to SPs in the same SM which share the same instruction unit, GPUs still do not allow varying instructions are executed simultaneously. Consequently, these instructions will be serialized during the warp execution. The problem of threads taking the branches in different directions is called \emph{warp divergence}. This leads to low SIMD throughput due to the underutilization of device resources. The reason of different execution paths can be \emph{if\_then\_else}, \emph{switch\_case} or different terminate conditions in \emph{loop} statements.

\textbf{Memory Access Irregularity:}
The memory system is also optimized for warp-based processing. If threads within a warp concurrently access words in the global memory that reside in the same 128-byte segment, the GPU merges 32 reads or writes into only one memory access transaction. In other words, the reading or writing speed is as fast as accessing to a single word. However, if the GPU accesses different 128-byte segments, threads within the warp experience diverse memory access latencies. In current organizations, the entire warp must wait until the last thread finishes its memory access. Graph related applications, in particular, tend to be irregular. Their memory access patterns highly depend on the structure of the input graphs such as the distribution of adjacency lists of nodes. The memory-access irregularity is, therefore, a common issue for graph algorithms that significantly decreases the performance. Thus, coalesced memory accesses are crucial to achieve a higher level of performance for device kernels. 

\textbf{Workload Imbalance:}
Graph-based algorithms implemented on GPUs suffer from the issue of workload imbalance due to uneven work distribution across different threads in the running kernel. The imbalance is often related to the irregular structure of the input graph. In most graphs, there exists a large variance in the degrees of nodes. Some nodes have a large number of neighbors while some nodes are connected to only a few nodes. Therefore, the amount of works that needs to be processed by each vertex is varying and the workload is imbalanced in the graph applications consequently. In other words, threads assigned to high degree nodes suffer from the heavy workload while the other threads have nothing to do. The total running time is dominated by threads with the heavy workload which leads to an inefficient use of GPU threads.

At runtime, warps currently running in an SM are called active warps. Due to the resource constraints, each SM only allows a certain number of active warps running at a time. Occupancy is the number of currently running warps divided by the maximum number of active warps. At runtime, when a warp stalls on a memory access operation, the SM switches to another active warp for arithmetic operations in order to hide the data access latency. Due to the problem of workload imbalance, some warps have to run heavy tasks which take a large amount of time, while the others only have very few works to do. After a while, only warps with heavy jobs still work. This leads to the low occupancy problem in which the number of active warps is not enough to adequately hide access latency. The problem also reduces the efficiency of GPU parallelism.

\section{Scope, Contributions and Organization} 
\label{chap1:contrib}

In the thesis, we adopt the GPU as an accelerator to leverage the performance of reasoning and query answering tasks on large-scale commonsense knowledge bases. First, the thesis focuses on solving the subgraph similarity detection problem which is one of the most important functions of commonsense reasoning using GPUs. Second, we propose a fast inference and query answering GPU-based systems on large-scale RDF data stores with rule-based entailment regimes.  After that, we exploit our GPU-based commonsense reasoning and querying framework to address the problem of multimodal sentiment analysis.

The organization and the main contributions of the thesis are summarized as follows.

\begin{itemize}
	%%%%%%%%%%%%%
	\item \textbf{Chapter 1:}
	\begin{itemize}
		\item We introduce the fundamental definitions of commonsense knowledge and commonsense reasoning problems.
		
		\item We provide the architecture of a typical modern GPU, its memory hierarchy and the programming model. In addition, the most significant challenging issues for implementing parallel graph-based algorithms on GPUs are given. 
	\end{itemize}
	%%%%%%%%%%%%%	
	\item \textbf{Chapter 2:}
	\begin{itemize}
		\item We discuss research works related to problems of commonsense reasoning and query answering. Then, we provide recent attempts to implement parallel algorithms on GPUs.
		\item We highlight the significance and novelty of our works compared with other related works in the literature.  
	\end{itemize}
	%%%%%%%%%%%%%
	\item \textbf{Chapter 3:}
	\begin{itemize}
		\item We discuss how a commonsense KB can be naturally represented as a graph and how such a KB can be directly transformed to a graph representation.
		
		\item We introduce a GPU-friendly algorithm, termed GpSense, to deal with large-scale subgraph similarity search which is the core function of commonsense reasoning.
		
		\item We propose a novel Joining-and-Filtering strategy which is specially designed for massively parallel computing architectures of modern GPUs. Our approach takes advantages of efficient GPU techniques of coalescence, warp-based and shared memory utilization, and a recursive refinement function for pruning irrelevant candidate nodes.
		
		\item For large-scale graphs, we present a multi-level graph compression method to reduce the size of data graphs which cannot fit into GPU memory, but still preserve query answering correctness.
		
		\item To highlight the efficiency of our solution, we perform an extensive evaluation of GpSense against state-of-the-art subgraph similarity search algorithms. Experiment results show that our solution outperforms the existing methods on large-scale commonsense knowledge graphs.
	\end{itemize}
	%%%%%%%%%%%%%
	\item \textbf{Chapter 4:}
	\begin{itemize}
		\item  We introduce a parallel approach for backward-chaining reasoning and query processing on modern GPUs. The main purpose of our system is to enhance the performance of on-the-fly reasoning at query time.
		
		\item We present a triple store layout which is able to immediately make comparisons between numeric data in the FILTER clauses without further requesting actual values from the dictionary. In addition, our method can directly return the results of unary functions on RDF terms such as $isIRI$, $isLiteral$, or $isNumeric$. Then, we discuss how to execute inference rules using the GPU primitives and the triple store layout.
		
		\item We introduce a GPU implementation of Bloom Filter algorithm for detecting triple duplication when performing the inference rules which generate a large number of new triples.
		
		\item We present extensive performance evaluation to show the efficiency of the proposed method against state-of-the-art single machine systems.
	\end{itemize}
	%%%%%%%%%%%%%
	\item \textbf{Chapter 5:}
	\begin{itemize}
		\item For the problem of multimodal sentiment analysis, we develop an ensemble application of ELM and GPU for real-time applications.
		
		\item Our method employs various GPU-friendly techniques to enhance the performance of the feature extraction process from different modalities, namely visual, audio and text. In addition, powerful ELM classifiers are applied to build the sentiment analysis model based on the extracted features.
		
		\item We discuss the experimental results on the YouTube dataset using both feature-level and decision-level fusions. Then, we describe the performance of GPU-based feature extraction as well as analyze the importance of each feature used in the classification task.
		
	\end{itemize}
	%%%%%%%%%%%%%
	\item \textbf{Chapter6:}
	\begin{itemize}
		\item  We present conclusions for the thesis and propose some potential research directions.
	\end{itemize}
\end{itemize}	% introduction
% !TEX spellcheck = en_US
\chapter{Literature Review}
\graphicspath{{Chapter2/fig/EPS/}{Chapter2/fig/}}
\label{tag:chap2}

In this chapter, we first discuss problems studied in the area of commonsense knowledge reasoning and query answering using both rule-based and graph-based representation strategies. Then, we introduce a review of recent GPU-accelerated research works including graph-based processing, rule-based reasoning algorithms, database query answering. Finally, recent studies related to the multimodal sentiment problems are mentioned in the last subsection.

\section{Commonsense Reasoning and Querying}

In this section, we discuss recent studies which solve the problem of reasoning and retrieving information on commonsense knowledge bases as well as common knowledge bases in general. Due to the fact that there are two primary strategies to represent commonsense knowledge, i.e. logic- and graph-based representation, many query answering and inference systems have been recently proposed to handle such strategies. For graph-based approach, this section mainly focuses on the subgraph similarity search problem which is the core function of all reasoning and query answering tasks.

\subsection{Logic-Based Reasoning}

In the earlier stage of commonsense knowledge representation and reasoning, most studies focus on building powerful and expressive extensions of classical logics. 
The first complete study which gave the formal definition of commonsense knowledge has been proposed by McCarthy and Lifschitz \cite{mccarthy1986applications} in 1990. In this work, the authors formalized commonsense knowledge by using mathematical logic and made inferences by deductive reasoning (DR). DR is a logical process in which a conclusion is based on the concordance of multiple assumptions that are usually assumed to be true. An important property of DR is monotonicity which is described as follows: if $A$ is a consequence of $S$, $A$ is also a consequence of $S \cup \{B\}$. General speaking, a conclusion inferred from a set of assumptions will be preserved even if additional information is added to that assumption set.

McCarthy also described a circumscription method of a non-monotonic reasoning for formalizing common sense knowledge \cite{mccarthy1987circumscription}. This approach is reasoning to conclusions on the basis of incomplete information.
To implement the circumscription, McCarthy augmented FOL to allow the minimization of the extension of some predicates. This minimization is similar to the closed world assumption which states that all unknown things are set to be false.
Ernest Davis \cite{davis1990representations} provided an ad hoc language for expressing commonsense knowledge and inference techniques for carrying out commonsense knowledge. In this work, his reasoning procedures focused on multiple domains including space, time, belief, plans, goals, and actions.

Doug Lenat et. al. \cite{lenat1989building} introduced a logic-based repository of commonsense knowledge called Cyc. The project attempts to assemble a comprehensive ontology and commonsense knowledge base with the purpose of supporting AI systems to make human-like reasoning. The Cyc knowledge base consists of more than 1.5 million facts, rules of thumb, and heuristics for reasoning about the objects and events of everyday life. The knowledge is represented in a declarative language called CycL based on FOL. To answer queries on the knowledge base, Cyc employs an inference engine that performs general logical deduction including modus ponens, modus tollens, and universal and existential quantification. The inference engine is based on a best-first search, proprietary heuristics and micro-theories to significantly reduce the search space.

Recently, the W3C has presented ontology web language OWL which is a computational logic-based language such that knowledge expressed in OWL can be exploited by reasoners to verify the consistency of the knowledge or to make implicit facts explicit. OWL comprises different dialects corresponding to various levels of expressiveness and reasoning capabilities, namely OWL-Lite, OWL-DL, and OWL-Full. RDF and OWL have been widely used to represent knowledge bases including common knowledge bases such as DBPedia \cite{bizer2009dbpedia}, Freebase \cite{bollacker2008freebase}, or Yago \cite{suchanek2007yago} and commonsense knowledge bases such as OpenCyc\footnote{http://sw.opencyc.org/}, SenticNet 3 \cite{cambria2014senticnet}. In order to support retrieving information on such knowledge bases, the W3C recommendation introduced SPARQL which is considered as the most popular RDF query language. An SPARQL query is composed of two clauses. The first one specifies the kind of the query. The second clause, the WHERE clause, consists in defining triple patterns through variables to identify the target data. The SPARQL queries may include conjunctions, disjunctions, or optionality. The results of SPARQL queries can be results sets or RDF graphs. In most reasoning and query answering systems, SPARQL queries are converted to logical forms before being executed by OWL reasoners to return the complete results.

In order to efficiently make inferences on OWL knowledge bases, many reasoners have been introduced in recent years. HermiT  is the first publicly-available OWL reasoner for ontologies written using the Web Ontology Language (OWL) \cite{shearer2008hermit}. HermiT can be used to determine the consistency of a given ontology, and identify subsumption relationships between classes. In this work, the authors propose a novel hypertableau calculus which makes inferences more efficiently in comparison with previously reported methods. HermiT also provides a faster process when classifying complex ontologies.

Pellet is an open-source OWL-DL reasoner developed in Java by The Mind Swap group. It is based on the tableau algorithm and supports expressive description logics. Pellet is the first reasoner that supported all of the OWL-DL which is based on SHOIN(D) logic. Then, it is extended to OWL 2 provides the expressiveness of SROIQ(D). Pellet. al.so supports other OWL2 profiles including OWL2-EL. In addition, it reasons ontologies through Jena as well as OWL-API interfaces. Pellet. al.so supports the explanation of bugs.

The Fast Classification of Terminologies, in short FaCT, reasoner has been introduced by Horrocks in 1998 \cite{horrocks1998fact}.  It is a sound and complete tableaux-based reasoner for expressive DL and is applied for testing modal logic satisfiability. FaCT can also be used as a description logic classifier. The improved version of FaCT, called FaCT++ \cite{tsarkov2006fact++}, is implemented in the same way as the original one. It covers OWL, OWL2 and DL-based ontology languages, but lacks support for key constraints and some datatypes.

In order to support retrieving information from knowledge bases, Jena \cite{carroll2004jena} builds a query parsing layer to accept queries such as SPARQL from users. The parsed queries are then transferred to the lower layers such as inference layer and query evaluation layer to produce the results. The default rule engine of Jena is based on the standard RETE algorithm \cite{forgy1982rete} which can support a wide range of rule sets including RDFS and OWL. In the latest version of Jena, some generic inference rules can be applied in combination with some RDFS or OWL inference. 
Kollia et. al. \cite{kollia2011sparql} propose a reasoning and query answering system which is built upon the HermiT inference engine. In this work, they present some new optimization techniques to efficiently rewrite the queries and determine a good execution order. In addition, their system examines the class and property hierarchy to reduce the query execution time.
Sesame \cite{broekstra2002sesame} and OWLIM \cite{bishop2011owlim} also offer an inference layer on top of the query layer to support in-memory and native inference.

Recently, Zhou et. al. \cite{zhou2014pay} have introduced OWL query answering system can provide scalable reasoning and querying for a wide range of OWL 2 ontologies. The system first employs an OWL 2 RL reasoner to find the lower bound answer set. The input ontology is then rewritten to produce the upper bound answer set by using the same OWL 2 RL reasoner. If lower and upper bound answers are similar, the system obviously can return a sound and complete answer. Otherwise, the gap between the two answer sets is verified by an OWL 2 reasoner such as HermiT and Pellet.

Urbani et. al. \cite{urbani2011querypie} propose a parallel and distributed backward-chaining method, called QueryPIE, to deal with very large knowledge bases. Their method is able to support the reasoning process using the OWL Horst ruleset at query time on the knowledge bases whose sizes are up to 1 billion triples. To enhance the performance of real-time reasoning, QueryPIE introduces some optimization techniques such as precomputing frequent triple patterns and early pruning reasoning branches based on precomputed data.

Unlike the above methods which make inferences at query time, some studies apply the forward-chaining approaches. These methods make explicit all implicit triples in the off-line pre-processing step. The resulting data are then maintained in scalable triple stores such as RDF-3X \cite{neumann2008rdf}, TripleBit \cite{yuan2013triplebit} to execute SPARQL queries. The process of inferring all implicit triples is usually time-consuming. For fast and scalable reasoning, Urbani et. al. \cite{urbani2010owl} introduce the WebPIE inference engine which is built on top of the Hadoop platform. Their experimental results show that WebPIE can scale-up to billions of triples, and outperforms all published approaches, both in terms of triple throughput and maximum system size. Subercaze et. al. \cite{subercaze2016inferray} present an in-memory inference engine, called Inferray, to efficiently make reasoning on RDFS, $\rho$df, and RDFS-Plus ontologies. The method is based on the vertical partitioning storage layout and  efficient sort-merge join inference.

\subsection{Subgraph Similarity Search}

Due to the fact that common and commonsense knowledge bases can naturally be represented as semantic graphs, graph-based processing approaches have been applied to implement to make inferences and queries on knowledge base systems. In those approaches, subgraph similarity search is usually the main function which consumes the majority processing time during the query answering. Bonstrom et. al. \cite{bonstrom2003storing} store the RDF triples as a graph in an object-oriented database. The advantage of this method is that they can directly execute queries on the graph structure without data reorganization.

gStore \cite{zou2011gstore} proposes a novel concept of signature graphs in which entities and classes are encoded into fixed-length bit strings. In the method, the edge labels connected to an entity and class node $u$ are also encoded and are combined with the encoded value of $u$ to represent the node. As a result, the neighborhood information of each node is captured in the encoding which can be efficiently exploited during subgraph matching process. In order to reduce the search space of the time-consuming subgraph matching step, the authors build a multi-resolution summary graph in which a graph of the high level is the summary graph of the graphs in the lower level. Then, they propose a filtering technique to take advantages of the summary graphs. gStore also supports the wildcard queries by hashing n-gram sets of string values.

The recent graph-based approach Turbo$_{HOM}$++ \cite{kim2015taming} modifies the state-of-the-art subgraph isomorphism algorithm Turbo$_{ISO}$ \cite{han2013turbo} to solve the problem of reasoning and query answering on RDF knowledge bases. Instead of directly transforming the RDF data to a directed graph, Kim et. al. introduce a type-aware graph transformation by embedding the classes into entities as node labels. This transformation method helps Turbo$_{HOM}$++ dealing with the subsumption relation and reducing the amount of graph exploration. In this work, they also adopt the NUMA architecture to efficiently execute the RDF query processing in parallel.

The core problem of subgraph similarity search, i.e. subgraph isomorphism, has attracted the interest of researchers in many years. Most existing algorithms are generally based on the \emph{filtering-and-verification} framework. First, they filter out all candidate nodes which cannot contribute to the final solutions. Then the verification phase follows, in which \emph{backtracking}-based algorithms are applied to find results in an incremental fashion. In this phase, different node matching order and candidate refinement techniques are applied to reduce the search space and consequently decrease the processing time. 

The very first practical algorithm following this approach was proposed by Ullmann \cite{ullmann1976algorithm} in 1976. In the filtering phase, Ullmann simply prunes out all nodes in the data graph which cannot be considered as candidates of query nodes in a linear fashion. A data node is said to be a candidate of a query node if it has the same label as and a greater degree than the query node. Because of its simplicity, the method cannot reduce as many candidates as other approaches, which are discussed later. However, the major advantage of the approach is the matching candidates of nodes can be found very fast. In the backtracking phase, after mapping a data node $v$ to a query node $u$, the algorithm refines the candidates of the remaining query nodes by iterating through all adjacent nodes of $u$. 

The Ullmann's algorithm does not pay attention to node matching order. It uses the input order of query nodes without any change. The performance of the algorithm depends on the node order of the input query graph. This is one of the reasons that make the algorithm become the slowest among recent algorithms. VF2 \cite{cordella2004sub}  makes an improvement by selecting a node which is adjacent to the already matched nodes. However, they do not introduce any effective method to choose the next node among nodes connected to the already matched nodes. In fact, the node is randomly selected. In addition,  VF2 employs three feasibility rules which are based on the candidate sets and adjacency of nodes to prune the search tree. QuickSI \cite{shang2008taming} proposes a new approach to choose the next node. Their node matching order selection is based on the label frequency. The method chooses nodes having infrequent node labels and adjacent to infrequent edge labels. The limitation of the method is it only works well with some data sets. For other data sets, it suffers serious performance problems \cite{lee2012depth}. 

A common approach used to speed up the subgraph isomorphism search time is based on neighborhood indexes. GADDI \cite{zhang2009gaddi} introduces a concept of Neighboring Discriminating Substructure (NDS) distance to reduce candidates of the neighborhood of the node that has just been matched. The distance is calculated by counting the number of a common NDS in the neighborhood of two nodes. GADDI selects discriminating substructures from the data graph. Then, for each data node pair, the algorithm indexes the distance between them using these substructures. The index is used in the refinement procedure to filter out irrelevant candidates. For each node $v$, GraphQL \cite{he2008graphs} indexes the subgraph within the radius $r$ of $v$, namely \textit{profile}. The profile is defined as a sequence of the node labels in lexicographic order. Using the similar idea, SPath \cite{zhao2010graph} encodes the labels of nodes within the distance $k$ of $v$ into a neighborhood signature and then indexes the achieved signature. Unlike the previous methods which select a node at a time, SPath finds the best path starting from an already matched query node. In the work, they employ a selective function $sel(p)$ for a given path $p$. The method chooses a path  $p$ with the largest selectivity $sel(p)$.

To overcome the limitation of choosing node matching order in previous studies, Turbo$_{ISO}$ \cite{han2013turbo} converts the query graph to a Neighborhood Equivalence Class (NEC) tree in which nodes with similar roles are merged into a single node and non-tree edges are pruned. After that, the authors propose a hierarchy-based structure, called Candidate Region, to maintain the candidate nodes of the NEC tree. Based on this structure, they can efficiently find a good node matching order for the backtracking-based phase.

These algorithms, however, are designed to work only in small-graph settings. The number of candidates grows significantly in medium-to-large-scale graphs, resulting in an exorbitant number of costly verification operations. To deal with large graphs, Sun et al.~\cite{sun2012efficient} introduce a distributed approach based on Trinity memory cloud \cite{shao2013trinity}. In this study, they decompose the query graphs into 2-level trees, called STWigs, and search for their matchings by using a pre-defined matching order. The STWig matching process is executed individually in participating machines, each of which stores a segment of the data graph. The results are then combined in the joining step. The distributed method have been introduced to is able to deal with large graphs by utilizing parallelism, yet there remains the open problem of high communication costs between the participating machines.

\section{GPU-Accelerated System}

In recent years, GPUs have been widely adopted to accelerate the performance of data processing in many large-scale applications. In this section, we present a review of GPU-based algorithms including graph-based algorithms, query answering and reasoning systems.

\subsection{Graph-based Algorithm}

Despite the high computational throughput, GPUs might appear poorly suited for sparse graph computation. Therefore, designing algorithms which can obtain high performance from parallelism is a non-trial task for all graph applications. In recent years, many attempts have tried to implement fundamental graph algorithms on large-scale input graphs such as Breadth-First Search (BFS), Single-Source Shortest Paths (SSSP), All-Pairs Shortest Paths (APSP), Strongly Connected Component (SCC). They focus on optimizing  specific graph algorithms targeting the massively parallel platforms. This section presents a survey of existing works on the most popular domains such as graph traversal, shortest paths, SCC decomposition.

\begin{figure}[htp]
	\centering
	\includegraphics[width=0.7\textwidth]{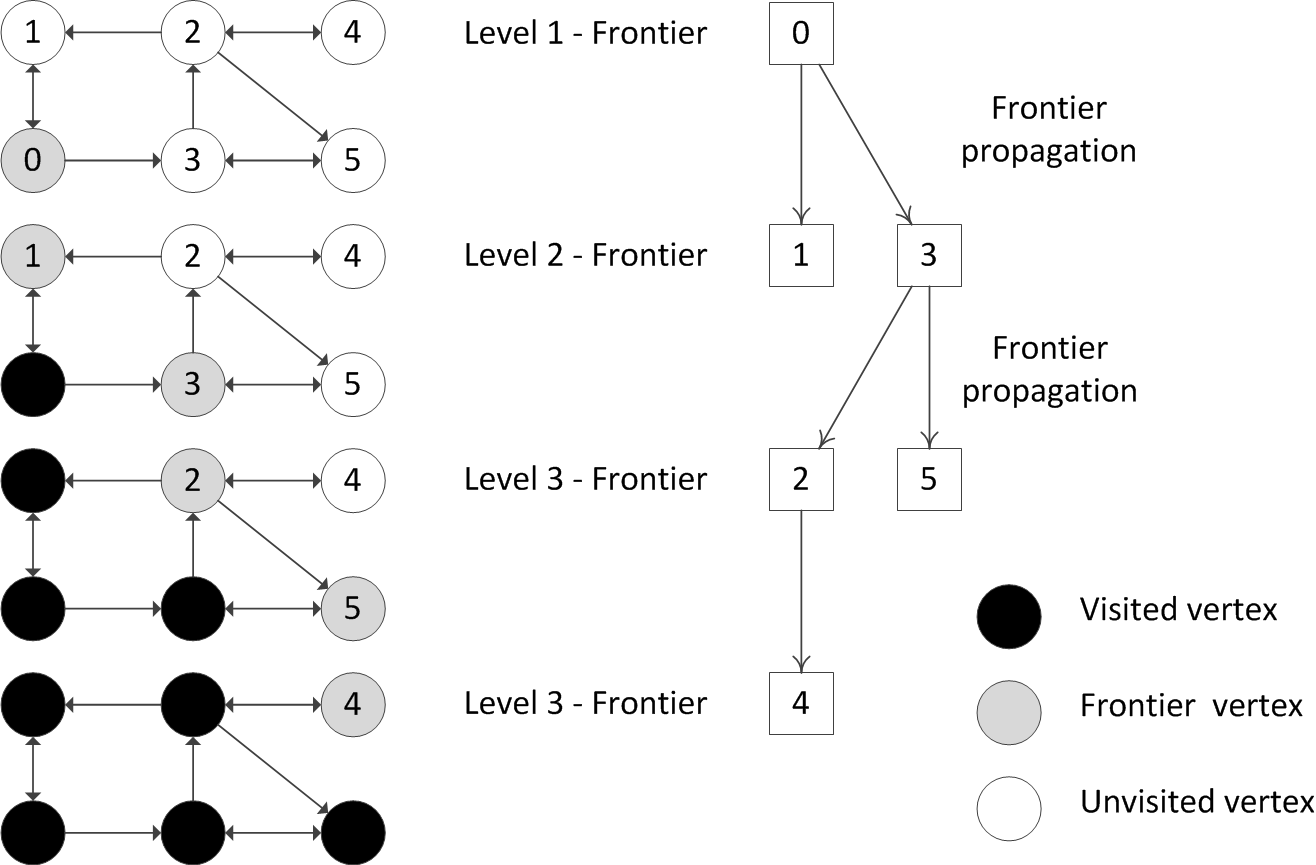}
	\caption{The BFS operation}
	\label{fig2:bfs}
\end{figure}

\textbf{Graph Traversal:} Graph traversal is considered as the building block of many high-performance graph analysis algorithms. Most of the previous graph traversal algorithms implemented on GPUs are based on Breadth-First Search. The traditional BFS algorithm, called Top-Down BFS, starts with a source node $s$ and labels all nodes in an increasing order of depth (Figure~\ref{fig2:bfs}). At each level, the algorithm identifies a \emph{frontier} which is a set of nodes being traversed. Frontier propagation checks the neighbors of all nodes in the frontier to see whether they are visited already; if not, the neighbors are inserted into the new frontier. The computational complexity is $O(|V|+|E|)$. For sparse graphs with $E$ = $O(|V|)$, the complexity of BFS is $O(|V|)$.

At the early stage, most of the parallel BFS implementations on GPUs follow \emph{quadratic parallelization} strategies. Instead of building a queue for the frontier, these algorithms use a boolean array $F$ of size $|V|$ to identify which nodes belong to the frontier of the current level. 
Harish et. al. \cite{harish2007accelerating} create another boolean array $X$ with the same size to mark visited nodes. The method is based on a thread-centric execution. In each iteration, a thread is mapped to an entry in the frontier array. If the entry has a $True$ value, the corresponding node is marked as visited and is removed from the frontier array. Then the algorithm explores its neighbors to label unvisited nodes and mark them as frontier nodes for the next iteration. The algorithm terminates when the frontier is empty. The work complexity is $O(|V|^2+|E|)$ since there may be $|V|$ iterations in the worst case. 
In order to take advantages of coalesced access to the global memory, Hussein et. al. \cite{hussein2007implementing} compact the frontier boolean array and construct a queue by using a parallel prefix sum operation. 
Deng et. al. \cite{deng2009taming} introduce an edge-oriented implementation on GPUs. The approach represents the graph as an adjacency matrix and then executes the BFS operation upon an efficient Sparse-Matrix Vector Product (SMVP) kernel.

Rather than threads, Hong et. al. \cite{hong2011accelerating} map nodes to warps to address the workload imbalance problem between threads which is witnessed by the previous thread-centric approaches. A group of threads within an actual physical warp explores the adjacency list of a node in the frontier in parallel. The thread group is called a \emph{virtual warp}. The number of threads in a virtual wap can be either 4, 8 or 16. The advantage of using virtual warps is that the method can overcome the problem of underutilization within a warp. 

For large graphs, the accelerated algorithms based on quadratic parallelization are still slower than the fast traditional sequential BFS. Therefore, work-efficient parallel BFS algorithms should perform \emph{linear parallelization} on the GPUs. In order to execute $O(|V|+|E|)$ work complexity, Luo et. al. \cite{luo2010effective} introduce an efficient three-level thread hierarchy strategy in order to quickly build the frontier queue. All threads in the kernel work together to create local queues at warp, block and grid levels. The local frontier queues generated at the lower level are used to build the queues at the higher level. At grid level, the method employs an inter-block synchronization to synchronize threads across blocks by communicating through global memory. 

Merrill et. al. \cite{merrill2012scalable} gather the frontier queue in two phases: neighbor expansion and contraction. The expansion collects the neighbors of nodes in the current frontier. The step employs a $scan+warp+CTA$ gathering strategy which can overcome all forms of workload imbalance. The contraction takes advantage of bitmask to filter previously-visited and duplicated neighbors. It then uses prefix sum to calculate the output offsets and finally writes the qualified nodes to the next frontier queue. 

Beamer et. al. \cite{beamer2013direction} observe that after several iterations of exploring a low-diameter graph, the number of qualified nodes which will be added to the frontier queue is very small compared to the number of edges to be examined. This leads to plenty of wasted attempts to check the statuses of neighbors during the frontier gathering. The authors propose a novel approach called \emph{bottom-up BFS} to deal with the problem. Instead of exploring the neighbors of frontier nodes, their approach investigates the unvisited nodes to check whether their parents are on the frontier. Once a node finds a parent in the current frontier, the checking operation will terminate immediately and the node will be inserted to the next frontier. As a result, they can reduce useless attempts to check the rest of its neighbors. In addition, each node writes to the frontier queue itself. Therefore, the approach does not require atomic operations to avoid redundant nodes in the queue. 

To overcome the fact that the bottom-up approach is only efficient when a large fraction of nodes is in the frontier, the authors propose a \emph{hybrid} algorithm on multi-core processors which combines both top-down and bottom-up approaches. The method chooses top-down or bottom-up for each iteration by using heuristic switching conditions based on the number of nodes and edges of the frontier, and the edge count from the unvisited nodes. 
Hiragushi et. al. \cite{hiragushi2013efficient} implemented the hybrid approach on GPUs. They propose several optimization techniques to enhance the performance of bottom-up BFS such arranging the adjacency matrix, removing unreachable nodes, and using the texture cache. Their implementation achieved a speedup of up to 29x compared to Merrill et. al.'s implementation.
Enterprise \cite{liu2015enterprise} employs a frontier queue construction which is optimized for top-down, direction-switching, and bottom-up while eliminating duplicated frontier nodes. The method also addresses the GPU workload imbalance by classifying the frontier into different queues based on the out-degrees of frontier nodes. For direction switching, they use the ratio of hub nodes in the queue as a switching condition. The out-degree of a hub node is greater than a threshold $\tau$. 

\textbf{Single-Source Shortest Path:} Most parallel solutions for SSSP are based on two classical Dijkstra and Bellman-Ford algorithms. The traditional serial Dijkstra implementation which utilizes a priority queue can run at $O(|V|log|V| + |E|)$. The algorithm, however, is known as an efficient sequential method which is poorly applied to parallel architecture. 
Although the Bellman-Ford algorithm runs in $O(|V||E|)$ which is much slower than Dijkstra SSSP, the algorithm is considered to be more well-suited to parallel execution because Bellman-Ford operates nodes individually on each iteration.

To implement Dijkstra SSSP on GPUs, Mart et. al. \cite{martin2009cuda} adapted a parallel reduction to find the minimum distance $m$ among all unresolved nodes whose shortest paths we have not found yet. The node $v$ with the tentative minimum distance $d(v)=m$ is called a \emph{frontier node}. The method then marks all nodes in the frontier set as resolved and updates the tentative minimum distances of their neighbors afterward, called \emph{relax operations}. To perform the task, the authors introduce two different kernels which are based on predecessors and successors. 

Instead of using the minimum distance $m$ in each iteration $i$, Crauser et. al. \cite{crauser1998parallelization} introduce a greater limited distance $\Delta_i$. All nodes which have the tentative distances smaller than $\Delta_i$ are inserted to the frontier set. The approach aims to add more unsolved nodes to the frontier set for efficient parallel execution. \cite{ortega2015comprehensive} implement the idea to accelerate SSSP on GPUs. In the preprocessing step,  the minimum weight of edges connected to a node $u$ is calculated, $\omega(u)=min\{w(u,v) | (u,v) \in E \}$. For each iteration, the limited distance is defined as $\Delta_i = min\{d(u) + w(u)\}$, for all unsolved nodes $u$.

Burtscher et. al. \cite{burtscher2012quantitative} implement the Bellman-Ford algorithm on the GPU using the traditional compressed sparse row (CSR) data format. They make an improvement by re-arranging neighboring nodes closer to each other in the data structure. With the improved memory layout, the performance of the SSSP algorithm can be enhanced by several factors.

Harish et. al. \cite{harish2007accelerating} introduce the first GPU implementation of the simple queue-based Bellman-Ford SSSP using CUDA. In each iteration, the approach executes two parallel steps. The first step explores the $marked$ nodes to calculate the distances of their neighbors and update the tentative minimum distances to a temporary array $U$. For all nodes in $V$, the next step compares the current distances of in the distance array $C$ and the new distances in the array $U$. If the distance of a node $v$ in $U$ is smaller than its distance in $C$, the method updates the minimum distance of $v$ to $C$ and marks $v$ as a node to be explored in the next iteration.
Busato et. al. \cite{15_Busato} adopt the idea of the sequential queue-based Bellman-Ford to reduce the number of relax operations. The authors propose a series of optimization techniques to decrease the duplicate nodes in the frontier and deal with the workload imbalance for the GPU-based implementation.

Davidson et. al. \cite{davidson2014work} propose three work-efficient solutions for SSSP on the GPUs. \emph{Workfront Sweep} implements a parallel queue-based Bellman-Ford which eliminates the replicated nodes in the frontier queue. \emph{Near-Far} splits the queue into two sets, namely \emph{Near Set} and \emph{Far Set}, based on an incremental weight $\Delta$ which is similar to the $\Delta$-stepping algorithm. \emph{Bucketing} refines the original $\Delta$-stepping for GPU execution. The method partitions active nodes into a fixed number of buckets. Then it applies Thrust radix sort \cite{bell2011thrust} to re-arrange the buckets by distance in each iteration.

\textbf{All-Pairs Shortest Path:} There are two approaches to solve the APSP problem. For each iteration, the first approach selects a node $s$ in $V$ as the source node and then runs an efficient SSSP algorithm for the source $s$. As a result, this approach executes at $O(|V|^2log|V| + |E||V|)$. The second solution is based on a dynamic programming algorithm, namely Floyd-Warshall (FW), which works at $O(|V|^3)$.

The former approach is first implemented on GPUs by Harish et. al. \cite{harish2007accelerating}. At each iteration of SSSP, the method generates a vector with size $|V|$ to maintain the distances from the source node. The vector then is copied back to the main memory in order to reduce the global memory space. 
Okuyama et. al. \cite{okuyama2012task} propose an improved version of Dijkstra-based APSP by caching data in on-chip shared memory. The method exploits the coarse-grained task parallelism in addition to the fine-grained data parallelism exploited by the previous method in order to share graph data between processing elements in the GPU. As a result, they can reduce data accesses to off-chip memory.
Hector et. al. \cite{ortega2014optimizing} solve the APSP problem based on their SSSP implementation \cite{ortega2015comprehensive}. The authors extend the kernel characterization criteria proposed by \cite{13_Torres} to optimize the GPU-based APSP algorithm. Their method achieves a performance improvement up to 62\% compared with baseline configurations.

Harish et. al. \cite{harish2007accelerating} also propose a Floyd-Warshall algorithm on GPUs. However, their GPU implementation only works well with small-scale graphs due to the high time complexity $O(|V|^3)$ and space complexity $O(|V|^2)$. 
Katz et. al. \cite{katz2008all} adapt the blocked APSP solution proposed by Venkataraman et. al. \cite{venkataraman2003blocked} for GPUs using CUDA. The method represents the data graph as an adjacency matrix and then partitions it into sub-matrices of size $B$x$B$, called \emph{blocks}. At each iteration, a block is picked as the primary block. The method first computes the FW algorithm within the primary block. After that it processes all blocks which share the same row or column as the primary block. Based on these blocks and the primary block, the method updates the distances in parallel. Because of the shared memory and cache-efficient strategy, the tiled FW provides a speedup of 5.0-6.5x compared to the GPU-based method of \cite{harish2007accelerating}. Matsumoto et. al. \cite{matsumoto2011blocked} extend the method to deal with a large graph which cannot fit into the memory of a single GPU. The approach utilizes CPU computation to enhance the GPU-based algorithm. The communication latency between the CPU and the GPU is hidden by utilizing pinned memory and re-using the matrix data within the GPU.

Buluc et. al. \cite{bulucc2010solving} introduce a GPU implementation of a recursively partitioned FW algorithm based on Gaussian elimination. In their method, the $N$x$N$ matrix is recursively split into four equal-size $N/2$x$N/2$ sub-matrices. The main subroutine of the algorithm is related to matrix multiplication between these sub-matrices. The CUDA kernel of the matrix operations is proposed by \cite{volkov2008lu}. For efficient computation on the global memory, the authors modify R-Kleen algorithm \cite{d2007r} for in-place APSP to avoid the expensive cost for copying data.

\textbf{Spanning Tree Construction:} Harish et. al. \cite{harish2009large} propose a modified version of parallel Boruvka MST on CUDA. Their method creates partial minimum spanning trees, called \emph{supernodes}, from all nodes. Supernodes can grow individually and are merged when they come in contact. They employ atomic operations to execute kernels such as finding minimum weighted edges, finding and removing cycles, merging supernodes and assigning colors to nodes.
Vineet et. al. \cite{vineet2009fast} improve the performance of GPU-based Boruvka MST by avoiding atomic operations. They use a series of basic primitives during parallel execution. They apply a scan primitive \cite{sengupta2007scan} to find the minimum weighted edge. They then merge supernodes by using a scalable split primitive \cite{patidar2009scalable}. The split and scan pair is also used to remove edge duplication and assign supernode colors. Their method gains a speed up of 8-10x over \cite{harish2009large}.

Based on an observation that the performance of parallel Boruvka MST relies on the process of minimum edge contraction, which is related to the task of merging the adjacency list of the edge's endpoints. Nasre et. al. \cite{nasre2013morph} propose efficient techniques to execute subgraph addition, deletion, conflict detection and some optimization techniques to improve the performance of the MST algorithm such as employing an adaptive schema to flexibly change kernel configurations. 
Arefin et. al. \cite{arefin2012knn} solve the MST problem on GPUs by proposing a solution which combines the classical Boruvka's algorithm and the k nearest neighbor ($k$NN) graph data structure. The method first generates a $k$NN graph based on the original graph. After that, they employ a divide and conquer approach similar to the work of \cite{harish2009large} to produce the MST using the $k$NN graph.

Rostrup et. al. \cite{rostrup2013fast} introduce a data-parallel Kruskal algorithm. They divide the problem into subproblems with lower memory requirements to address the scalability issue. The subproblems are solved in parallel on the GPU by using Boruvka MST. They also use sort and split primitives and show a possible way to partition the graph into subgraphs.

Nobari et. al. \cite{nobari2012scalable} introduce a parallel version of the serial Prim algorithm on the GPU. The method first executes a tailored version of Prim's algorithm, called \emph{Partial Prim (PP)}, in parallel. It then compacts each connected set of subtrees generated by PP into a single node and removes self-loops. The process is conducted until no more edges are left. This approach can minimize the communication cost between different processors by allowing them produce partial MSTs independently while solving the conflicts without raising extra communication overhead.

\textbf{SCC Decomposition:} Implementing the problem of decomposing a directed graph into strongly connected components on GPUs and multi-core CPUs is not trivial because the state-of-the-art serial algorithm is based on a depth-first search manner. Barnat et. al. \cite{barnat2011computing} have made the first attempt to implement SCC decomposition on the GPUs. They choose CPU-based SCC approaches such as Forward-Backward \cite{mclendon2005finding}, Coloring Head-off \cite{sminia2004distributed} and Recursive OBF \cite{barnat2008improved} and modify them in order to support the GPU execution. Their implementations are based a thread-centric approach which uses a direct mapping between threads and nodes. The drawback of the implementations is the underutilization of the GPU capacity.

Li et. al. \cite{li2014efficient} address the issue by applying a hybrid strategy of coarse-grained expansion and fine-grained expansion to fully utilize the GPU memory and instruction throughput in the core tasks of computing the forward and backward reachability closure. They also present a partition procedure that can greatly reduce the recursion depth and enhance the performance of the divide-and-conquer approach.

FW-BW-Trim algorithm is implemented by Hong et. al. \cite{hong2013fast} for small-world graphs.
Some optimization techniques such as parallelizing the Trim step, avoiding directly modify the graphs with additional data structures, and making use of a work queue are employed during the execution. They propose three extensions of two-phase parallelization, finding weakly connected components and fast detecting size-2 SCCs to the parallel algorithm that take advantage of small-world graph properties to address the deficiencies in existing algorithms.

Wijs et. al. \cite{wijs2014gpu} improve the performance of a GPU-based SCC algorithm by decreasing the memory access latency. Before executing the main routine of the SCC decomposition, they reconstruct the input graph in order to take advantage of coalesced memory accesses. Vertices with the same index in the adjacency lists of the nodes in a group whose size equals to the number of threads in a warp are moved close to each other. As a result, threads within a warp can access consecutive elements in the input array. In the forward and backward reachability closure computation, data is cached in shared memory to reduce the global memory accesses. They also present a parallel pivot selection by taking use of a hashing method.

\subsection{GPU-based Reasoning and Query Processing}

Some studies have recently adopted GPUs to accelerate the time-consuming rule-based reasoning process. All of them, however, follow the forward-chaining inference scheme. Heino and Pan \cite{heino2012rdfs} introduce a GPU implementation of RDFS reasoning. In this work, they solve the problem on six main rules which are divided into two classes: 1) rules that execute only on the schema, and 2) rules that run on both schema and instance triples. The GPU-based algorithm, then, operates these rules based on a pre-defined processing order. The authors also introduce some optimization techniques such as hashing, and parallel sort to eliminate the global and local duplicated triples during execution.

Peters et. al. \cite{peters2013rule} propose a GPU-based method for performing arbitrary rulesets. Their approach is based the \textit{Rete Match} algorithm \cite{forgy1982rete} and adapts vector-based operations for efficient execution on the massively parallel architectures of GPUs. The authors \cite{peters2014scaling} further reduce duplicated triples inferred during beta-matching by using a HashMap. They also partition the workload by splitting the large datasets into smaller trunks for concurrently processing on GPUs.

To support query processing on relational database systems, He et. al. \cite{he2008relational} implement GPU-based joining operations including indexed, non-indexed nested-loop, sort-merge, and hash joins. These operations are based on a set of primitives such as split, sort, map, scatter and gather specially designed for parallel execution. YDB \cite{yuan2013yin} is a GPU-based system for data warehouse query analytics. YDB is built upon a column-based storage format. It creates query plans that run in a push-based, batch-oriented scheme. In this work, they also propose an analytical model to understand and predict the query performance on GPUs. Wang et. al. \cite{wang2014concurrent} present a GPU query engine, called MultiQx-GPU, which is able to support concurrent query processing efficiently. To efficiently share GPUs among concurrent queries for high throughput, they design the system based on GPU query scheduling and device memory swapping policies.

\section{Multimodal Sentiment Analysis}

In this section, we present an overview of state-of-the-art studies in the fields of 1) textual sentiment analysis; 2) audio-video emotion identification; and 3) recent approaches for fusing multimodal features.

\subsection{Text-based Sentiment Analysis}

The techniques developed so far for subjectivity and sentiment analysis have mainly focused on textual data \cite{camcls,camacsa}. These approaches consist of either rule-based techniques that take advantage of opinion lexicon \cite{wiebe2005creating}, or statistical methods that employ a large annotated dataset labeled with emotion \cite{pang2004sentimental}.

For rule-based classifiers, significant works have been conducted to automatically recognize three basic classes of sentiment, namely positive, negative, and neutral, which are associated with different levels, from opinion words \cite{turney2002thumbs} to more linguistically complex phrases \cite{riloff2003learning,wilson2005recognizing}. 
Several methods \cite{alm2005emotions} go beyond this to further explore how to automatically identify sophisticated emotions (for example, anger, sadness, happiness, fear, and surprise) that are either explicitly or implicitly expressed within the textual data.
On the other hand, the data-driven methods make use of large datasets manually annotated for opinions which might cover the domain of product reviews, news articles or newspaper headlines \cite{xiafea}. 
Many supervised as well as unsupervised classifiers are constructed to recognize emotion from textual data \cite{yang2007building}. The SNoW architecture \cite{chaumartin2007upar7} is one of the widely applied frameworks to handle the problem. 

Over the last few years, a large number of researchers \cite{hu2004mining,lin2007emotions,pak2010twitter} have been working on extracting sentiment from texts of various formats, e.g.,  blogs \cite{lin2007emotions}, Twitter messages \cite{pak2010twitter}, and customer reviews \cite{hu2004mining}, etc. Sentiment analysis from social media allows us to make useful predictions in many aspects such as the customers' reactions of a newly released product, or the predicted results of a voting. To accomplish this, there have been some emotion \cite{balahur2012building} and knowledge-­based sentiment \cite{esuli2006sentiwordnet} lexicons for emotion and sentiment analysis at both word and phrase levels. Cambria et. al. \cite{cambria2014senticnet} introduced a novel commonsense knowledge lexicon, SenticNet, which assigns polarity values to 30,000 commonsense knowledge multi-word expressions for concept-level sentiment analysis.

\subsection{Audio-Video Emotion Analysis}

Over the last few years, we have witnessed a lot of researches \cite{busso2007interrelation,song2008robust} in emotion recognition which address the problems of facial expression detection and/or audio affect recognition. Audio affect recognition of speech signal aims to identify the emotional states of humans by analyzing their voices. Many speech-based emotion analyses \cite{johnstone1996emotional,murray1993toward} have given high attention to identifying some important audio features such as bandwidth, duration, fundamental frequency (pitch), intensity of utterance, and Mel frequency coefficients \cite{el2011survey}. To accelerate the performance of audio feature extraction, Michalek et. al. \cite{michalek2014open} released an open-source tool which is based on parallel processing on General Purpose GPUs.

There are also studies that analyze the visual cues such as facial expression and body movement. One of the most significant works on facial expressions has been done by Ekman et. al. \cite{ekman1974universal} in 1974. 
According to this study, universal facial expressions can be sufficiently employed as clues for detecting human emotions. 
They focused on six basic classes of emotions, namely joy, sadness, anger, fear, disgust, and surprise. They also claimed that such categories can sufficiently describe most of the emotions expressed by facial expressions. Ekman et. al. \cite{ekman1977facial} also introduced a coding system, in short FACS, to encode facial expressions by dismantling such an expression into a collection of action units (AUs), where the AUs are identified by using particular movements of facial muscles. 
Some examples of recent methods that utilize FACS to recognize expressed facial expressions are Active Appearance Model \cite{lanitis1995unified} and Active Shape Model \cite{cootes1995active,li2009accelerating}.
Several artificial neural networks based research works \cite{ueki1994expression} have successfully managed to identify sentiments from facial expressions by using the AUs as features.

In addition to the above studies which only focused on individual audio or video modalities, there is a growing body of works that include both video and audio emotion recognition \cite{busso2007interrelation,song2008robust}. The features used by those methods are mainly low level features, such as tracking points for collection visual data, or extracting audio features at pitch level.

\chapter{Commonsense Reasoning via Subgraph Similarity Search}
\graphicspath{{Chapter3/fig/EPS/}{Chapter3/fig/}}
\label{tag:chap3}
\setcounter{equation}{0}

\section{Motivation}

In the context of sentic computing \cite{cambria2013sentic}, commonsense is represented as a semantic network of natural language concepts interconnected by semantic relations. Besides the methodological problem of relevance (selection of relevant nodes during spreading activation), this kind of representation presents two major implementation issues: performance and scalability, both due to the many new nodes, or natural language concepts learnt through crowdsourcing, continuously integrating into the graph. These issues are also crucial problems of querying and reasoning over large-scale commonsense knowledge bases (KBs).

The core function of commonsense reasoning is subgraph similarity search which is defined as finding all embeddings of a small graph in a large database graph. Subgraph similarity search is usually a bottleneck for the overall performance as it involves subgraph isomorphism which is known as an NP-complete problem. Previous methods for subgraph similarity search are backtracking algorithms \cite{han2013turbo,ullmann1976algorithm,cordella2004sub,he2008graphs}, with novel techniques for filtering candidates sets and re-arranging visit order. These algorithms, however, are designed to work only in small-graph settings. The number of candidates grows significantly in medium-to-large-scale graphs, resulting in an exorbitant number of costly verification operations. Recently, Graphics Processing Units (GPUs) have become popular computing devices owing to their massive parallel execution power. Many fundamental graph-based algorithms have been efficiently implemented on GPUs. The previous backtracking methods for subgraph similarity search, however, cannot be straightforwardly applied to GPUs due to their inefficient use of GPU memories and SIMD-optimized GPU multi-processors.

In this chapter, we introduce a GPU-friendly method for subgraph similarity search in the context of commonsense reasoning, called \emph{GpSense}. We first discuss how a commonsense KB can be naturally represented as a graph and how such a KB can be directly transformed to a graph representation. Then, we propose a novel \emph{filtering-and-joining} strategy for the subgraph similarity search problem. The method is specially designed for the massively parallel architecture of GPUs. In order to optimize the performance in depth, we utilize a series of optimization techniques which contribute towards increasing GPU occupancy, reducing workload imbalances and in particular speeding up subgraph similarity search on commonsense graphs.

Most of the commonsense knowledge graphs, however, contain millions to billions of nodes and edges. These huge graphs cannot be stored on the memory of a single GPU device. We may thus have to use main memory and even hard-disk, if necessary, as the main storage of the knowledge graphs. To address the issue, we propose a \emph{multiple-level graph compression} technique to reduce graph sizes while preserving all subgraph search results. The graph compression method converts the data graph to a weighted graph which is small enough to be maintained in GPU memory. We then present a complete GpSense solution which exploits the weighted graph to solve the subgraph similarity search problem.

\section{Commonsense Knowledge as a Graph}
\label{chap3:commonsense}

In this section, we discuss how a commonsense KB can be naturally represented as a graph and how such a KB can be directly transformed to a graph representation.

\subsection{Commonsense Knowledge Graph}

\begin{figure}[htp]
	\centering
	\includegraphics[width=0.7\textwidth]{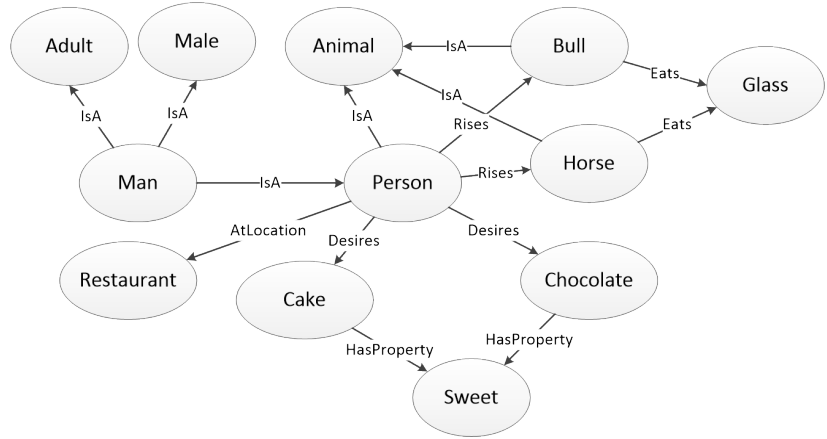}
	\caption{Commonsense knowledge graph}
	\label{fig3:csg}
\end{figure}

Instead of formalizing commonsense reasoning using mathematical logic \cite{mccarthy1986applications}, some recent commonsense KBs, e.g., SenticNet \cite{cambria2014senticnet}, represent data in the form of a semantic network and make it available for use in NLP applications. In particular, the collected pieces of knowledge are integrated into the semantic network as triples, using the format: $<concept$-$relation$-$concept>$. By considering triples as directed labeled edges, the KB naturally becomes a directed graph. Figure~\ref{fig3:csg} shows a semantic graph representation for part of a commonsense knowledge graph.

\subsection{Commonsense Graph Transformation}

\begin{figure}[htp]
	\centering
	\includegraphics[width=0.6\textwidth]{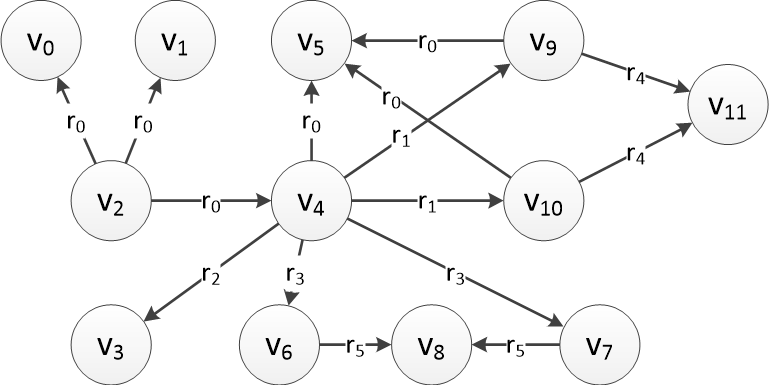}
	\caption{Direct transform of Commonsense KB}
	\label{fig3:transgrp}
\end{figure}

This subsection describes how to directly transform a commonsense KB to a directed graph. The simplest way for transformation is to convert the KB to a flat graph using direct transformation. This method maps concepts to node IDs and relations to labels of edges. Note the obtained graph contains no node labels as each node is mapped to a unique ID. Table~\ref{tab3:node} and~\ref{tab3:edge} show the mapping from concepts and relations of the commonsense KB in Figure~\ref{fig3:csg} to node IDs and edge labels. The transformed graph from the KB is depicted in Figure~\ref{fig3:transgrp}.

\begin{table}[ht]
	\begin{minipage}[b]{0.45\linewidth}\centering
		\centering
		\begin{tabular}{|c|c|}
			\hline
			\textbf{Concept} & \textbf{Node ID} \\ \hline
			Adult            & $v_0$             \\ \hline
			Male             & $v_1$             \\ \hline
			Man              & $v_2$             \\ \hline
			Restaurant       & $v_3$             \\ \hline
			Person           & $v_4$             \\ \hline
			Animal           & $v_5$             \\ \hline
			Cake             & $v_6$             \\ \hline
			Chocolate        & $v_7$             \\ \hline
			Sweet            & $v_8$             \\ \hline
			Bull             & $v_9$             \\ \hline
			House            & $v_{10}$             \\ \hline
			Glass            & $v_{11}$             \\ \hline
		\end{tabular}
		\vspace{5mm}
		\caption {Node Mapping Table}
		\label{tab3:node}
	\end{minipage}
	\hspace{0.5cm}
	\begin{minipage}[b]{0.45\linewidth}
		\centering
		\begin{tabular}{|c|c|}
			\hline
			\textbf{Relation}      & \textbf{Edge Label} \\ \hline
			IsA           & $r_0$   \\ \hline
			Rises     & $r_1$   \\ \hline
			AtLocation    & $r_2$   \\ \hline
			Desires       & $r_3$   \\ \hline
			Eats       & $r_4$   \\ \hline
			HasProperty   & $r_5$   \\ \hline
		\end{tabular}
		\vspace{5mm}
		\caption {Edge Label Mapping Table}
		\label{tab3:edge}
	\end{minipage}
\end{table}

In the general subgraph similarity search problem, all nodes of a query graph $q$ are variables. In order to produce the subgraph isomorphisms of $q$ in a large data graph $g$, we must find the matches of all query nodes. Unlike the general problem, query graphs in commonsense querying and reasoning tasks contain two types of nodes: concept nodes and variable nodes. 

\begin{figure}[htp]
	\centering
	\subfigure[Commonsense query]{\includegraphics[width=0.35\textwidth]{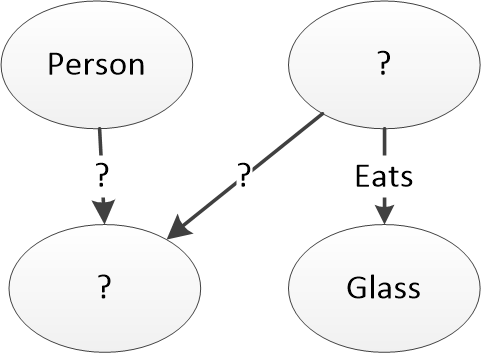}}
	\quad
	\quad
	\subfigure[Transformed query]{\includegraphics[width=0.3\textwidth]{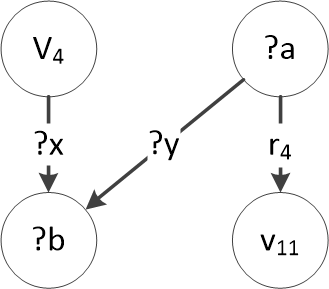}}
	\caption{Direct transformation of Commonsense query}
	\label{fig3:csq}
\end{figure}

A concept node can only be mapped to one node ID in the data graphs while a variable node may have many candidate nodes. Similarly, query edges are also categorized into variable and labeled edges. Figure~\ref{fig3:csq} illustrates the conversion of a commonsense query to a directed query graph.

In the sample query transformation, the query concepts \emph{Person} and \emph{Glass} correspond to two data nodes with IDs of $v_4$ and $v_{11}$. The relation \emph{Eats} is mapped to the edge label $r_4$. The query graph also contains 2 variable edges: $?x$, $?y$ and 2 variable nodes: $?a$, $?b$. The direct transformation is a simple and common approach to naturally convert a semantic network to a directed graph.

\section{Subgraph Similarity Search}
\label{chap3:submatch}

In order to make it easier to follow, we give a formal problem statement using undirected labeled graphs, though our method can be applied to other types of graphs as shown in Section~\ref{chap3:eval}.

\begin{definition}
	A \textbf{\emph{labeled graph}} is a 4-tuple $G = (V, E, L, l)$, where V is the set of nodes, $E \subseteq V \times V$ is the set of edges, $L$ is the set of labels and $l$ is a labeling function that maps each node to a label in $L$.
\end{definition}

\begin{definition}
	A graph $G = (V, E, L, l)$ is \textbf{\emph{subgraph isomorphic}} to another graph $G' = (V', E', L', l')$, denoted as $G \subseteq G'$, if there is an injective function (or a \textbf{\emph{match}}) $f: V \to V'$, such that $\forall (u, v) \in E$, $(f(u), f(v)) \in E'$, $l(u) = l'(f(u))$, and $l(v) = l'(f(v))$.
\end{definition}

\begin{figure}[htp]
	\centering
	\subfigure[Query graph Q]{\includegraphics[width=0.45\textwidth]{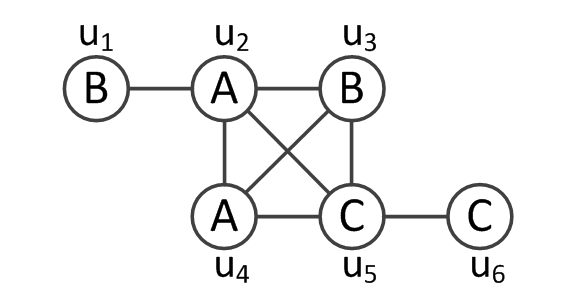}}
	\quad
	\subfigure[Data graph G]{\includegraphics[width=0.45\textwidth]{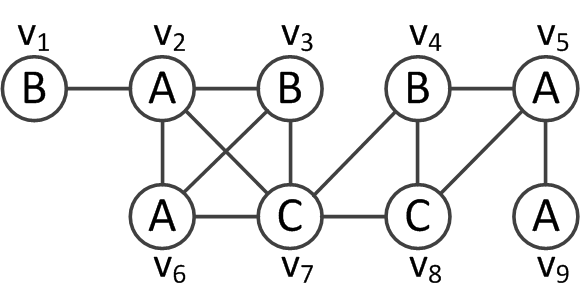}}
	\caption{Sample query and data graph}
	\label{fig3:example}
\end{figure}

Subgraph similarity search problem is defined as follows: Given a large data graph $G$ and a query graph $Q$, we find all matches of $Q$ in $G$. For example, the subgraph similarity search solution of the query graph $Q$ in the data graph $G$ in Figure~\ref{fig3:example} is \{($u_1, v_1$), ($u_2, v_2$), ($u_3, v_3$), ($u_4, v_6$), ($u_5, v_7$), ($u_6, v_8$)\}.

\begin{definition}
	Given a query graph $Q =(V, E, L, l)$ and a data graph $G = (V', E', L', l')$, a node $v \in V'$ is called a \textbf{\emph{candidate}} of a node $u \in V$ if $l(u) = l'(v)$, $degree(u) \le degree(v)$ where $degree(u)$, $degree(v)$ are the number of nodes connected to edges starting node $u$ and $v$ respectively. The set of candidates of $u$ is called \textbf{\emph{candidate set}} of $u$, denoted as $C(u)$.
\end{definition}

The query node $u_3$ in Figure~\ref{fig3:example} has a label of B and a degree of 3. For the data graph node $v_3$ in Figure 1b, the label is also B and the degree is 3 which is equal to the degree of $u_3$. Therefore, $v_3$ is a candidate of $u_3$. The candidate set of $u_3$ is C($u_3$) = \{$v_3$, $v_4$\}.

An \emph{adjacency list} of a node \emph{u} in a graph $G$ is a set of nodes which are the destinations of edges starting from \emph{u}, denoted as \emph{adj(u)}. For example, the adjacency list of $u_3$ is $adj(u_3)$ = \{$u_2$, $u_4$, $u_5$\}.

\subsection{Filtering-And-Joining Strategy}

In this subsection, we introduce a parallel approach to solve the subgraph similarity search problem on General-Purpose Graphics Processing Units (GPGPUs). Before describing the algorithm in detail, we explain how a data graph is represented in memory. In order to support graph query answering on GPUs, we use two arrays to represent a graph $G=(V,E)$: \emph{nodes array} and \emph{edges array}. The edges array stores the adjacency lists of all nodes in $V$, from the first node to the last. The nodes array stores the start indexes of the adjacency lists, where the $i$-th element of the nodes array has the start index of the adjacency list of the $i$-th node in $V$. These arrays have been used in previous GPU-based algorithms \cite{harish2007accelerating,hong2011accelerating,merrill2012scalable}. Two additional arrays with the lengths of $|V|$ and $|E|$ are used to stored the labels of nodes and edges (if capable). 

\begin{figure}[htp]
	\centering
	\includegraphics[width=0.7\textwidth]{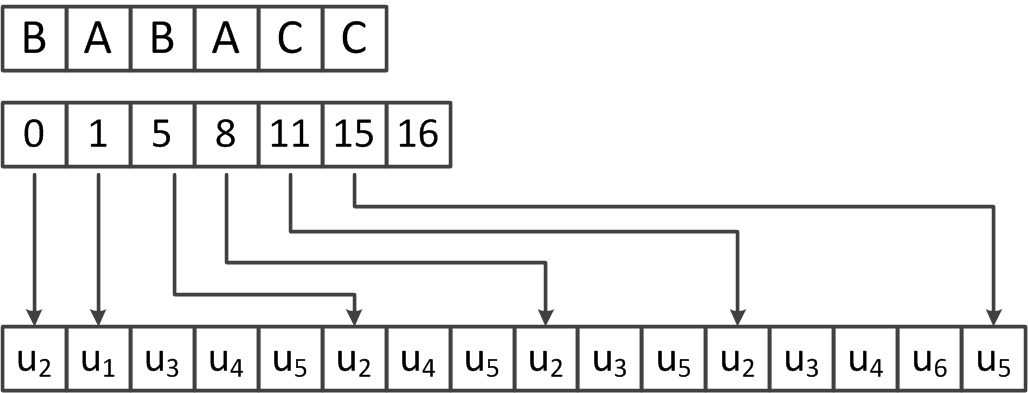}
	\caption{Graph representation of the data graph in Figure~\ref{fig3:example}b}
	\label{fig:repgrp}
\end{figure} 

Based on the above graph structure, we propose a simple and efficient subgraph similarity search algorithm. The approach is based on a novel \emph{filtering-and-joining} strategy which is specially designed for massively parallel computing architectures of modern GPUs. The main routine of the GPU-based method is depicted in Algorithm~\ref{alg3:gpsm}.

The inputs of the algorithm are a query graph $q$ and a data graph $g$. The output is a set of subgraph isomorphisms (or matches) of $q$ in $g$. In the method, we present a match as a list of pairs of a query node and its mapped data node. Our solution is the collection M of such lists. Based on the input graphs, we first generate a query plan for the subgraph similarity search task (Line 1). The query plan contains the order of query nodes which will be processed in the next steps. The query plan generation is the only step that runs on the CPU. The main procedure will then be executed in two phases: filtering phase (Line 2-3) and joining phase (Line 4-6). In the filtering phase, we filter out candidate nodes which cannot be matched to any query nodes (Line 2). 

\begin{algorithm}[htp]
	\label{alg3:gpsm}
	
	\caption{FilteringAndJoining ( q(V, E, L), g(V', E', L') )}
	\SetKwFunction{planer}{generate\_query\_plan}
	\SetKwFunction{init}{initialize\_node\_candidates}
	\SetKwFunction{filter}{identify\_node\_candidates}
	\SetKwFunction{collect}{collect\_node\_candidates}
	\SetKwFunction{explore}{filter\_neighbor\_candidates}
	\SetKwFunction{refine}{refine\_node\_candidates}
	\SetKwFunction{edge}{collect\_edge\_candidates}
	\SetKwFunction{combine}{combine\_edge\_candidates}
	\KwIn{query graph q, data graph g}
	\KwOut{all matches of q in g}
	\BlankLine
	
	P := \planer{q, g};
	
	c\_set := \init{q, g};
	
	\refine{c\_set, q, g};
	
	\ForAll {edge e (u,v) $\in$ E} {
		EC(e) := \edge{e, c\_set, q, g};
	}
	
	M := \combine{EC, q, g};
	
	\KwRet{M}
	
\end{algorithm}

Upon completion of this task, there still exists a large set of irrelevant candidate nodes which cannot contribute to subgraph similarity search solutions. The second task continues pruning this collection by calling the refining function \emph{refine\_node\_candidates}. In such a function, candidate sets of query nodes are recursively refined until no more can be pruned. The joining phase then finds the candidates of all data edges (Line 4-5) and merges them to produce the final subgraph similarity search results (Line 6).

\subsection{Query Plan Generation}

The purpose of this step is to create a node visiting order such that we can reduce the number of candidate nodes during the Filtering phase. Initially, we take a spanning tree generated from the query graph as the input. This subsection presents a heuristic approach to selecting a good spanning tree among many spanning trees of the query graph. The approach is based on the observation that if the filtering starts from the query nodes with the smallest number of candidates, its intermediate results can be kept to the minimum. Since we do not know the number of candidates in the beginning, we estimate it by using a node ranking function $f(u)=\frac{deg(u)}{freq(u.label)}$ \cite{han2013turbo,sun2012efficient}, where \emph{deg(u)} is the degree of \emph{u} and \emph{freq(u.label)} is the number of data nodes having the same label as \emph{u}. 

\begin{figure}[htp]
	\centering
	\subfigure[Spanning tree]{\includegraphics[width=0.4\textwidth]{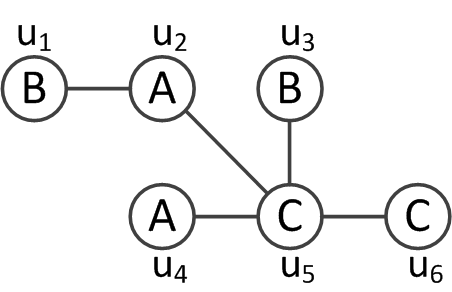}}
	\caption{Spanning tree of Figure~\ref{fig3:example}a graph}
	\label{fig3:spantree}
\end{figure}

We find a spanning tree $T$ and a visit order $O$ for a query graph as follows: Initially, we pick a query edge $(u,v)$ such that $f(u) \geq f(v)$ and $f(u)$ + $f(v)$ is the maximum among all query edges. We add $u$ to the visit order $O$, and add the edges connected to $u$ to the spanning tree $T$, except those whose endpoints are already in the nodes set of $T$, i.e., $V(T)$. The process continues to pick up another query edge connected to $T$ and add to $O$ and $T$ until no edge remains. Figure~\ref{fig3:spantree} depicts the spanning tree of the Figure~\ref{fig3:example}a graph. Also, the visit order is $u_5$, $u_2$.

\subsection{Filtering Phase}

This subsection describes the implementation of the Filtering phase on GPUs. The purpose of this phase is to reduce the number of candidate nodes and thus decrease the number of candidate edges as well as the running time of the Joining phase. The Filtering phase consists of two tasks: initializing candidate nodes and refining candidate nodes.

\subsubsection{Candidate Nodes Initialization}

Algorithm~\ref{alg3:init} outlines the former task of searching for candidate nodes of each query node from the data graph, following the visit order obtained earlier.  For each query node $u$, the algorithm first checks if each of data node is a candidate of $u$ and keeps the candidacy information in the Boolean array $c\_set[u]$ in parallel ($kernel\_check$\footnote{Note that all functions whose names start with $kernel$ are device functions that run on GPUs.}; Line 7) in the case that its candidate set is not initialized (Line 6). It then creates an integer array ($c\_array$) that collects the indexes of candidates of $u$ from $c\_set[u]$ ($kernel\_collect$; Line 9). The algorithm calls another device function ($kernel\_explore$; Line 10) that prunes out all candidate nodes $u'$ of $u$ such that there is a node $v \in adj(u)$ which has no candidate node in $adj(u')$ (Lines 16-18), and explores the adjacency list of $u$ in the spanning tree in order to filter the candidates of the nodes in $adj(u)$ (Lines 19-22). Thus, the final outputs are $Boolean$ arrays $c\_set$, which represent the filtered candidate sets of query nodes.

\begin{algorithm}[ht]
	\label{alg3:init}
	\KwIn{spanning tree $T$, data graph $g$}
	\KwOut{candidate sets of nodes $c\_set$}
	\SetKwFunction{algo}{CandidateNodesFilter}
	\SetKwFunction{bound}{IsBound}
	\SetKwFunction{valid}{IsMatched}
	\SetKwFunction{candidate}{IsCandidate}
	\SetKwFunction{getnode}{GetCandidate}
	\SetKwFunction{getedge}{GetAdjacentVertex}
	\SetKwFunction{kcheck}{kernel\_check}
	\SetKwFunction{kcollect}{kernel\_collect}
	\SetKwFunction{kexplore}{kernel\_explore}
	
	\SetKwProg{myalg}{Algorithm}{}{}
	
	\myalg{\algo{T, g}}{
		\ForEach {node u $\in$ T} {
			$c\_set[u][v]$ := $false$; $\forall v \in V_g$
			
			$initialized[u]$ := $false$;
		}
		
		\ForEach {u $\in$ T in the visit order} {
			\If{initialized[u] = false} {
				\kcheck{c\_set[u], g};
				
				$initialized[u]$ := $true$;
			}
			
			$c\_array$ := \kcollect{u, c\_set[u]};
			
			\kexplore{u, c\_array, c\_set, T, g};    
			
			\ForEach{v $\in$ adj(u) }{
				$initialized[v]$ := $true$;
			}
		}
		
		\KwRet{c\_set};
	}{}
	
	\SetKwProg{myproc}{Procedure}{}{}
	
	\myproc{\kexplore{u, c\_array, c\_set, T, g}}{
		$u'$ := \getnode{c\_array, warp\_id};
		
		\If {exist $v \in adj(u)$ such that no $v' \in adj(u')$ is a candidate of $v$} {
			$c\_set[u][u']$ := $false$;
			
			\KwRet{};
		}
		
		\ForEach{v $\in$ adj(u) }{
			$v'$ := \getedge($u', thread\_id$);
			
			\If {$v'$ is a candidate of $v$} {
				$c\_set[v][v']$ := $true$;
			}
		}
	}
	\caption{Candidate nodes initialization}
\end{algorithm}

\textbf{GPU implementation:} We implement the two GPU device functions \emph{kernel\_collect} and $kernel\_explore$ in the first step of the filtering phase, based on two optimization techniques: \emph{occupancy maximization} to hide memory access latency and \emph{warp-based execution} to take advantage of the coalesced access and to deal with workload imbalance between threads within a warp. We skip details of the device function $kernel\_check$ since its implementation is straightforward.

1) \textit{$kernel\_collect$}. This function is to maximize the occupancy of the \emph{kernel\_explore} execution. At runtime, warps currently running in an SM are called active warps. Due to the resource constraints, each SM allows a maximum number of active warps running concurrently at a time. Occupancy is the number of concurrently running warps divided by the maximum number of active warps. At runtime, when a warp stalls on a memory access operation, the SM switches to another active warp for arithmetic operations. Therefore, high-occupancy SM is able to adequately hide access latency.

\begin{figure}[htp]
	\centering
	\includegraphics[width=0.8\textwidth]{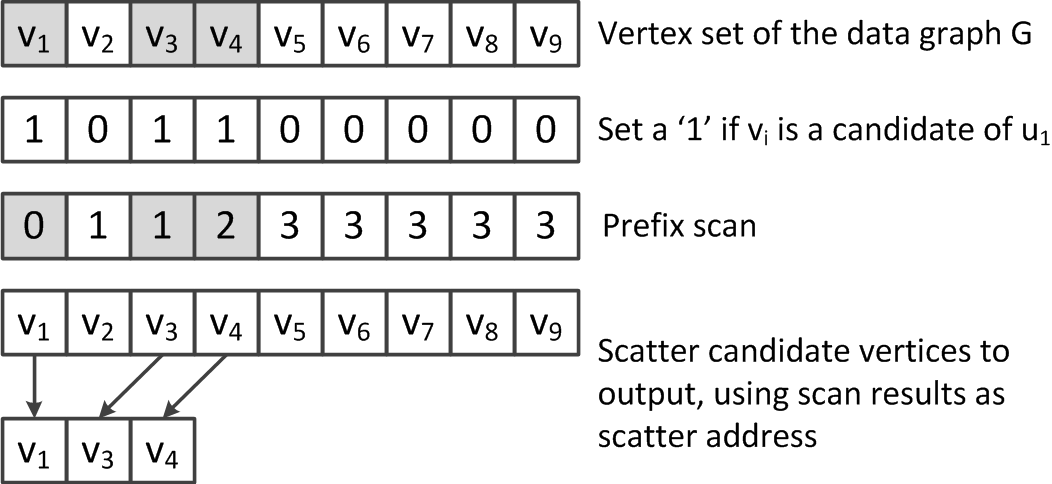}
	\caption{Collect candidate nodes of $u_1$}
	\label{fig3:collect}
\end{figure}

A naive approach to executing $kernel\_explore$ is that only the warps corresponding to the $true$ elements of $c\_set[u]$ continue filtering nodes in $adj(u)$. However, the approach suffers from the low-occupancy problem since warps with the $false$ elements are idle. For example, we assume that the maximum number of active warps on the multiprocessor is 3. In the first 3 active warps, the occupancy is 66.66\% because only the warps corresponding to $v_1$ and $v_3$ execute $kernel\_explore$ while the warp with $v_2$ is idle. For the next 3 concurrently running warps, the occupancy is only 33.33\%. Our method resolves the issue by adopting a stream compaction algorithm \cite{harris2007gpu} to gather candidate nodes into an array $c\_array$ for those $c\_set[u]$ with true values. The algorithm employs prefix scan to calculate the output addresses and to support writing the results in parallel. The example of collecting candidate nodes of $u_1$ is depicted in Figure~\ref{fig3:collect}. By taking advantage of $c\_array$, all 3 active warps are used to explore the adjacency lists of $v_1$, $v_3$ and $v_4$. As a result, our method achieves a high occupancy.

\begin{figure}[htp]
	\centering
	\includegraphics[width=0.9\textwidth]{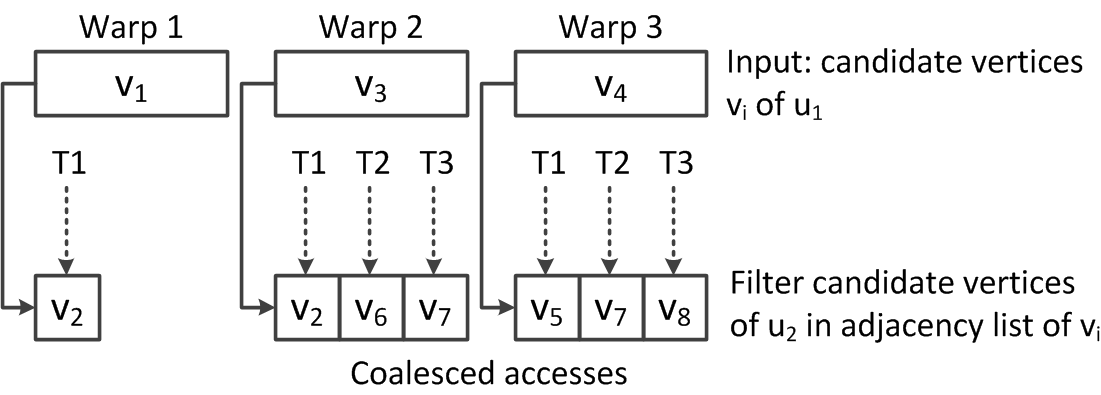}
	\caption{Filter candidate nodes of $u_2$ based on adjacency lists of $C(u_1)=\{v_1, v_3, v_4\}$}
	\label{fig3:explore}
\end{figure}

2) \textit{$kernel\_explore$}. Inspired by the warp-based methods used in BFS algorithms for GPUs \cite{hong2011accelerating}, we assign to each warp a candidate node $u' \in C(u)$ (or $c\_array$ from $kernel\_collect$). Within the warp, consecutive threads find the candidates of $v \in adj(u)$ in $adj(u')$. This method takes advantage of coalesced access since the nodes of $adj(u')$ are stored next to each other in memory. It also addresses the warp divergence problem since threads within the warp execute similar operations. Thus, our method efficiently deals with the workload imbalance problem between threads in a warp. Figure~\ref{fig3:explore} shows an example of filtering candidate nodes of $u_2$ based on the candidate set of $u_1$, $C(u_1)=\{v_1, v_3, v_4\}$. 

If a data node has an exceptionally large degree compared to the others, out method deals with it by using an entire block instead of a warp. This solution reduces the workload imbalance between warps within the block.

\subsubsection{Candidate Nodes Refinement}

After filtering out candidate nodes for the first time, there can be still a large number of candidate nodes which cannot be parts of final solutions. To address this issue, we propose a recursive filtering strategy to further prune irrelevant candidate nodes. The size of candidate edges and intermediate results are then reduced consequently.

\begin{figure}[htp]
	\centering
	\subfigure[Simplified graph]{\includegraphics[width=0.4\textwidth]{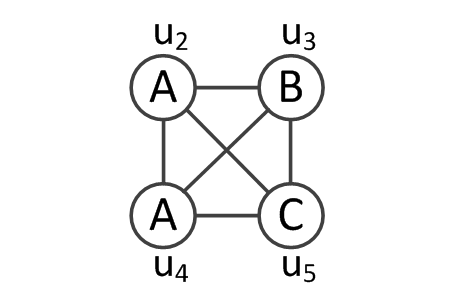}}
	\caption{Simplified graph of Q}
	\label{fig3:query}
\end{figure}

We observe the followings: 1) Exploring non-tree edges (i.e., those that form cycles) can reduce the number of irrelevant candidates significantly; and 2) the more edges a node has, the more irrelevant candidates of the node the filtering techniques aforementioned can filter out. Based on the first observation, from the second round of the filtering process, our method uses the original query graph for exploration rather than a spanning tree of the query graph. Based on the second observation, our method ignores query nodes connected to a small number of edges, called \emph{low connectivity nodes}. For small-size query graphs, a low connectivity node has the degree of 1. As for big query graphs, we can increase the value of degree threshold to ignore more low connectivity nodes. The query graph obtained after removing low connectivity nodes from $Q$ is shown in Figure~\ref{fig3:query}.

\textbf{GPU implementation:} The main routine of the refining task is similar to the filtering in the previous section. The differences are as follows: 1) $kernel\_check$ is not necessary for the refining process and 2) we only use the pruning task (Lines 16-18) in the $kernel\_explore$ function. By taking advantage of the $c\_set$ array generated in the initialization step, the refinement can verify the candidate conditions easily and reduce the random accesses during the candidate verification.

Ideally, the optimal candidate sets of query nodes are obtained when the refinement is recursively invoked until no candidate is removed from the candidate sets. However, our experiments show that most of the irrelevant candidates are pruned in the first few rounds. The later rounds do not prune out many candidates, but lead to inefficiency and reduce the overall performance. Therefore, the refining task terminates after a limited number of rounds. 

In the tasks of initializing and refining candidate sets of query nodes, our method requires $O(|V_q| \times |V_g|)$ space to maintain $Boolean$ arrays which are used to collect candidate nodes and $O(|V_g|)$ space to keep the collected set. Let $S$ be the number of SMs. Each SM has $P$ active threads. For each visited node, the prefix scan in $kernel\_collect$ executes in $O(|V_g|\times log(|V_g|)/(S\times P))$ time while $kernel\_explore$ runs in $O(|V_g| \times |d_g| / (S\times P))$, where $d_g$ is the average degree of the data graph. Assume that the candidate refinement stops after $k$ rounds, the total time complexity of the filtering phase is $O(|V_q| \times k \times (|V_g|\times log(|V_g|) + |V_g| \times |d_g|)/(S\times P))$.

\subsection{Joining Phase}

In the joining phase, our method first gathers candidate edges in the data graph and then combines them into subgraph similarity search solutions.

\begin{figure}[htp]
	\centering
	\includegraphics[width=0.5\textwidth]{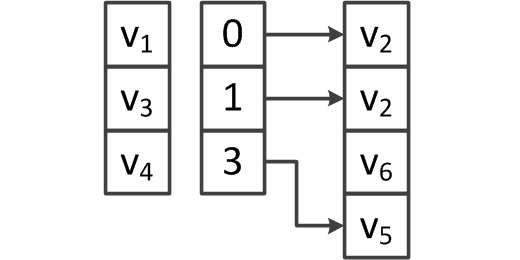}
	\caption{Candidate edges of $(u_1, u_2)$}
	\label{fig3:hashtable}
\end{figure}

The output of each query edge $(u,v)$ in the task of gathering candidate edges is represented as a hash table, as depicted in Figure~\ref{fig3:hashtable}. The keys of this table are candidate nodes $u'$ of $u$, and the value of a key $u'$ is the address of the first element of the collection of candidate nodes $v'$ of $v$ such that $(u',v') \in E_g$. An issue of the step is that the number of the candidate edges is unknown, and thus that we cannot directly generate such a hash table. To address this issue, we employ the \emph{two-step output scheme} \cite{he2008mars} as follows: 1) Given a query edge $(u,v)$, each warp is assigned to process a candidate node $u'$ of $u$ and counts the number of candidate edges starting with $u'$ (designated as $(u',v')$). The system then computes the address of the first $v'$ for $u'$ in the hash table of $(u,v)$. 2) It then re-examines the candidate edges and writes them to the corresponding addresses of the hash table. 

After finding the candidate edges, our method combines them to produce subgraph similarity search solutions as follows: Initially, we pick a query edge $(u,v)$ with the smallest number of candidate edges, and mark as \emph{visited} the nodes $u$, $v$ and the edge $(u,v)$. Here the candidates of $(u,v)$ are partial subgraph similarity search solutions. We select the next edge  among the unvisited edges of the query graph, denoted by $(u',v')$, such that 1) both $u'$ and $v'$ are visited nodes, or 2) if there is no such edge, either $u'$ or $v'$ is a visited node. If there are multiple such edges, we select the one with the smallest number of candidates. Candidate edges of $(u',v')$ are then combined with the partial solutions. The procedure is conducted repeatedly until all query edges are visited.

\textbf{GPU implementation:} The GPU implementation for the task of gathering candidate edges is similar to that of the filtering phase, except introducing the two-step output scheme. For the task of combining partial subgraph similarity search solutions, we apply the warp-based approach as follows: Each warp $i$ is responsible for combining a partial solution $M_i(q)$ with candidate edges of $(u,v)$, where $u$ is already visited. First, the warp retrieves the candidate node of $u$ from $M_i(q)$ (e.g., $u'$). It looks up the hash table storing candidate edges of $(u,v)$ to find the key $u'$ and retrieve the candidate nodes $v'$ of $v$ from the hash table. By using our data structure of candidate edges, this task can be done in logarithmic time. Threads within the warp then verify whether $(u',v')$ can be merged to $M_i(q)$, in which our method again follows the two-step output scheme to write the merged results.

\textit{Shared memory utilization.} The threads within the warp $i$ should share the partial solution $M_i(q)$ and access them frequently. We thus store and maintain $M_i(q)$ in the shared memory instead of the device memory, which efficiently hides the memory stalls.

Let $C(e_i)$ be the candidate edges of the edge $e_i$. The joining phase is done in $O(\prod_{i=1}^{|E_q|} |C(e_i)| \times log(|V_g|) / (S \times P)$ time. Note that the running time of the joining phase highly depends on the number of candidates of query edges. Therefore, reducing the number of candidate nodes in the filtering phase plays an important role in decreasing both the running time and the memory used to maintain partial solutions.

\textbf{Issues with large-scale commonsense reasoning:} Despite the fact the algorithm can deal with subgraph similarity search on general graphs efficiently, there still remain a number of issues for applying the approach to commonsense reasoning, specifically: 1) Unlike query graphs in general subgraph similarity search problems, commonsense query graphs contain concept nodes and variable nodes. We only need to find the matches of nodes in a subset of variable nodes, termed \emph{projection}; 2) Many commonsense knowledge graphs contain millions to billions of nodes and edges. These huge graphs cannot be stored on the memory of a single GPU device. We may thus have to use main memory and even a hard-disk, if necessary, as the main storage of knowledge graphs.

To overcome these issues, the next section introduces a graph compression method to decrease the size of data graphs. Following this, we describe the complete implementation of our method and a series of optimization techniques to enhance the performance of commonsense reasoning.

\section{Multi-level Graph Compression}
\label{sec:compress}

Due to the large size of the data graph, it cannot be maintained within the memory of a single GPU device. The next offline computation aims to reduce the data graph size such that we can fit it into GPU memory while still preserving the subgraph similarity search solutions of any query graphs in the original data graph. In a randomly labeled graph, the distribution of nodes and edges are unpredictable. However, a commonsense knowledge graph contains a lot of similar nodes which share the same group of nodes in their adjacency lists. For example, $v_0$ and $v_1$ of the data graphs in Figure~\ref{fig3:transgrp} are similar nodes  they have the same adjacency list. As a result, the two nodes play the same role in the data graphs and can be combined into one hyper-node.

Based on the above observation, we apply a \emph{multi-level compression} technique to compress the data graph. During the graph compressing process, a sequence of smaller graphs $G_i = (V_i,E_i)$ are constructed from the original graph $G=(V,E)$. At each level $i$, similar nodes are combined to form a weighted node which is defined later. The set of nodes which are combined into the weighted node $u$ after $i$ levels called the \emph{mapping list} of $u$, denoted as $M(u)$. The compressing task terminates when the size of $G_i$ is small enough to be maintained in GPU memory, as depicted in Figure~\ref{fig3:compress}. The final mapping lists are stored in main memory. At each label $i$, graph $G_i$ is a weighted graph which is defined as follows:

\begin{figure}[htp]
	\centering
	\includegraphics[width=0.6\textwidth]{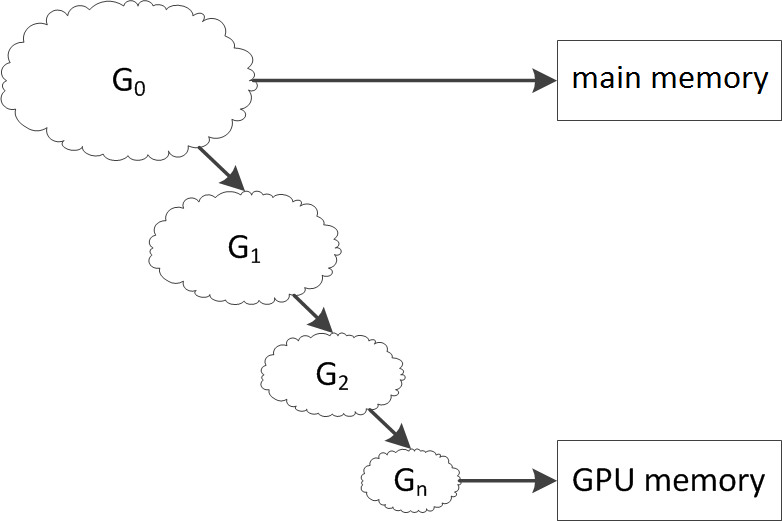}
	\caption{Multi-level graph compression}
	\label{fig3:compress}
\end{figure}

\begin{definition}
	A \emph{weighted graph} at level $i$ is a 5-tuple $G_i=(V_i,E_i,L,l,w)$ where $V_i$ is the set of nodes, $E_i$ is the set of edges, $L$ is the set of labels, $l$ is a labeling function that maps each node to a label in $L$ and $w$ is a weighting function that maps each node or edge to an integer value.
\end{definition}

Each \emph{weighted node} $u \in V_i$ is a combination of $p,q \in V_{i-1}$ and $w(u)$ = max($|\{adj(x)$ $\bigcap$ $(M(p)$ $\bigcup$ $M(q))$ $|$ $x$ $\in M(p)$ $\bigcup$ $M(q)\}|$). Generally, the weight of node $u$ is the maximum degree among nodes in the graph constructed by $M(p)$ $\bigcup$ $M(q)$.

For each \emph{weighted edge} ($u,v$) starting from $u$, if $v \in V_i$ is a combination of $n,m \in V_{i-1}$ then $w(u,v)$ = max($w(p,n), w(q,n)$) + max($w(p,m), w(q,m)$). Note the initial weight of all edges in the original graph is 1.

An edge to/from $v$ is called a \emph{common edge} two nodes $u_1$ and $u_2$ if there exists two edges $(u_1, v)$ and $(u_2, v)$ such that $l(u_1, v)$ = $l(u_2, v)$ = $l_u$, denoted as $e(l_u, v)$. In the Figure~\ref{fig3:transgrp}, $e(r_0, v_2)$ is a common edge of $v_0$ and $v_1$. The list of common edges between $u_1$ and $u_2$ is denoted as $common(u_1, u_2)$.

Given a user-defined threshold $\delta$ such that 0 $< \delta \leq$ 1, $u$ and $v$ are called \emph{similar nodes} if $max(|adj(u)|/|common(u, v)|$,  $|adj(v)|/|common(u, v)|) \geq \delta$. These similar nodes, thus, can be combined into a hyper-node in the next graph compression level. By using $\delta$, we can easily adjust the ratio of graph compression at each level.

\begin{table}[htp] \centering
		\centering
		\begin{tabular}{|c|c|}
			\hline
			\textbf{Weighted Nodes} & \textbf{Mapping List} \\ \hline
			$u'_0$                      & $v_0, v_1$                     \\ \hline
			$u'_1$                      & $v_2$                     \\ \hline
			$u'_2$                      & $v_3$                     \\ \hline
			$u'_3$                      & $v_4$                     \\ \hline
			$u'_4$                      & $v_5$                     \\ \hline
			$u'_5$                      & $v_6, v_7$                     \\ \hline
			$u'_6$                      & $v_8$                     \\ \hline
			$u'_7$                      & $v_9, v_{10}$                     \\ \hline
			$u'_8$                      & $v_{11}$                     \\ \hline
		\end{tabular}
		\vspace{5mm}
		\caption{Mapping list of nodes}
		\label{tab:map}
\end{table}

\begin{figure}[htp]
	\centering
	\includegraphics[width=0.5\textwidth]{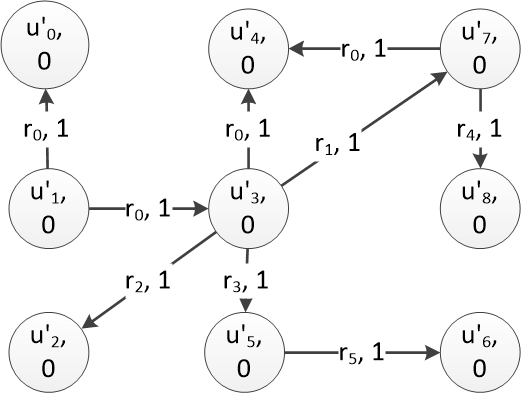}
	\caption{A sample weighted data graph}
	\label{fig3:weightgrp}
\end{figure}

Assume that the data graph is the commonsense knowledge graph in Figure~\ref{fig3:transgrp}. After the first level of data graph compression with $\delta$ of 1, we obtain a sample weighted data graph $G_1$ as in Figure~\ref{fig3:weightgrp} and a mapping list in Table~\ref{tab:map}. Each node is presented as a circle with a label and a weight. At this level, we combine the following pairs of nodes into weighted nodes: $(v_0,v_1)$, $(v_6,v_7)$, $(v_9,v_{10})$. The mapping lists of nodes in $G_1$ are illustrated in Table~\ref{tab:map}. For the real commonsense knowledge graph, i.e., \emph{SenticNet}, the compression ratio is illustrated in Table~\ref{tab:ratio}. The ratio is calculated as the total number of nodes and edges of the compressed graph divided by that of the original graph.

\begin{table}[htp]
	\centering
	\begin{tabular}{|c|c|c|}
		\hline
		\textbf{Level} & \textbf{Threshole $\delta$} & \textbf{Ratio} \\ \hline
		1              & 0.8                & 61.4\%         \\ \hline
		2              & 0.7                & 46.2\%         \\ \hline
		3              & 0.7                & 32.2\%         \\ \hline
	\end{tabular}
	\vspace{5mm}
	\caption{Compression ratio of SenticNet}
	\label{tab:ratio}
\end{table}

The weighted graph $G_w$, which is obtained after reducing the size of the original data graph, is used for checking subgraph similarity search solutions of given query graphs. Due to the differences in graph structures of $G_w$ and the original data graph $G$, we can re-define candidates (or matches) of a query node, as follows:

\begin{definition}
	Given a query graph $Q =(V, E, L, l)$ and a weighted data graph $G_w=(V_w,E_w,L_w,l_w,w)$, a node $v \in V_w$ is considered as a \emph{candidate} of a node $u \in V$ if $l(u) = l_w(v)$, degree(u) $\le$ w(v) + $\sum$ w(v,z) where $z \in adj(v)$, denoted as weight(z).
\end{definition}

For example, node $u_7'$ is a candidate of $?a$ in the query graph in Figure~\ref{fig3:csq}b since $degree(?a)$ = 2 which is smaller than $w(u_7')$ + $w(u_7',u_4')$ + $w(u_7',u_8')$ = 2. Similarly, $u_1'$ and $u_3'$ are also candidate nodes of $?a$.

\begin{theorem}
	Given a query graph $Q=(V_q,E_q)$, a data graph $G=(V,E)$ and a weighted graph $G_w=(V_w,E_w)$ which is the compression result of $G$. If a node $v \in V$ is a candidate of node $u \in V_q$ then node $x \in V_w$ such that $v \in M(x)$ is also a candidate of $u$.
\end{theorem}

\emph{Proof.} We need to prove two conditions: 1) $u$, $v$, and $x$ have the same label because $v$ is a candidate of $u$ and $v \in M(x)$. 2) Based on the definition of weighted graphs, we can see that degree(v) $\le$ w(x) + $\sum$ w(x,z) where $z \in adj(x)$ or weight(x). Therefore, degree(u) $\le$ weight(x). As a result, $x$ is a candidate of $u$.

\begin{theorem}
	For each node $u \in V_q$, if node $z \in W_w$ is not a match of $u$ in any subgraph similarity search solution of $Q$ in $G_w$ then all nodes $v \in M(z)$ are not  matches of $u$ in any subgraph similarity search solution of $Q$ in $G$.
\end{theorem}

\emph{Proof.} We prove by contradiction. Suppose that there exists a node $v \in Q$ which is in a subgraph similarity search solution of $Q$ in $G$, but node $z$ is such that $v \in M(z)$ is not. According to the definition of the above subgraph isomorphism, there is an injective function f: $V_q \to V$ such that $\forall$ (x, y) $\in$ $E_q$, (f(x), f(y)) $\in$ $E$, l(x) = l(f(x)), l(y) = l(f(y)), and v = f(u). We can see that $\forall$ $(a,b) \in E$, $a \in M(p)$, $b \in M(q)$, $(p,q) \in E_w$. Let a function g: $V_q \to V_w$ such that $\forall$ $x \in V_q$. $x \in M(g(x))$. Clearly, $f \circ g$ is a subgraph isomorphism from $Q$ to $G_w$ and $z = f \circ g (u)$. This contradicts that $z$ is not in any subgraph similarity search solution.

\section{GpSense Framework}
\label{sec:gpsense}

\begin{figure}[htp]
	\centering
	\includegraphics[width=0.6\textwidth]{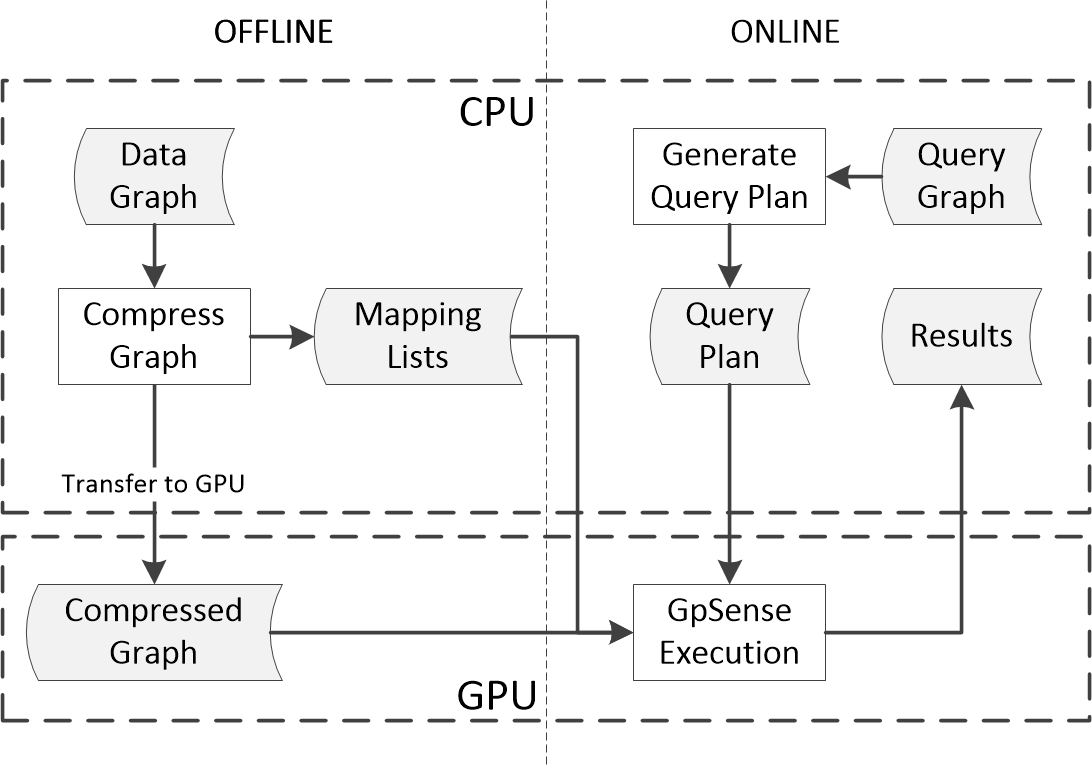}
	\caption{GpSense overview}
	\label{fig:gpsense}
\end{figure}

Based on the multi-level graph compression method introduced in the previous section, we propose a complete algorithm for subgraph similarity search on large-scale commonsense knowledge graphs using GPUs. Figure~\ref{fig:gpsense} gives us an overview of the proposed method, termed \emph{GpSense}, for subgraph similarity search on large commonsense graphs, which cannot fit the global memory of a single GPU device, using both GPUs and CPUs. Rectangles denote tasks while the others represent data structures used in the method.

Our \emph{GpSense} subgraph similarity search solution comprises two separate tasks: an offline task containing graph compression, and online query answering. Initially, the data graph $G$ is stored in the main memory due to its large size. For offline processes, we start by creating a data structure for the input data graph, as described in Section~\ref{chap3:submatch}. The data graph can be maintained in a hard-disk or main memory depending on the size of the data graph and main memory. Assuming we use main memory as the storage of the created index, we then compress the data graph using a multiple-level approach until the obtained graph $G'$ can fit into GPU memory. All mapping lists are also maintained in the main memory. The compressed data graph $G'$ , then, is transferred to GPU memory and stored for GPU execution.

In the online query answering task, after receiving a graph query $Q$, $GpSense$ generates a query plan for the input query graph. The obtained query plan is then transferred to GPU memory. Following this, our method applies the Algorithm~\ref{alg3:gpsm} on the weighted graph achieved by the graph compression step, to find the subgraph similarity search results on the GPU. If no solution is found, we can conclude there is no subgraph similarity search solution from $Q$ to $G$. Otherwise, based on the achieved solutions and the in-memory mapping lists, we continue searching for the final subgraph similarity search solutions of $Q$ in $G$.

Algorithm~\ref{alg3:gpsm}, however, is designed for solving the subgraph similarity search on a general graph. In order to adapt the algorithm to commonsense reasoning, we introduce some optimization techniques to enhance the performance of \emph{GpSense} on large-scale commonsense knowledge graphs as follows:

\textbf{Modify the query plan} based on the properties of commonsense queries. First, unlike query graphs in general subgraph similarity search problems, commonsense query graphs contain concept nodes and variable nodes. We only need to find the matches of nodes in a subset of variable nodes, termed \emph{projection}. Second, nodes of a commonsense knowledge graph are not labeled but mapped to node IDs. Therefore, the frequency of a concept node in a query is 1 and that of a variable node is equal to the number of data nodes. As a result, the ranking function used for choosing the node visiting order cannot work for commonsense subgraph similarity search.

Based on the above observations, we can make a modification to generate the node order as follows: we prefer picking a concept node $u$ with the maximum degrees as the first node in the order. By choosing $u$, we can minimize the candidates of variable nodes connected to $u$. The next query node $v$ will be selected if $v$ is connected to $u$ and the adjacency list of $v$ consists of the maximum number of nodes which is not in the order among the remaining nodes. We continue the process until edges connected to nodes in the node order can cover the query graph.

\textbf{Employ both incoming and outgoing graph representations:} An incoming graph is built based on the incoming edges to the nodes while an outgoing graph is based on the outgoing edges from the nodes. The representation of Commonsense graph in Figure~\ref{fig:repgrp} is an example of outgoing graph representation. Given a query graph in Figure~\ref{fig3:csq}, we assume using only an outgoing graph as the data graph. Based on the above query plan generator, node $v_4$ is the first node in the order. We then filter the candidates of $?b$ based on $v_4$. Since $?b$ does not have any outgoing edges, we have to pick $?a$ as the next node and find its candidates by scanning all the data graphs. There are, however, some issues with this approach: 1) We need to spend the time to scan all the data graph nodes. 2) The number of candidates can be very large as the filtering condition is weak. To overcome this problem, we use an incoming graph along with the given outgoing graph. By using the additional graph, candidates of $?a$ can be easily filtered based on the candidate set of $?b$. The number of candidates of $?a$, therefore, is much smaller than that in the previous approach. Consequently, GpSense can reduce many of the intermediate results during execution, which is a key challenge for GPU applications.  

\textbf{Only use one-time refinement:} Ideally, the optimal candidate sets of query nodes are obtained when the refinement is recursively invoked until no candidate is removed from the candidate sets. However, our experiments show most irrelevant candidates are pruned in the first round. The later rounds do not prune out many candidates, but lead to inefficiency and reduce the overall performance. Also, we observe that if the node visiting order is reversed during the refinement, GpSense is more efficient in terms of minimizing the intermediate data, as well as in improving performance.

\section{Performance Evaluation}
\label{chap3:eval}

We evaluate the performance of GpSense in comparison with state-of-the-art subgraph similarity search algorithms, including VF2 \cite{cordella2004sub}, QuickSI (QSI) \cite{shang2008taming}, GraphQL (GQL) \cite{he2008graphs} and Turbo$_{ISO}$ \cite{han2013turbo}. The experiments are conducted on SenticNet and its extensions \cite{cambria2014senticnet,poria2012merging}. The query graphs are extracted from the data graph by picking a node in SenticNet and following breadth-first search (BFS)  to select other nodes. We choose nodes in the dense area of SenticNet to ensure the obtained queries are not just trees.

The runtime of the CPU-based algorithms is measured using an Intel Core i7-870 2.93 GHz CPU with 8GB of memory. Our GPU algorithms are tested using the CUDA Toolkit 6.0 running on NVIDIA Tesla C2050 GPU with 3 GB global memory and 48 KB shared memory per Stream Multiprocessor. For each of those tests, we execute 100 different queries and record the average elapsed time. In all experiments, algorithms terminate only when all subgraph similarity search solutions are found.

\subsection{Comparison with state-of-the-art CPU algorithms}

The first set of experiments is to evaluate the performance of GpSense on SenticNet and compare it with state-of-the-art algorithms. SenticNet is a commonsense knowledge graph of about 100,000 nodes, which is primarily used for sentiment analysis. In this experiment, we extract subsets of SenticNet with the size varying from 10,000 to 100,000 nodes. All the data graphs can fit into GPU memory. The query graphs contain 6 nodes.

\begin{figure}[htp]
	\centering
	\includegraphics[width=0.7\textwidth]{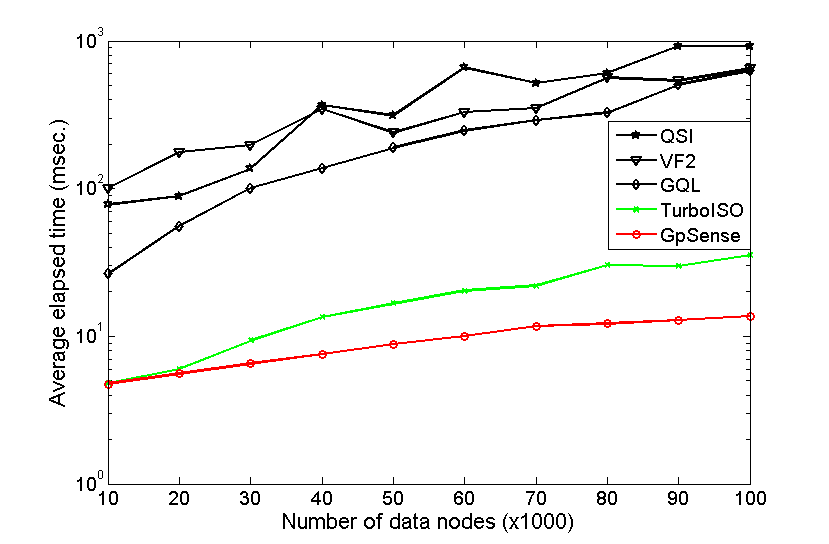}
	\caption{Comparison with state-of-the-art methods}
	\label{fig:compare}
\end{figure}

Figure~\ref{fig:compare} shows that GpSense clearly outperforms VF2, QuickSI, and GraphQL. Compared to Turbo$_{ISO}$, our GPU-based algorithm obtains similar performance when the size of the data graphs is relatively small (i.e., 10,000 nodes). However, when the size of data graphs increases, GpSense is more efficient than Turbo$_{ISO}$.

We also perform out method using other real datasets, namely Gowalla and Enron. Gowalla network consists of 196,591 vertices and 950,327 edges while Enron network has 36,692 vertices and 183,831 edges. In these experiments, we use 20 labels for Gowalla network and 10 labels for Enron network. The number of query vertices varies from 6 to 13.

\begin{figure}[htp]
	\centering
	\includegraphics[width=0.7\textwidth]{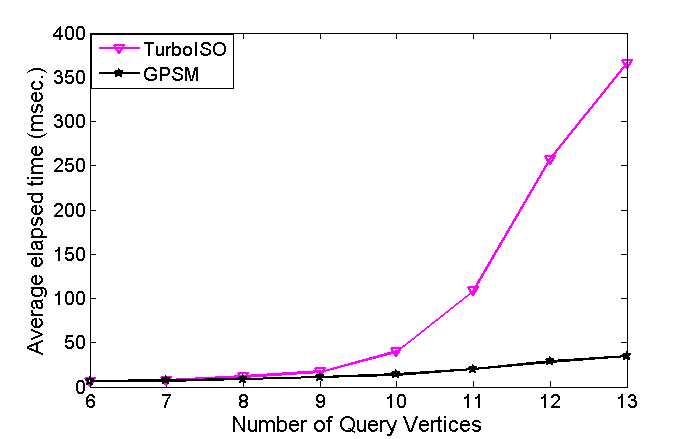}
	\caption{Experiment on Gowalla datasets}
	\label{fig3:Gowalla}
\end{figure}

\begin{figure}[htp]
	\centering
	\includegraphics[width=0.7\textwidth]{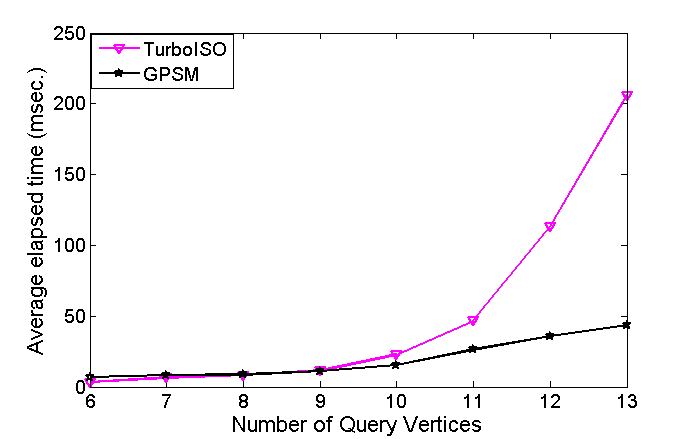}
	\caption{Experiment on Enron datasets}
	\label{fig3:Enron}
\end{figure}

Figure~\ref{fig3:Gowalla} shows that Turbo$_{ISO}$ answers the subgraph matching queries against the Gowalla network efficiently when the size of query graphs is small. As the number of vertices increases, however, the processing time of Turbo$_{ISO}$ grows exponentially. In contrast, GpSM shows almost linear growth. The two methods show similar performance difference when evaluated against the Enron network, as plotted in Figure~\ref{fig3:Enron}.

\begin{figure}[htp]
	\centering
	\subfigure[Varying query sizes]{\includegraphics[width=0.7\textwidth]{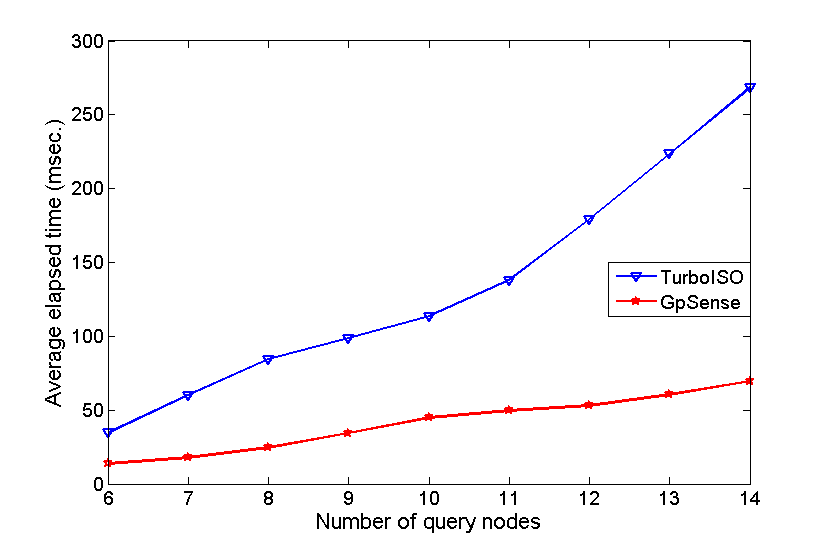}}
	\quad
	\subfigure[Varying average degrees]{\includegraphics[width=0.7\textwidth]{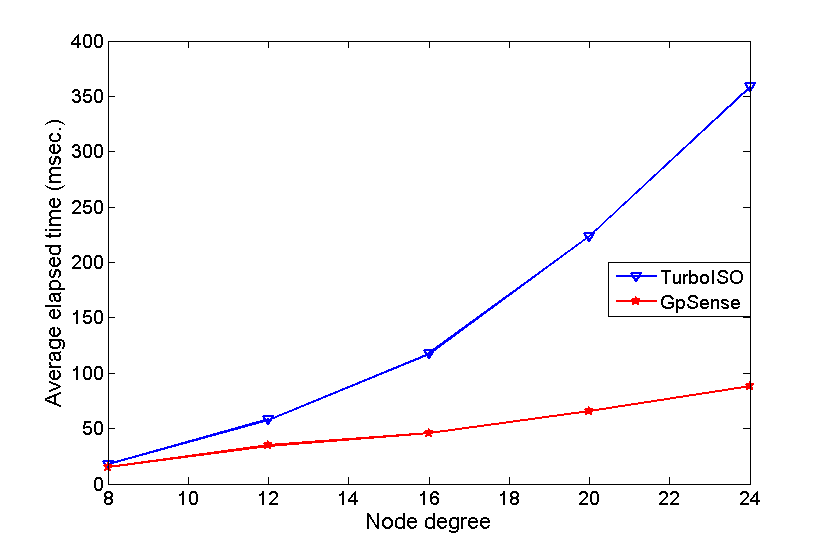}}
	\caption{Comparison with Turbo$_{ISO}$}
	\label{fig:turbo}
\end{figure}

Figure~\ref{fig:turbo}a shows the performance results of GpSense and Turbo$_{ISO}$ on the query graphs whose numbers of nodes vary from 6 to 14. Figure~\ref{fig:turbo}b shows their performance results when the node degree increases from 8 to 24, where the number of query nodes is fixed to 10. As can be seen in the two figures, the performance of Turbo$_{ISO}$ drops significantly while that of GpSense does not. 

This may be due to the number of recursive calls of Turbo$_{ISO}$ growing exponentially with respect to the size of query graphs and the degree of the data graph. In contrast, GpSense, with a large number of parallel threads, can handle multiple candidate nodes and edges at the same time, thus its performance remains stable.

\subsection{Effect of Optimization Techniques}

\begin{figure}[htp]
	\centering
	\subfigure[Refinement runing time]{\includegraphics[width=0.7\textwidth]{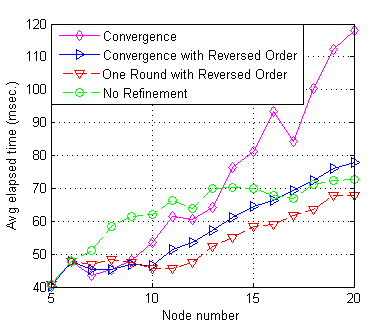}}
	\quad
	\subfigure[Intermediate results reduction]{\includegraphics[width=0.7\textwidth]{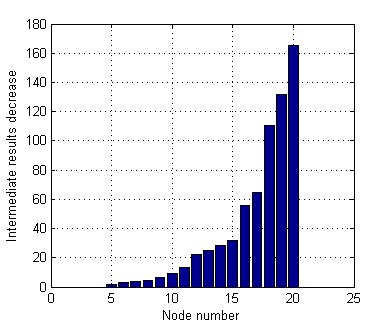}}
	\caption{Effect of optimization techniques}
	\label{fig:refine}
\end{figure}

Here, we carry out a series of experiments to demonstrate improvements of the proposed refinement function. Figure \ref{fig:refine}a shows a comparison between GpSense with and without the Candidates Refinement function in terms of average elapsed time. We compare four different versions of GpSense. The first version implements the refinement function until convergence. The second version is identical to the first apart from reversing the node visit order after the candidates sets initialization. The third version stops refining after the first round, and also reverses the node visit order. The fourth version does not employ the refinement function. As shown in Figure \ref{fig:refine}a, the response time is faster when using reversed node visiting order, compared to the original order, and GpSense with a limited number of iterations (i.e., the 3rd version) exhibits the best performance among the four implemented versions.

Figure \ref{fig:refine}b illustrates the effect of optimization techniques for refinement and two-data graphs utilization. In terms of intermediate results size, when the size of query graph is 20 nodes, the amount of memory that GpSense needs to maintain the intermediate results, without the use of these techniques, is up to 150 times more than GpSense using refinement and two-data graphs utilization.

\subsection{Scalability Test}

\begin{figure}[htp]
	\centering
	\includegraphics[width=0.7\textwidth]{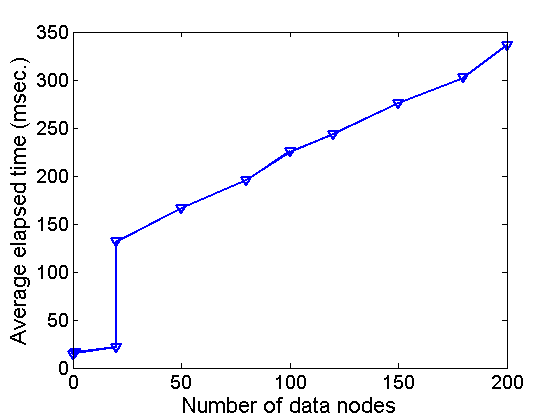}
	\caption{Scalability tests}
	\label{fig:scalable}
\end{figure}

We tested GpSense's scalability against SenticNet. The number of data nodes varies from 100,000 to 200 million nodes. The data graph is stored as follows: When the data graph is small, i.e., from 100,000 to 20 million nodes, we store it in the GPU global memory. If the node number of the data graph is between 20 million and 200 million, CPU memory is used to maintain the data graph. The number of query nodes is 6.

When the data graph size is 20 million nodes, we perform two experiments. The first maintains the whole data graph in GPU memory and the second uses CPU memory. As shown in Figure~\ref{fig:scalable}, the second experiment answers subgraph similarity search queries slower than the first experiment, due to the time taken for data transfer from CPU memory to GPU memory.

\section{Summary}

In this chapter, we introduced an efficient GPU-friendly method for answering subgraph similarity search queries over large-scale commonsense KBs. Our proposed method, GpSense, is based on a \emph{filtering-and-joining} approach which is shown to be suitable for execution on massively parallel GPU architectures. Along with efficient GPU techniques of coalescence, warp-based and shared memory utilization, GpSense provides a series of optimization techniques which contribute to enhancing the performance of subgraph similarity search-based commonsense reasoning tasks. We also present a \emph{multi-level graph compression} method to reduce the size of data graphs which cannot fit into GPU memory, but still preserve query answering correctness. Simulation results show that our method outperforms state-of-the-art backtracking-based algorithms on CPUs, and can efficiently answer subgraph similarity search queries on large-scale commonsense KBs. 
	% common-sense reasoning
% !TEX spellcheck = en_US

%%%%%%%%%%%%%%%%%%%%%%%%%%%%%%%%%%%%%%%%%%%%%%%%%%%%%%%%%%%%%%
%%%%%%%%%%%%%%%%%%%%%%%%%%%%%%%%%%%%%%%%%%%%%%%%%%%%%%%%%%%%%%
\chapter{Rule-based Query Processing on Commonsense Knowledge}
\graphicspath{{Chapter4/fig/EPS/}{Chapter4/fig/}}
\label{tag:chap4}

\section{Motivation}
\label{chap4:motivate}

In recent years, the Resource Description Framework (RDF) and Web Ontology Language (OWL) are widely applied to model knowledge base systems including both common and commonsense KBs. Query answering and retrieving the complete set of information on such knowledge bases is a complicated and time-consuming task. The crucial issue is that the querying process requires a reasoning step which derives implicit facts from the original datasets based on a predefined set of rules. In practice depending on the structures of commonsense knowledge bases and the requirements of the reasoning tasks, different rulesets are utilized for information retrieval systems. The W3C also provides some standard profiles of OWL, namely OWL EL, OWL RL, and OWL QL, each of which presents a particular set of rules and reasoning ability.

With the immense volume of data crawled daily from the Internet sources, the sizes of many knowledge bases have exceeded millions of concepts and facts. Real-time inference on such huge datasets with various user-defined rulesets is a non-trivial task which faces the challenging issues in term of system performance. As a consequence, efficiently retrieving and reasoning information on large-scale rule-based systems have attracted an increasing interest from researchers recently. Query answering systems such as Jena \cite{carroll2004jena}, Sesame \cite{broekstra2002sesame}, and OWLIM \cite{bishop2011owlim} integrate an inference layer on top of the query layer to perform the reasoning process and retrieve the complete set of results. These methods are also designed to support in-memory execution. Most of them, however, are facing the problems of scalability and execution time. Another set of studies is based on the forward-chaining approach which makes explicit all implicit facts in the pre-processing phase \cite{subercaze2016inferray,peters2013rule}. The resulting facts could then be explicitly written into the data storages of query engines including relational databases, RDF(S) triple stores, and graph-based query engines. The benefits of the forward-chaining inference scheme are 1) the time-consuming materialization is an off-line computation; 2) the inferred facts can be consumed as explicit ones without integrating the inference engine with the runtime query engine. However, the drawback of this approach is that we only can reason and query on the knowledge bases with a pre-processed rule-sets. In addition, the amount of inferred facts could be very large in comparison with the original dataset.

To address the requirements of scalability and execution time for real-time reasoning and query answering systems over OWL-based commonsense knowledge bases, this chapter introduces a parallel method, called $gSparql$, which utilizes the massive computation power of General Purpose GPUs. Our method is based on the backward-chaining approach which performs inference at query time.

\section{RDF and SPARQL}
\label{chap4:background}

The Resource Description Framework (RDF)\footnote{https://www.w3.org/TR/rdf-primer/}, a W3C recommendation, is used for representing information about Web resources. Resources can be anything, including documents, people, physical objects, and abstract concepts. RDF data model enables the encoding, exchange, and reuse of structured data. It also provides the means for publishing both human-readable and machine-processable vocabularies. RDF data is represented as a set of triples $<S,P,O>$, as in Table~\ref{tab:triple}, where each triple $<s,p,o>$ consists of three components, namely $subject$, $predicate$, and $object$. Each component of the RDF triple can be represented in either URI (Universal Resource Identifier) or literal form. For brevity, an URI is usually written along with a prefix (e.g., $<http://dbpedia.org/resource/isPartOf>$ is written as $x:isPartOf$), while a literal is written with double quotes (e.g., ``brad'').

\begin{table}[htp]
	\centering
	\begin{tabular}{|c|c|c|}
		\hline
		\textbf{Subject (s)} & \textbf{Predicate (p)} & \textbf{Object (o)} \\ \hline
		x:Alice              & y:isSisterOf           & x:Andy              \\ \hline
		x:Bob                & y:isSiblingOf          & x:Andy              \\ \hline
		x:Bob                & y:hasParent            & x:Brad              \\ \hline
		x:Alice              & y:liveIn               & x:London            \\ \hline
		x:Bob                & y:liveIn               & x:Paris             \\ \hline
		x:Andy               & y:liveIn               & x:England           \\ \hline
		x:Alice        	 	 & rdf:type               & x:Female            \\ \hline
		x:Alice              & y:age                  & 23              \\ \hline
		x:Andy               & y:age                  & 17              \\ \hline
		x:Bob                & y:age                  & 25              \\ \hline
		x:London             & y:isPartOf             & x:England           \\ \hline
		y:hasParent          & rdfs:domain            & x:Person            \\ \hline
		x:Female        	 & rdfs:subClassOf        & x:Person            \\ \hline
		x:isSiblingOf        & rdf:type       			& owl:SymmetricProperty            \\ \hline
	\end{tabular}
	\caption{RDF triples}
	\label{tab:triple}
	
\end{table}

RDF data is also represented as a directed labeled graph. The nodes of such a graph represent the subjects and objects, while the labeled edges are the predicates. We give the formal definition of an RDF graph as follow:

\begin{figure}[htp]
	\centering
	\includegraphics[width=0.6\textwidth]{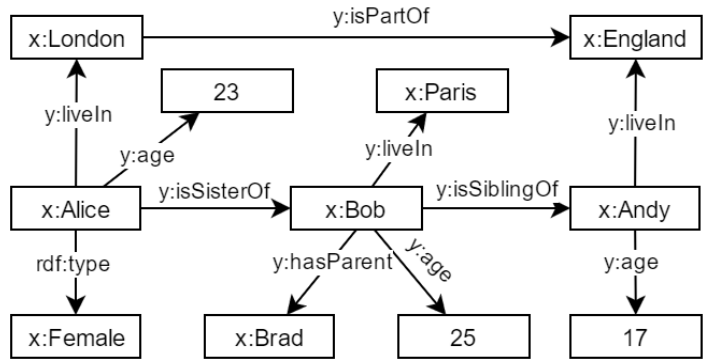}
	\caption{An example RDF knowledge graph}
	\label{fig:abox}
\end{figure}

\begin{definition} 
	An RDF graph is a finite set of triples (subject, predicate, object) from the set $T = U \times U \times (U \cup L)$, where $U$ and $L$ are disjoint, $U$ is the set of URIs, and $L$ the set of literals.
\end{definition}

\begin{figure}[htp]
	\centering
	\includegraphics[width=0.5\textwidth]{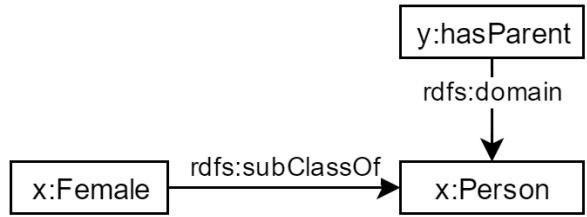}
	\caption{An example RDF schema graph}
	\label{fig:tbox}
\end{figure}

For example, Figure~\ref{fig:abox} illustrates the RDF graph based on the RDF triples in Table~\ref{tab:triple}. RDF graphs are further classified into two sub-types, namely \emph{RDF knowedge graph} and \emph{RDF schema graph}. The set of nodes in an RDF knowledge graph includes entities, concepts, and literals, as can be seen in Figure~\ref{fig:abox}. In the other hand, the RDF schema graph describes the relationships between types/predicates. Each edge labeled with $subClassOf$/$subPredicateOf$ connects two types or predicates (Figure~\ref{fig:tbox}).

Similar to an RDF graph, an SPARQL query\footnote{https://www.w3.org/TR/rdf-sparql-query/} also contains a set of triple patterns. The subject, predicate and object of a triple pattern, however, could be a variable, whose bindings are to be found in the RDF data. 

\begin{definition} 
	A SPARQL triple pattern is any element of the set $T = (U \cup V) \times (U \cup V) \times (U \cup L \cup V)$, where $V$ is the variable set.
\end{definition}

A SPARQL triple pattern can also be recursively defined as follows:

1) If $P_1$ and $P_2$ are SPARQL triple patterns, then expressions with the forms of $P_1$ . $P_2$, $P_1$ \textit{OPTIMAL} $P_2$, and $P_1$ \textit{UNION} $P_2$ are also SPARQL triple patterns.

2) If $P$ is a SPARQL triple pattern and $C$ is a supported condition, then $P$ \textit{FILTER} $C$ is also a SPARQL triple pattern.

In a SPARQL query, the \textit{SELECT} keyword is used to identify the variables which appear in the result set. For example, one wants to list all people whose parent is Brad and whose ages are greater than 20. The SPARQL query for this question is illustrated below:

\begin{verbatim}
SELECT ?a ?b
FROM {
   ?a rdf:type x:Person.
   ?a y:hasParent x:Brad.
   ?a y:age ?b
   FILTER (?b > 20)
}
\end{verbatim}

This query returns an empty result set because we cannot find any matches in the RDF data triples in Table~\ref{tab:triple}. However, if a ruleset $R=\{R_1,R_2,R_3,R_4,R_5,R_6,R_7\}$, which is given below, is applied to the original data triples, the result set of the query will be (?a, ?b) = \{(x:Alice, 23), (x:Bob, 25)\}.

\vspace{2mm}
$R_1:$ (?x y:isSisterOf ?y) $\to$ (?x y:isSiblingOf ?y)

\vspace{2mm}
$R_2:$ (?x y:isSiblingOf ?y) (?y y:isSiblingOf ?z) $\to$ (?x y:isSiblingOf ?z)

\vspace{2mm}
$R_3:$ (?x  rdf:type ?y) (?y rdfs:subClassOf ?z) $\to$ (?x rdf:type ?z)

\vspace{2mm}
$R_4:$ (?x y:isSiblingOf ?y) (?y y:hasParent ?z) $\to$ (?x y:hasParent ?z)

\vspace{2mm}
$R_5:$ (?x ?p ?y) (?p rdfs:subPropertyOf ?q) $\to$ (?x ?q ?y)

\vspace{2mm}
$R_6:$ (?x ?p ?y) (?p rdf:type owl:SymmetricProperty) $\to$ (?y ?p ?x)

\vspace{2mm}
$R_7:$ (?x ?p ?y) (?p rdfs:domain ?z) $\to$ (?x rdf:type ?z)

\vspace{2mm}
These results can be explained as follows: Based on rules $R_1$, $R_2$, and $R_6$, we can infer a triple (x:Alice y:isSiblingOf x:Bob). Then we obtain a triple (x:Alice y:hasParent x:Brad) by applying $R_4$ to that triple. Finally, $R_7$ generates two other triples relevant to the query, i.e. (x:Alice rdf:type x:Person) and (x:Bob rdf:type x:Person).

\section{System Overview}
\label{chap4:overview}

In this section, we introduce an overview of our rule-based reasoning and query processing system, called $gSparql$. Our method is based on backward-chaining reasoning and is accelerated by the massive parallel computing power of GPUs. Then, we discuss the triple store layout used in gSparql.

\subsection{Overview of gSparql}

\begin{figure}[htp]
	\centering
	\includegraphics[width=0.8\textwidth]{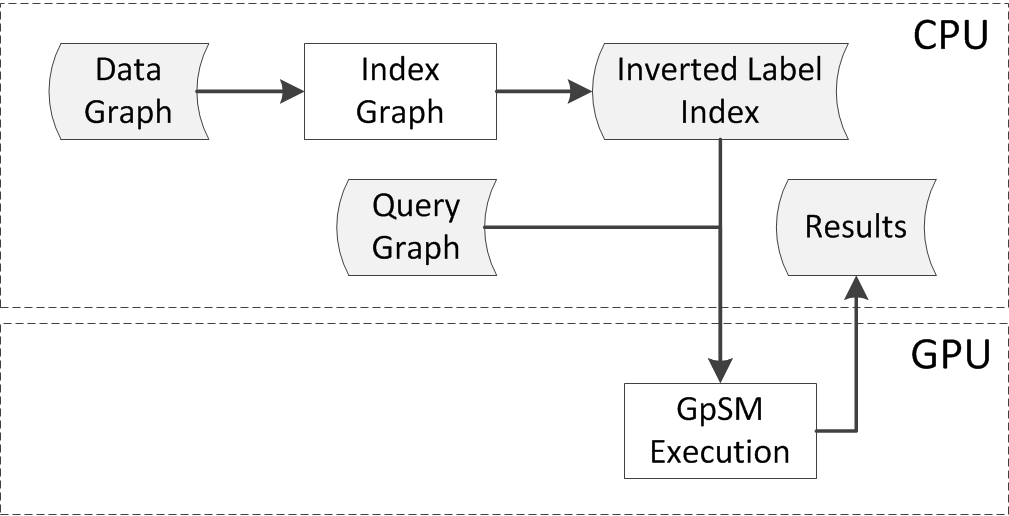}
	\caption{An overview of gSparql}
	\label{fig:gsparql}
\end{figure}

In this section, we present an overview of our backward-chaining method for reasoning and query processing on RDF-based commonsense knowledge bases, i.e. $gSparql$. As can be seen in Figure~\ref{fig:gsparql}, the input of our system is an SPARQL query and the output is a result set which contains a collection of tuples in which the selected variables are bound by RDF terms. There are three main modules in our method, namely Query Extension, Query Planner, and Query Evaluation.

The Query Extension module first parses the incoming SPARQL query $Q$ into single triple patterns. After that, the module extends the query based on the ruleset given by users. For each SPARQL triple pattern $p$ in $Q$, our system generates a reasoning graph which is a rooted DAG (directed acyclic graph) with the root of $p$. The details of the reasoning graph generation and the structure of the graph will be discussed in Section~\ref{chap4:inference}.
A query plan is then created in two consecutive steps. In the first step, the Query Planner module builds an execution plan for each reasoning graph of a single SPARQL triple pattern. We take the sample SPARQL query in the previous section as our input query. The triple pattern \emph{(?a rdf:type x:Person)} can be directly derived from the triple pattern \emph{(?a y:hasParent x:Brad)}, rule $R_5$ and the RDF schema graph. Therefore, the result set of \emph{(?a rdf:type x:Person)} is the superset of the matched solution of \emph{(?a y:hasParent x:Brad)}. As a consequence, the joining operation between the result sets of the two single triple patterns is redundant. The Query Planner module will prune out \emph{(?a rdf:type x:Person)} as well as its reasoning graph before generating the execution plan for evaluating the conjunctive triple pattern query. The two modules are executed on the CPU because their processing time contributes a very small fraction of the total working time. 

The Query Evaluation module then executes matching the extended query to the RDF data. The matching process takes the majority of the querying time due to time-consuming operations such as joining, filtering, and duplication elimination. Thus, our method takes advantages of massively parallel computing ability of GPUs to accelerate such operations. In particular, we employ an efficient parallel scheme which combines GPU-friendly primitives such as sort, merge, prefix scan. A data buffer in the device memory is utilized to temporarily maintain the required data and intermediate results during execution. After performing the query evaluation phase, the final results are transferred back to the main memory.

\subsection{Triple Store}

In gSparql, the triple store layout is based on the \textit{property tables} approach in which all triples with the same predicate name are stored in the same table \cite{abadi2007scalable}. In the traditional method, a property table consists of two columns $<s,o>$ and is sorted by \textit{subject}. In order to support efficient merging and joining operations during reasoning and query processing, we maintain another $<o,s>$ column table, which is sorted by \textit{object}, for each predicate name. Our method uses a dictionary to encode URI and literal RDF terms into numeric values (Table~\ref{tab:dict}). This encoding is commonly used in large-scale triple stores such as RDF-3X \cite{neumann2008rdf} and Hexastore \cite{weiss2008hexastore} to reduce tuple space and faster comparison. The numeric values are stored in their native formats.

\begin{table}[htp]
	\centering
	\begin{tabular}{|c|c|}
		\hline
		\textbf{URI/Literal} & \textbf{Numeric ID} \\ \hline
		y:liveIn             & 0                   \\ \hline
		y:age                & 1                   \\ \hline
		rdfs:subClassOf      & 2                   \\ \hline
		...                  & ...                 \\ \hline
		x:Alice              & 10                  \\ \hline
		x:Andy               & 11                   \\ \hline
		x:Bob                & 12                   \\ \hline
		x:England            & 13                  \\ \hline
		x:London             & 14                  \\ \hline
		x:Paris              & 15                  \\ \hline
		...                  & ...                 \\ \hline
	\end{tabular}
	
	\caption{URI/Literal dictionary}
	\label{tab:dict}
\end{table}

\begin{figure}[htp]
	\label{fig:layout}
	\centering
	\includegraphics[width=0.5\textwidth]{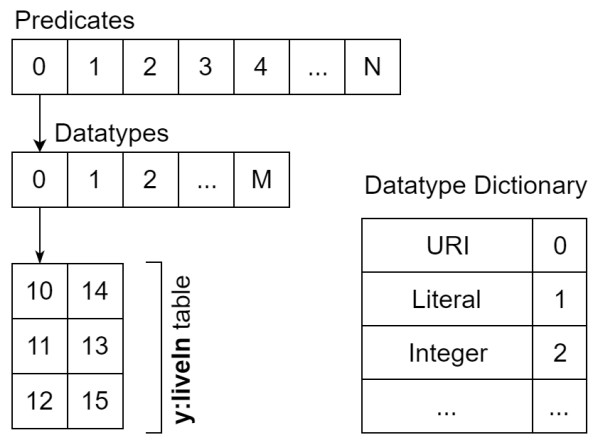}
	\caption{RDF triple store layout}
\end{figure}

In practice, the objects of the same predicate name might have different datatypes. The objects in the predicate related to \textit{Born} in Wikipedia, for example, consist of Literal values (e.g. "179-176 BC") , Integer values (e.g. 1678), and Datetime values (e.g. 06 June 1986). For each predicate name, we further divide the column tables $<s,o>$ and $<o,p>$ into smaller ones based on the object datatypes. Figure~\ref{fig:layout} illustrates the triple store layout used in our method.

The advantages of our RDF data representation are: 1) gSparql is able to immediately make comparisons between numeric data in the FILTER clauses without further requesting actual values from the dictionary; 2) Our method can directly return the results of unary functions on RDF terms such as $isIRI$, $isLiteral$, or $isNumeric$; 3) We only need to execute joining operations on columns with the same datatype. Thus, unnecessary joins can be pruned out; 4) The vertical partitioning approach enables GPU kernels to retrieve the table content in a coalesced memory access fashion which significantly improves the performance of GPU-based systems.

\section{Backward-Chaining Inference}
\label{chap4:inference}

In comparison to materialization-based method, the reasoning and query processing system based on backward-chaining is usually required to perform more computation at the query time. The real-time inference is considered as the bottleneck of this approach which decreases the overall response time. Thus, the objective of gSparql is to enhance the performance of the backward-chaining reasoning process.

\begin{figure}[htp]
	\centering
	\includegraphics[width=0.8\textwidth]{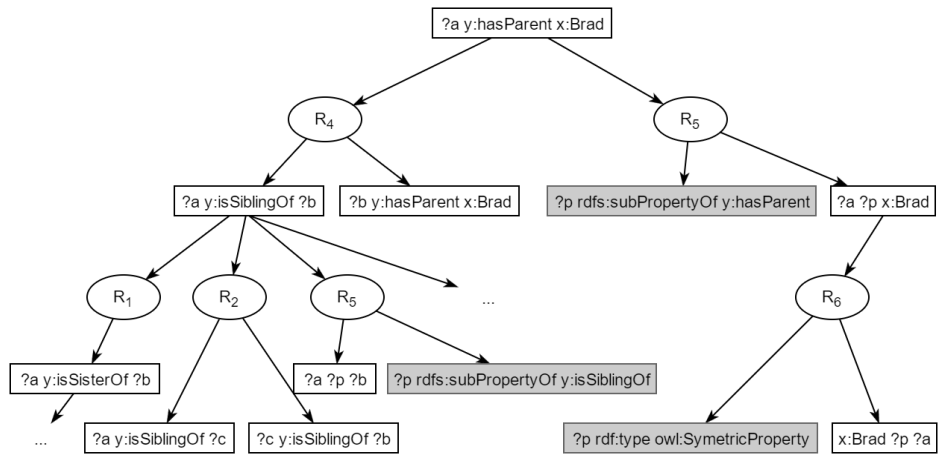}
	\caption{Reasoning tree of (?a y:hasParent x:Brad)}
	\label{fig:tree}
\end{figure}

The general idea of the backward-chaining approach is constructing \textit{reasoning trees} for all triple patterns in the given query. In fact, if we combine nodes which refer to the same triple pattern, the reasoning tree becomes a rooted DAG (directed acyclic graph). Figure~\ref{fig:tree} shows a part of the reasoning tree of the \textit{(?a y:hasParent x:Brad)} triple pattern in the sample query. A reasoning tree comprises two types of nodes, namely pattern nodes and rule nodes. A pattern node is established by connecting rule nodes using \textit{or} operations. A rule node, in contrast, is created by applying \textit{and} operations between pattern nodes. General speaking, the parent of a rule node $R_i$ is the consequent of the rule and its child nodes are the antecedents of $R_i$. The reasoners build those trees by recursively applying rules to pattern nodes in BFS or DFS fashions. The reasoning tree construction terminates when no more rules can be further applied. 

The real-time backward-chaining inference is executed by matching and joining RDF triples based on the reasoning trees using bottom-up approaches. To match a rule node, we first search for the matches of its child nodes in the RDF triple store. Then, joining operations are applied to return the rule node's results. This is the common procedure to make inferences on \textit{general rules}. The matches of a pattern node are obtained by merging the results of its child rule nodes. In rules $R_2$ and $R_4$ of the example ruleset, the triple patterns related to column tables \textit{y:isSiblingOf} and \textit{y:hasParent} appear on both sides of the rules respectively. In these cases, new triples are potentially generated when we continuously apply the same rules to the derived triples. We call such rules \textit{recursive rules}.

The crucial requirement for backward-chaining approach is an efficient response time when we execute reasoning tasks at query time. In the next subsections, we present our approach to accelerate the reasoning performance on the massively parallel hardware architecture of modern GPUs.

\subsection{General Rules}

This subsection discusses GPU implementation of reasoning on general rules. At the beginning, we give brief descriptions of some important GPU primitives which significantly outperform the CPU-based counterparts. Then, we discuss how to map these primitives to different groups of inference rules.

\textbf{Prefix scan:} A prefix scan (in short, \textit{scan}) employs a binary operator to the input array of size $N$ and generates an output of the same size. An important example of prefix scan is prefix sum which is commonly used in database operations. In gSparql, we apply the GPU implementation from \cite{harris2007gpu}.

\textbf{Sort:} Our system employs Bitonic Sort algorithm for sorting relations in parallel. The bitonic sort merges bitonic sequences, which are in monotonic ascending or descending orders, in multiple stages. We adapt the standard bitonic sort algorithm provided by NVIDIA library.

\textbf{Merge:} Merging two sorted triples is a useful primitive and is a basic building block for many applications. To perform the operation on GPUs, we apply an efficient algorithm called GPU Merge Path \cite{green2012gpu}.

\textbf{Sort-Merge Join:} Following the same procedure as the traditional sort-merge joins, we execute sorting algorithms on two relations and after that merge the two sorted relations. Due to the fact that our triple store layout is based on vertical partitioning approach, the sort-merge join is well-suited for reasoning and query processing execution.

Next, we discuss how to apply these operations on various groups of rules. Our considered inference rules can be divided into some major groups, namely copy rules, subject/object join rules, and predicate join rules.

\textbf{Copy rules:} For this group of rules, the joining operations are not required. We simply copy the whole column table into a new one. The rule $R_1$ is an example of the rule group. At implementation level, we do not perform actual copy operations. 

\textbf{Subject$\backslash$Object join rules:} Performing this rule group requires joining two predicate tables in the positions of subject or object. Since our triple store maintains both sorted predicate tables $<s,o>$ and $<o,s>$, these join rules are straightforwardly executed by the standard sort-merge join.

\textbf{Predicate join rules:} This kind of rules joins two triple patterns in which one join attribute is located at the predicate position of a triple pattern and the other attribute is in the subject or object position of the remain triple pattern. We reduce the joining operation of the rules to $scan$ and $merge$. First, we scan the triple store to collect the predicate names of the join attribute in the first triple pattern. Then, we merge column tables of the obtained predicate names.

\subsection{Recursive Rules}

The main routine of making inferences on recursive rules is illustrated in Algorithm~\ref{algo:inference}. The general idea of the algorithm is recursively applying the rule $R$ to derived triples until no new triple is found.

\begin{algorithm}[htp]
	\label{algo:inference}
	\caption{Reasoning procedure for recursive rules}
	\SetKwFunction{infer}{apply\_rule}
	
	\KwIn{set of triples T, rule R}
	\KwOut{set of triples T}
	\BlankLine
	
	NewT := T;
	
	\While{NewT \textbf{not empty}}{
		InferT := \infer{NewT, R};
		
		NewT := T $\backslash$ InferT;
		
		T := T $\cup$ NewT;
	}
	
	\KwRet{T}
\end{algorithm}

The algorithm often generates a large number of duplicated triples in each iteration. A popular approach to overcome the problem is based on sorting and then removing duplications \cite{heino2012rdfs,subercaze2016inferray}. The method, however, needs to request all existing triples of the related column table to identify triple duplications and new derived triples from the inferred set. Unlike in-memory systems, such as Inferray, which are able to maintain all data in the main memory, the storage capability of a typical GPU device is very limited. For the property table whose size cannot fit into the global memory, we must frequently transfer data between GPU and CPU memory during execution. This might become the bottleneck which significantly reduces the overall reasoning performance.

To achieve the high performance in such cases, we present a GPU implementation of Quotient Filter \cite{bender2012don} to resolve the problem of triple duplication. 

\subsection{Optimization Techniques}

In this subsection, we present some optimizations that aim to reduce the number of expensive joining operations and accesses to column tables when we make inferences using the reasoning tree. These optimization techniques are described below:

\textbf{Pre-computation of RDF schema graphs:} If we look at the reasoning tree in Figure~\ref{fig:tree}, we notice that the execution of some reasoning branches depends less on the input than others. For example, the pattern
\textit{(?a rdfs:subPropertyOf y:hasParent)} is more generic than \textit{(?a ?p x:Brad)},
because the latter refers to a specific term from the query. Another difference between these two patterns is that the first corresponds to only schema triples while the second can match any triple. In practice, schema triples are far less than the others on Web-data,
therefore, the first pattern will match with many fewer triples than the other.

We make a distinction between these two types of patterns, calling the first
schema triple patterns. A schema triple pattern is a triple pattern
which has as predicate or object a term from either the RDFS or the OWL
vocabularies.
These schema patterns are responsible for a notable computational
cost that affects many queries. If we precompute all the ground instances of
these triple patterns that are entailed by the knowledge base, then whenever
the reasoner needs such patterns it can use the precomputed results avoiding to perform additional computation. This would simplify our task to only have to perform reasoning on the non-schema patterns.

\textbf{Branch pruning using RDF schema graphs:} The pre-calculation of the schema closure allows us to implement another
optimization that can further reduce the size of the reasoning tree by identifying
beforehand whether a rule can contribute to derive facts for the parent branch.
In this case, the triples in the schema closure can be used for the
purposes of inducing early failures: the truth of these triples is easy to verify
since they have been precomputed and no inference is needed. Therefore, when
scheduling the derivation of rule-antecedents, we give priority to antecedents that
potentially match these precomputed triples so that if these cheap antecedents
do not hold, the rule will not apply anyway, and we can avoid the computation
of the more expensive antecedents of the rule for which further reasoning would
have been required.
To better illustrate this optimization, we proceed with an example. Suppose
we have the reasoning tree described in Figure~\ref{fig:tree}. In this tree, the reasoner fires
rule $R_6$ (concerning symmetric properties in OWL) to be applied on the second antecedent of rule $R_5$.
In this case, Rule $R_6$ will fire only if some of the subjects of the triples
part of \textit{(?p rdfs:subPropertyOf y:hasParent)} will also be the subject of triples
part of \textit{(?p rdf:type owl:SymmetricProperty)}. Since both patterns are more
specific than schema patterns, we know beforehand all the possible ``?p", and therefore we can immediately perform an intersection between the two sets
to see whether this is actually the case. If there is an intersection, then the reasoner proceeds executing rule $R_6$, otherwise it can skip its execution since it will never fire.
It is very unlikely that the same property appears in all the schema
patterns, therefore by performing such intersections we are able to further reduce the tree size not considering rules that will derive no conclusion.

\section{Experiment Results}

In this section, we compare our gSparql algorithm
against an existing system, Jena, that also answers
queries via a combination of an in-memory backward chaining
reasoner with basic knowledge base retrievals.

The comparison was carried out using two LUBM
benchmarks consisting of one knowledge base describing a
single university and another describing 10 universities. Prior
to the application of any reasoning, these benchmarks contained
100,839 and 1,272,871 triples, respectively.

\begin{table}[htp]
	\centering
	\caption{comparison with the Jena system}
	\label{tab:jena}
	\begin{tabular}{|l|l|l|l|l|}
		\hline
		\textbf{LUBM}    & \textbf{1 University} & \textbf{100,839 triples} & \textbf{10 Universities} & \textbf{1,272,871 triples} \\ \hline
		& \textbf{gSpaql}       & \textbf{Jena}            & \textbf{gSpaql}          & \textbf{Jena}              \\ \hline
		& \textbf{Time}         & \textbf{Time}            & \textbf{Time}            & \textbf{Time}              \\ \hline
		\textbf{Query1}  & 0.20                  & 0.32                     & 0.43                     & 0.86                       \\ \hline
		\textbf{Query2}  & 0.50                  & 130                      & 2.1                      & n/a                        \\ \hline
		\textbf{Query3}  & 0.026                 & 0.038                    & 0.031                    & 1.5                        \\ \hline
		\textbf{Query4}  & 0.52                  & 0.021                    & 1.1                      & 0.41                       \\ \hline
		\textbf{Query5}  & 0.098                 & 0.19                     & 0.042                    & 1.0                        \\ \hline
		\textbf{Query6}  & 0.43                  & 0.49                     & 1.9                      & 3.2                        \\ \hline
		\textbf{Query7}  & 0.29                  & 45                       & 2.2                      & 8,100                      \\ \hline
		\textbf{Query8}  & 0.77                  & 0.91                     & 3.7                      & 52                         \\ \hline
		\textbf{Query9}  & 0.36                  & n/a                      & 2.5                      & n/a                        \\ \hline
		\textbf{Query10} & 0.18                  & 0.54                     & 1.8                      & 1.4                        \\ \hline
		\textbf{Query11} & 0.24                  & 0.011                    & 0.18                     & 0.032                      \\ \hline
		\textbf{Query12} & 0.23                  & 0.0020                   & 0.33                     & 0.016                      \\ \hline
		\textbf{Query13} & 0.025                 & 0.37                     & 0.21                     & 0.89                       \\ \hline
		\textbf{Query14} & 0.024                 & 0.58                     & 0.18                     & 2.6                        \\ \hline
	\end{tabular}
\end{table}

We evaluated these using a set of 14 queries taken from
LUBM \cite{guo2005lubm}. These queries involve properties associated
with the LUBM university-world ontology, with none of the
custom properties/rules whose support is actually our end goal). Answering these queries requires,
in general, reasoning over rules associated with both RDFS
and OWL semantics, though some queries can be answered
purely on the basis of the RDFS rules.

Table~\ref{tab:jena} compares our algorithm to the Jena system using
a pure backward chaining reasoner. Our comparison focuses
on response time, as our optimization algorithm should
be neutral with respect to result accuracy, offering no more
and no less accuracy than is provided by the interposed reasoner.
As a practical matter, however, Jena’s system cannot
process all of the rules in the OWL semantics rule set, and
was therefore run with a simpler ruleset describing only the
RDFS semantics. This discrepancy accounts for the differences
in result size for several queries. Result
sizes in the table are expressed as the number of tuples returned
by the query and response times are given in seconds.
An entry of “n/a” means that the query processing had not
completed (after 1 hour).

Despite employing the larger and more complicated rule
set, our algorithm generally ran faster than Jena, sometimes
by multiple orders of magnitude. The exceptions to this trend
are limited to queries with very small result set sizes or queries
10-13, which rely upon OWL semantics and so could not
be answered correctly by Jena. In two queries (2 and 9), Jena
timed out.

\section{Summary}

To retrieve information on rule-based commonsense knowledge bases, the section introduces gSparql, a fast and scalable inference and querying method on mass-storage RDF data with custom rules.  Our method focuses on dealing with backward-chaining reasoning which makes inferences at query time when the inferred triples are determined by the set of triple patterns defined in the query. To efficiently answer SPARQL queries in parallel, we first build reasoning trees for all triple patterns in the query and then execute those trees on GPUs in a bottom-up fashion. In particular, we convert the execution tree into a series of primitives such as sort, merge, prefix scan, and compaction which can be efficiently done on GPU devices. We also utilize a GPU-based Bloom Filter method and sort algorithms to overcome the triple duplication. Extensive experimental evaluations show that our implementation scales in a linear way and outperforms current optimized CPU-based competitors.	% rule-based common-sense reasoning
% !TEX spellcheck = en_US

%%%%%%%%%%%%%%%%%%%%%%%%%%%%%%%%%%%%%%%%%%%%%%%%%%%%%%%%%%%%%%
%%%%%%%%%%%%%%%%%%%%%%%%%%%%%%%%%%%%%%%%%%%%%%%%%%%%%%%%%%%%%%
\chapter{Multimodal Sentiment Analysis using Commonsense Knowledge}
\graphicspath{{Chapter5/fig/EPS/}{Chapter5/fig/}}
\label{tag:chap5}

\section{Motivation}

Sentiment analysis is the automatic identification of private states of the human mind (i.e. opinions, emotions, sentiments, behaviors and beliefs), along with their polarity (positive or negative). 
The field of sentiment classification has recently become an attractive research direction due to a large number of real-world applications which require identifying human opinions for better decision-making. Some noticeable examples of those applications are branding and product analysis \cite{hu2004mining}, tracking sentiment timelines in on-line forums and news \cite{lloyd2005lydia} and conversation summarization \cite{carenini2008summarizing}. 

Most of the recent works on sentiment analysis are based on natural language processing techniques in which the detection of emotions is conducted from humanly created textual data and resources including lexicons \cite{wiebe2005creating} or large annotated datasets \cite{pang2004sentimental}.
With the rapid growth of social media websites such as Facebook and YouTube, people are expressing their opinions in various forms which include videos, images and audios.
Compared to the textual data, these resources might provide more valuable information through richer channels such as the tones of speakers and facial expressions. 
As a result, the necessity of analyzing and understanding on-line generated data from multimodal cues has arisen in recent years \cite{cambria2013sentic,morency2011towards}.

Collecting and processing such information, however, are very challenging tasks as they involve in dealing with a huge amount of information that is changing at a very high speed.
In the last few years, Graphic Processing Units (GPUs) have become popular computing devices owing to their massively parallel processing architectures. General Purpose GPUs have been successfully used as an efficient accelerator to leverage the performance of big data analytics \cite{he2008mars}.
Extreme Learning Machine (ELM) \cite{huang2015new,huang2006extreme} has also been an emerging learning technique that provides efficient unified solutions to generalized feed-forward networks including single or multiple hidden-layer neural networks, radial basis function networks, and kernel learning. ELM offers significant advantages such as fast learning speed, ease of implementation, and minimal human intervention.
Inspired by these promising results, we investigate whether and how GPU and ELM can be leveraged to accelerate the task of real-time multimodal sentiment analysis, i.e., harvesting sentiments from Web videos by taking into account audio, visual and textual modalities as sources of the information.   

In the chapter, we propose an ensemble application of ELM and GPU for real-time multimodal sentiment analysis. 
Our method takes advantages of the massively parallel processing power of modern GPUs to enhance the performance of feature extraction from different modalities. 
In addition, powerful ELM classifiers are applied to build the sentiment analysis model based on the extracted features. 
To highlight the efficiency of our solution, we conducted our experiments on the YouTube dataset and achieved an accuracy of 78\% which outperforms all previous systems.
In term of processing speed, our method shows improvements of several orders of magnitude for feature extraction compared to CPU-based counterparts.

\section{Multimodal Sentiment Analysis}
\label{sec:overview}

In this section, we give an overview of our multimodal sentiment analysis approach. Then, we describe the datasets that we employed to build our system as well as to conduct our experiments. 

\subsection{GPU-ELM Multimodal Sentiment Analysis}

\begin{figure}[htp]
	\centering
	\includegraphics[width=0.8\textwidth]{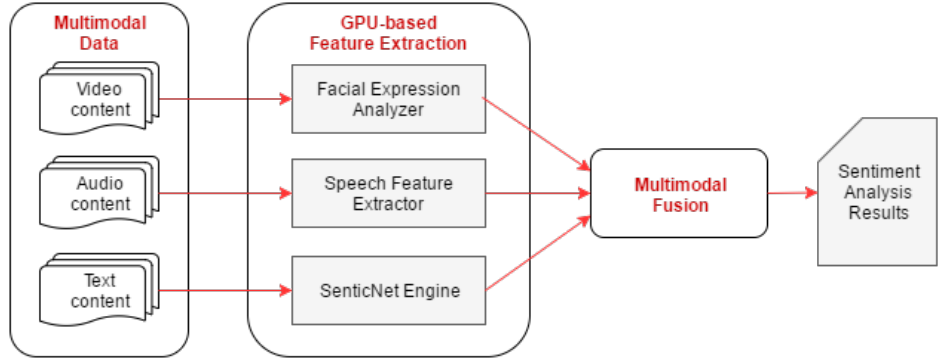}
	\caption{Overview of GPU-based multimodal sentiment analysis}
	\label{fig:gpufusion}
\end{figure}

Figure~\ref{fig:gpufusion} shows an overview of our real-time GPU-ELM multimodal sentiment analysis approach. At the beginning, our system receives the multimodal data from users. The data might be created by these users expressing their opinions in front of the camera. Later, our system splits the video into several segments. A transcriber is used to obtain the text transcription of the audio.

In the vector stream generation section, the multimodal data consisting of visual, audio and text content are processed on the GPU. 
In the facial expression analyzer, we employ a GPU-based Active Shape Model (ASM) method to track 66 facial points. 
In the algorithm, we first apply the sketch based facial expression recognition technique \cite{song2008robust} to speed up the convergence of ASM searching.
Then, we accelerate the performance of the ASM tracking by processing multiple sketched images simultaneously.
The relevant achieved points are then used to construct facial features. 
The important acoustic features such as MFCC, Spectral Centroid, and Pitch are also extracted on the GPU for emotion recognition by employing and extending an open source tool of Michalek et al. \cite{michalek2014open}.
In our system, we apply a block-based approach to process multiple windows concurrently.
For textual data, we propose a GPU-based SenticNet engine to select only important features and aspects of human opinions. In the engine, commonsense knowedge bases such as ConceptNet and SenticNet are employed to automatically construct the ontology containing domain specific commonsense knowledge. The obtained ontology is used to track the important review aspects as well as opinion words in the input texts. 

Later, the feature vectors of all three modalities are fused in the multimodal fusion module. The resemble vectors are then used for classifying each video segment into sentiment classes. To be more formal, let the input signal $S = (s_v, s_a, s_t)$ 
\[ R = sentiment\_analysis(S) = fusion(face(s_v), speech(s_a), sentic(s_t) ) \]

\subsection{YouTube Dataset}

In our work, we use an available dataset consisting of 47 videos which were collected from the social media web site YouTube \cite{morency2011towards}. Videos in the dataset were based on different topics (for instance politics, electronics product reviews etc). The videos were found using the following keywords: opinion, review, product review, best perfume, toothpaste, war, job, business, cosmetics review, camera review, baby product review, I hate, I like \cite{morency2011towards}. The final video set had 20 female and 27 male speakers randomly selected from YouTube, with their age ranging approximately from 14-60 years. Although they belonged to different ethnic backgrounds (e.g., Caucasian, African-American, Hispanic, Asian), all speakers expressed themselves in English. The videos were converted to mp4 format with a standard size of 360x480. The length of the videos varied from 2-5 minutes. All videos were pre-processed to avoid the issues of introductory titles and multiple topics. Many videos on YouTube contained an introductory sequence where a title was shown, sometimes accompanied with a visual animation. To address this issue, the first 30 seconds was removed from each video.
Morency et al. provided the transcriptions with the videos. Each video was segmented and each segment was labeled by a sentiment. Because of this annotation scheme of the dataset, textual data was available for our experiment.

We employ this YouTube dataset in our experiments to build the multimodal sentiment analysis system as well as evaluate the system's performance. 

\subsection{SenticNet Datasets}
As a priori polarity lexicon of concepts, we use SenticNet 3.0 \cite{cambria2014senticnet}, a lexical resource that contains 30,000 concepts along with their polarity scores in the range from –1.0 to +1.0. SenticNet 3.0 also contains all WordNet Affect (WNA) \cite{strapparava2004wordnet} concepts. The first 10 SenticNet concepts in lexicographic order along with the corresponding polarities are shown in Table \ref{table:senticnet}. 

\begin{table}[htp]
	\centering % centering table
	\begin{tabular}{|l|r|} % creating eight columns
		\hline
		\textbf{Concept} & \textbf{Polarity}\\
		\hline
		a lot & +0.258 \\ \hline
		a lot sex & +0.858 \\ \hline
		a little & +0.032 \\ \hline
		abandon & –0.566 \\ \hline
		Abase & –0.153 \\ \hline
		Abash & –0.174 \\ \hline
		Abashed & –0.174 \\ \hline
		abashment & –0.186 \\ \hline
		Abhor & –0.391 \\ \hline
		abhorrence & –0.391 \\ \hline
	\end{tabular}
	\vspace{3mm}
	\caption{Sample of SenticNet data} %title of the table
	\label{table:senticnet} % is used to refer this table in the text
\end{table}

The EmoSenticNet dataset \cite{poria2012merging} contains about 13,741 commonsense knowledge concepts, including those concepts that exist in the WNA list, along with their affective labels in the set {anger, joy, disgust, sadness, surprise, fear}.

The EmoSenticSpace dataset is a knowledge base for emotive reasoning. It is built by applying the so-called ``blending'' technique to ConceptNet \cite{liu2004conceptnet} and EmoSenticNet. Blending is a technique that performs inference over multiple sources of data simultaneously, taking advantage of the overlap between them. It linearly combines two sparse matrices into a single matrix, in which the information between the two initial sources is shared. Before performing blending, we represented EmoSenticNet as a directed graph similar to ConceptNet. For example, the concept birthday party was assigned an emotion of joy. We took them as two nodes, and added the assertion HasProperty on the edge directed from the node birthday party to the node joy. 
Then, we converted the graphs to sparse matrices in order to blend them. After blending the two matrices, we performed Truncated Singular Value Decomposition (TSVD) on the resulting matrix to discard those components representing relatively small variations in the data. We discarded all of them, keeping only 100 components of the blended matrix to obtain a good approximation of the original matrix. The number 100 was selected empirically: the original matrix could be best approximated using 100 components.

\section{Feature Extraction}
\label{sec:features}

In this section, we explain how visual, audio and textual features are automatically extracted from multimodal data using the GPU-based techniques. 

\subsection{Visual Feature Extraction}
\label{sec:visual}

Humans are known to express emotions in different ways through the face. Facial expressions, thus, can provide important clues for identifying emotions, which we use to complement the linguistic and acoustic features. 
A facial expression analyzer automatically identifies emotional clues associated with facial expressions, and classifies facial expressions in order to define sentiment categories (i.e. positive, negative and neutral) and to discriminate between them.
In the annotations provided with the YouTube dataset \cite{morency2011towards}, each video was segmented into some parts and each of the sub-segments had the length of few seconds. Every segment was annotated as either 1, 0 and -1 denoting positive, neutral and negative sentiments respectively.

\begin{table}[h]
	%title of the table
	\centering % centering table
	
	\begin{tabular}{|l|l|} % creating eight columns
		%\hline\hline %inserts double horizontal lines
		\hline
		\textbf{Features} & \textbf{Description} \\ [0.5ex] % inserts table
		\hline
		0 &	Left eye \\ \hline
		1 &	Right eye \\ \hline
		24 &	Left eye inner corner \\ \hline
		23 &	Left eye outer corner \\ \hline
		38 &	Left eye lower line \\ \hline
		35 &	Left eye upper line \\ \hline
		29 &	Left eye left iris corner \\ \hline
		30 &	Left eye right iris corner \\ \hline
		25 &	Right eye inner corner \\ \hline
		26 &	Right eye outer corner \\ \hline
		41 &	Right eye lower line \\ \hline
		40 &	Right eye upper line \\ \hline
		33 &	Right eye left iris corner \\ \hline
		34 &	Right eye right iris corner \\ \hline
		13 &	Left eyebrow inner corner \\ \hline
		16 &	Left eyebrow middle \\ \hline
		12 &	Left eyebrow outer corner \\ \hline
		14 &	Right eyebrow inner corner \\ \hline
		17 &	Right eyebrow middle \\ \hline
		54 &	Mouth top \\ \hline 
		55 &	Mouth bottom \\ \hline
	\end{tabular}
	\vspace{3mm}
	\caption{Some relevant facial feature points}
	\label{tab:ffp} % is used to refer this table in the text
\end{table}

\begin{figure}[htp]
	\centering
	\includegraphics[width=0.8\textwidth]{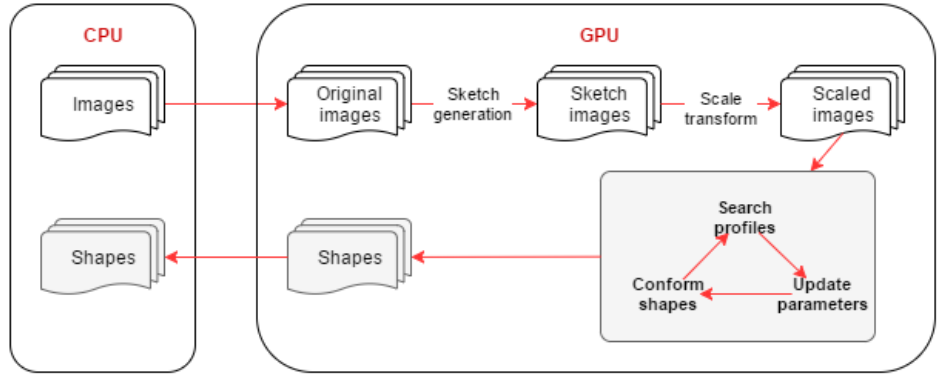}
	\caption{Overview of GPU-based feature extraction}
	\label{fig:videofeatures}
\end{figure}

We first convert all videos in the dataset to image frames. After that, we extract facial feature points from each image frame. Table~\ref{tab:ffp} lists some relevant facial feature points out of the 66 feature points that we use to construct facial features. These features are defined as the distances between feature points; see example in Table~\ref{tab:features}.
To detect feature points from the images, we use Active Shape Model (ASM) \cite{cootes1995active} which is a popular statistical model for facial feature detection. 
However, the performance of ASM searching might be very poor when the input image is noisy. 
In addition, the tracking result is very sensitive to the initial location and size. Some smoothing algorithms may need to be used to improve the tracking accuracy and efficiency. 
These operations are, however, in general computationally expensive. To overcome the issue, we introduce a GPU-based approach to enhance the performance of the ASM method for facial feature extraction.
At the beginning, images and models are transferred into the global memory of the GPU. 
After that, we perform GPU-based Active Shape Model in two main steps: (1) sketch generation and (2) ASM model matching.
The overview of the ASM-based feature points detection on the GPU is illustrated in Figure~\ref{fig:videofeatures}.

\begin{table}[htp]
	\centering % centering table
	
	\begin{tabular}{|l|} % creating eight columns
		\hline
		\textbf{Features}\\ [0.5ex] % inserts table
		\hline
		Distance between right eye and left eye \\ \hline
		Distance between the inner and outer corner of the left eye \\ \hline
		Distance between the upper and lower line of the left eye \\ \hline
		Distance between the left iris corner and right iris corner of the left eye \\ \hline
		Distance between the inner and outer corner of the right eye \\ \hline
		Distance between the upper and lower line of the right eye \\ \hline
		Distance between the left iris corner and right iris corner of the right eye \\ \hline
		Distance between the left eyebrow inner and outer corner \\ \hline
		Distance between the right eyebrow inner and outer corner \\ \hline
		Distance between top of the mouth and bottom of the mouth \\ \hline
	\end{tabular}
	\vspace{3mm}
	\caption{Some important facial features used for the experiment} %title of the table
	\label{tab:features} % is used to refer this table in the text
\end{table}

The sketch generation implemented on the GPU follows two consecutive operations, namely edge detection and tone mapping, which are similar to the GASM algorithm used in \cite{song2008robust}. 
A thread-based approach is applied to detect and sharpen the edge of contour of the face components. 
In this stage, we convert the color of each pixel to a luminance value, and then calculate the square diagonal differences which are summed up afterward. 
After reverting the results, we multiply them by a large number to make the values visible.
To achieve the best performance, we use the shared memory of GPUs to store data points during the computation.
However, the result obtained from the edge detection and enhancement still remains noisy, which makes the latter ASM tracking process slow and unreliable. 
The next step will further eliminate the high frequency noise by changing the tonal range \cite{02_Ashikhmin}.

In the ASM model matching, the sketch images (rather than the original images) are used as data inputs. Since the edges in the sketch images are much stronger than the original ones, the model matching process converges more quickly. 
To execute parallel ASM searching in many images synchronously in the GPU \cite{li2009accelerating}, we first need to transform the scale of the images according to the size of the detected face. 
The approach to accelerate the time-consuming image scaling is to make use of the texture memory of the GPU and bilinear interpolation of texture.
The GPU-based parallel ASM searching algorithm is similar to the iterative CPU-based method. 
First, a region of the image around each point is sampled and calculated to find the best match. Then, the model parameters are updated to fit the new search shape. Finally, the shapes are conformed by the parameters.
After each iterative computation, the appropriate shapes are transferred to CPU for displaying and further processing.
Unlike the traditional ASM, however, the searching process is calculated with GPU on multiple images concurrently.

\subsection{Acoustic Feature Extraction}

\begin{figure}[htp]
	\centering
	\includegraphics[width=0.6\textwidth]{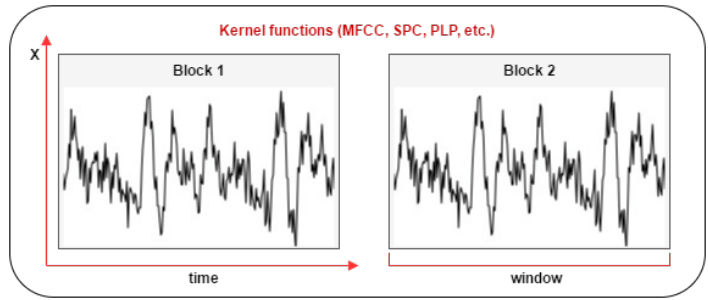}
	\caption{Block-based audio feature extraction}
	\label{fig:audiofeatures}
\end{figure}

Besides visual information, the audio data also reflects the speaker's emotion. Thus, our method automatically extracts the audio features from each annotated segments of the videos. For accelerating the audio feature extraction processing, we optimize and extend the GPU-based implementation of Michalek et al. \cite{michalek2014open}. In the approach, all windows are stored into the device memory and then copied to the fast shared on-chip memory at the beginning of each kernel. Due to the data independence between windows, we process the windows concurrently on different thread blocks, as can be seen in Figure~\ref{fig:audiofeatures}. The audio features include several statistic measures, e.g., max and min value, standard deviation, variance etc. of some key feature groups. In order to extract those features in parallel, we apply some CUDA core primitives such as reduction and scan \footnote{https://nvlabs.github.io/moderngpu/scan.html} to efficiently compute the results in the GPU.  Some of the useful key features are described below.

\begin{itemize}
	\item \textit{Mel frequency cepstral coefficients (MFCC): }
	Mel frequency cepstral coefficients are frequently used for speech recognition. The advantage of the features is to represent the amplitude of the spectrum in a compact form. The MFCC are  calculated based on short time Fourier transform (STFT). First, log-amplitude of the magnitude spectrum is taken, and the process is followed by grouping and smoothing the fast Fourier transform (FFT) bins according to the perceptually motivated Mel-frequency scaling. 
	
	\item \textit{Spectral Centroid (SPC):} 
	Spectral Centroid is the center of gravity of the magnitude spectrum of the STFT. Here, $M_{i}[n]$ denotes the magnitude of the Fourier transform at frequency bin $n$ and frame $i$. The centroid is used to measure the spectral shape. A higher value of the centroid indicates brighter textures with greater frequency. The spectral centroid $C_i$ is calculated as follows:
	
	\[C_{i} = \frac{\sum_{i=0}^{n}nM_{i}[n]}{\sum_{i=0}^{n}M_{i}[n]} \]
	
	\item \textit{Spectral Flux: }
	Spectral Flux is defined as the squared difference between the normalized magnitudes of successive windows:
	$F_{i} = \sum_{n=1}^{n}(N_{t}[n]-N_{t-1}[n])^{2}$
	where $N_{t}[n]$ and $N_{t-1}[n]$ are the normalized magnitudes of the Fourier transform at the current frame $t$ and the previous frame $t-1$, respectively. The spectral flux represents the amount of local spectral change.
	
	\item \textit{Beat Histogram: }
	Beat Histogram is a histogram showing the relative strength of different rhythmic periodicities in a signal. It is calculated as the auto-correlation of the RMS.
	
	\item \textit{Beat Sum: }
	This feature is measured as the sum of all entries in the beat histogram. It is a very good measure of the importance of regular beats in a signal.
	
	\item \textit{Strongest Beat: }
	This feature is defined as the strongest beat in a signal, in the beats per minute. It is found by finding the strongest bin in the beat histogram.
	
	\item \textit{Pause Duration: }
	Pause direction is the percentage of time the speaker is silent in the audio segment.
	
	\item \textit{Pitch: }
	The feature is computed by the standard deviation of the pitch level for a spoken segment.
	
	\item \textit{Voice Quality:}
	Harmonic to noise ratio in the audio signal.
	
	\item \textit{PLP: }
	The Perceptual Linear Predictive Coefficients of the audio segment.
\end{itemize}

During the feature extraction, we use a 30Hz frame-rate with windows of 100ms. The features are averaged over all the frames to obtain one feature vector for each utterance.

\subsection{Textual Feature Extraction}

\begin{figure}[htp]
	\centering
	\includegraphics[width=0.8\textwidth]{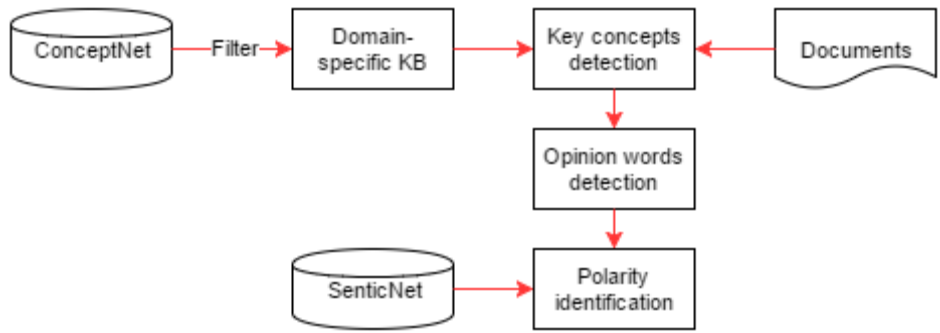}
	\caption{Common-sense-based text feature extraction}
	\label{fig:textfeatures}
\end{figure}

Identifying sentiments in text is a challenging task, because of ambiguity of words in the text, complexity of meaning and interplay of various factors such as irony, politeness, writing style, as well as variability of language from person to person and from culture to culture. In this work, we follow a commonsense-based method to identify both opinion targets and emotional words. Inspired by the promising results of GPU-based commonsense reasoning and commonsense-based sentiment analysis \cite{agarwal2015sentiment}, we introduce a method to extract important features based on commonsense knowledge bases in parallel on the GPU. Figure~\ref{fig:textfeatures} represents the flow diagram of the approach.

\textit{Domain-specific knowledge graph generation: } At first, we represent the commonsense knowledge bases as directed graphs. In addition to the node/edge arrays used in GpSense, we create a value array to maintain the string values of concepts and another array whose elements point to the offsets of values in the value array. The advantage of storing the value array is that we can directly search for the concept names on the GPU. Next, we transfer the data to the GPU memory. However, the commonsense knowledge graph still contains many irrelevant concepts which might not be used during the feature extraction process. In the initial step, we aim to filter out the concepts which are not relevant to the product reviews domain. We first take some key concepts as the inputs and then explore the knowledge graph to find the domain specific concepts. The GPU implementation of this graph exploration process is similar to the Filtering phase in GpSense algorithm. We extract the important concepts from commonsense knowledge bases up to 3-4 levels.

\textit{Key concepts extraction: } In order to extract important concepts from the given textual data, we check whether the data contains concepts in the generated domain-specific commonsense knowledge graph. To implement the task on the GPU, we detect words in the textual data in following steps: (1) The separate symbols are located in parallel; (2) The offsets of words are maintained in an intermediate array by using the CUDA compaction algorithm. After that, our system finds the segmented words in the commonsense knowledge graph. 

In the next step, we identify the opinion words associated with the domain-specific features. We then build the sentiment lexicon with the polarity values based on the publicly available resources, namely, SenticNet and EmoSenticNet. After identifying the polarities of the opinion words, we may need to change their values if any negative words such as ``not'' are discovered nearby. Thus, if we find ``not'' within two words from a positive opinion word $w$, we must reverse its polarity value from positive to negative.

\section{Fusion}
\label{sec:fusion}
This section discusses feature-level and decision-level fusion methods to use the information of the textual, audio and visual modalities. Multimodal fusion is the most important part of any multimodal sentiment analysis engine. There are two main fusion techniques: feature-level fusion and decision-level fusion.

\textbf{Feature-Level Fusion: } We follow feature-level fusion by concatenating the feature vectors of all three modalities in order to form a single long feature vector. This trivial method has the advantage of relative simplicity, yet is shown to produce significantly high accuracy.
We concatenate the feature vectors of all modalities into one feature vector stream. This feature vector is then used for classifying each video segment into sentiment classes. To estimate the accuracy, we use ten-fold cross validation.

\textbf{Decision Level Fusion: } In decision-level fusion we obtain the feature vectors from the above-mentioned methods. However, instead of concatenate the feature vectors like feature-level fusion we use a separate classifier for each modality. The output of each classifier is treated as classification scores. In particular, from each classifier we obtain a probability score for each sentiment class. In our case, as there are 3 sentiment classes, so we obtain 3 probability scores from each modality. We then calculate the final label of the classification using a rule based approach given below:

\[ l' = \arg \max_i(q_1s_i^a + q_2s_i^v + q_3s_i^t) , \quad i = {1,2,3,...,C} \]

$q_1, q_2$ and $q_3$ are weights for the three modalities. We use equal weighted scheme, so in our case $q_1 = q_2 = q_3 = 0.33$. $C$ is the number of sentiment classes. $s_i^a, s_i^v$ and $s_i^t$ denote the scores from the audio, visual and textual modality respectively.

\section{Experiments and Discussion}
\label{sec:eval}
In this section, we discuss the experimental results on the YouTube dataset, and make a comparison with the results obtained by the approach in \cite{morency2011towards}. Several supervised classifiers, i.e., Na\"{\i}ve Bayes, SVM, ELM, Neural Networks are employed on the fused feature vector to obtain the sentiment of each video segment. However, we obtain the best accuracy using ELM.

\begin{table}[h]
	\centering % centering table
	%\left
	\begin{tabular}{|l|r|r|} % creating eight columns
		\hline
		& \textbf{Precision} & \textbf{Recall} \\
		\hline
		Textual Modality    & 0.62 & 0.59 \\
		Audio Modality  & 0.65 & 0.67 \\ % Entering row contents
		Video Modality  & 0.68 & 0.68 \\
		Visual and Textual Modalities & 0.72 & 0.72 \\
		Visual and Audio Modalities & 0.73 & 0.73 \\
		Audio and Textual Modalities & 0.71 & 0.71 \\
		Visual, Audio and Textual Modalities & 0.78 & 0.77 \\
		\hline
	\end{tabular}
	\vspace{3mm}
	\caption{Results of Feature-Level Fusion} %title of the table
	\label{tbl:results3}
\end{table}

Table~\ref{tbl:results3} and \ref{tbl:results4} show the classification performances of four different models: text-only, visual-only, audio-only and trimodal integration.
Results on feature-level fusion are presented in Table \ref{tbl:results3} while Table \ref{tbl:results4} illustrates the experiment results of decision-level fusion. Compared to the experiment results reported in \cite{morency2011towards}, our method outperforms the approach in terms of accuracy.

\begin{table}[h]
	\centering % centering table
	
	\begin{tabular}{|l|r|r|} % creating eight columns
		\hline
		& \textbf{Precision} & \textbf{Recall} \\
		\hline
		Textual Modality    & 0.59 & 0.58 \\
		Audio Modality & 0.62 & 0.65 \\ % Entering row contents
		Video Modality & 0.67 & 0.66 \\
		Visual and Textual Modalities & 0.68 & 0.68 \\
		Visual and Audio Modalities & 0.71 & 0.70 \\
		Audio and Textual Modalities & 0.66 & 0.66 \\
		Visual, Audio and Textual Modalities & 0.75 & 0.73 \\
		\hline
	\end{tabular}
	\vspace{3mm}
	\caption{Results of Decision-Level Fusion} %title of the table
	\label{tbl:results4}
\end{table}

Clearly, the accuracy improves when we use audio, visual and textual modalities together in the experiment. These improvements are observed for both precision and recall in comparison with text-only, visual-only, and audio-only modalities.

\subsection{Feature Extraction Analysis}

In this section, we describe the performance of GPU-based feature extraction as well as analyze the importance of each feature used in the classification task. The best accuracy is obtained when all features are used together. The runtime of the CPU-based algorithms is measured using an Intel Core i7-870 2.93 GHz CPU with 8GB of memory. Our GPU algorithms are tested using CUDA Toolkit 6.0 running on the NVIDIA Tesla C2050 GPU with 3 GB global memory and 48 KB shared memory per Stream Multiprocessor.

For the visual feature extraction, our method is 42-50 times faster than CPU-based ASM algorithm when we run our experiments on 32 images concurrently. By using the sketch image generation technique, the convergence time of the ASM algorithm is reduced by approximately 50\%.

For audio feature extraction task, the overall speedup of GPU-based acceleration is around 12 times in comparison with the CPU-based method. MFCC and Spectral Centroid have lower importance on the overall accuracy of the sentiment analysis system. However, exclusion of those features causes the degradation of accuracy in the audio based sentiment analysis task. We also experiment the role of some audio features like {\it time domain zero crossing}, {\it root mean square}, {\it compactness}. However, we do not get higher accuracy using these features.

In the case of text based sentiment analysis, we find that concept-gram features play a major role compared to the SenticNet based feature. In particular, SenticNet based features mainly help to detect associated sentiment in a text using an unsupervised method. We aim to develop a multimodal sentiment analysis system where sentiment from the text will be extracted in an unsupervised way using SenticNet as a knowledge base.

\subsection{Performance Comparison of Different Classifiers}
In this section, we discuss the performance comparison of different classifiers in terms of both accuracy and training time.

\textbf{Accuracy:} 
On the same training and test sets, we run the classification experiment using SVM, Artificial Neural Network and ELM. ELM outperforms ANN by 12\% in term of accuracy, as seen in Table~\ref{table:results5}. However, we observe only little difference in accuracy obtained by ELM and SVM.

\begin{table}[h]
	\centering % centering table
	
	\begin{tabular}{|l|r|r|} % creating eight columns
		\hline
		\textbf{Classifiers} & \textbf{Recall} & \textbf{Training} Time \\
		\hline
		SVM & 77.03\% & 2.7 minutes \\
		ELM & 77.10\% & 25 seconds \\
		ANN & 57.81\% & 2.9 minutes \\
		\hline		
	\end{tabular}
	\vspace{3mm}
	\caption{Comparison of Classifiers} %title of the table
	\label{table:results5}
\end{table}

\textbf{Training Time:}
In terms of training time, ELM outperformed SVM and ANN by a large margin. As our goal is to develop a real-time multimodal sentiment analysis engine, we prefer ELM as a classifier because it helps to provide the best performance in terms of both accuracy and training time. 

\section{Summary}
\label{sec:end}
In this chapter, we have developed an ensemble application of ELM and GPU for real-time multimodal sentiment analysis. This work includes sets of relevant features for text and audio-visual data, as well as a simple technique for fusing the features extracted from different modalities. 
Our method employs various GPU-friendly techniques to enhance the performance of the feature extraction process from different modalities. 
In addition, powerful ELM classifiers are applied to build the sentiment analysis model based on the extracted features.  
In particular, our textual sentiment analysis module has been enriched by sentic-computing-based features, which have offered significant improvement in the performance of our textual sentiment analysis system. Visual features also play key role to outperform the state-of-the-art.
	% multimodal sentiment analysis
% !TEX spellcheck = en_US

\chapter{Conclusion and Future Work}
\graphicspath{{Chapter6/fig/EPS/}{Chapter6/fig/}}
\label{tag:chap6}

In this chapter, we recapitulate the chapters and highlight key points which are the major contributions for this thesis. We then present potential research directions of commonsense reasoning and its applications.

%%%%%%%%%%%%%%%%%%%%%%%%%%%%%%%%%%%%%%%%%%%%%%%%%%%%%%%%%%%%%%
%%%%%%%%%%%%%%%%%%%%%%%%%%%%%%%%%%%%%%%%%%%%%%%%%%%%%%%%%%%%%%
\section{Conclusion}
\label{chap6:Conclusions}

In this thesis we have presented commonsense reasoning and query processing systems on large-scale knowledge bases. Our systems are intensively implemented on massively parallel architecture of GPUs to solve the reasoning and querying problems on both graph-based and rule-based knowledge representations. Then, we have employed GPUs and commonsense knowledge bases to build a real-time multimodal sentiment analysis system.

To address the problem of reasoning and query processing on large-scale graph-based commonsense knowledge bases, the thesis presents GpSense algorithm to solve the subgraph matching problem which is the core function of commonsense reasoning systems. Our approach is based on a novel filtering-and-joining strategy which is specially designed to work on massively parallel architectures. In order to optimize the performance in depth, we utilize a series of optimization techniques which contribute towards increasing GPU occupancy, reducing workload imbalances and in particular speeding up subgraph matching on commonsense graphs. For large graphs whose sizes exceed the capacity of a typical GPU memory, we propose a multiple-level graph compression technique to reduce graph sizes while preserving all subgraph matching results. The graph compression method converts the data graph to a weighted graph which is small enough to be maintained in GPU memory. To highlight the efficiency of our solution, we perform an extensive evaluation of GpSense against state-of-the-art subgraph matching algorithms. Experiment results on both real and synthetic data show that our solution outperforms the existing methods on large graphs.

To retrieve information on rule-based commonsense knowledge bases, the thesis introduces gSparql, a fast and scalable inference and querying method on mass-storage RDF data with custom rules.  Our method focuses on dealing with backward-chaining reasoning which makes inferences at query time when the inferred triples are determined by the set of triple patterns defined in the query. To efficiently answer SPARQL queries in parallel, we first build reasoning trees for all triple patterns in the query and then execute those trees on GPUs in a bottom-up fashion. In particular, we convert the execution tree into a series of primitives such as sort, merge, prefix scan, and compaction which can be efficiently done on GPU devices. We also utilize a GPU-based Bloom Filter method and sort algorithms to overcome the triple duplication. Extensive experimental evaluations show that our implementation scales in a linear way and outperforms current optimized CPU-based competitors.

We utilize commonsense knowledge bases to address the problem of real-time multimodal analysis. In particular, we focus on the problem of multimodal sentiment analysis, which consists in the simultaneous analysis of different modalities, e.g., speech and video, for emotion and polarity detection. Our approach takes advantages of the massively parallel processing power of modern GPUs to enhance the performance of feature extraction from different modalities. In addition, in order to extract important textual features from multimodal sources we generate domain-specific graphs based on commonsense knowledge and apply GPU-based graph traversal for fast feature detection. Then, powerful ELM classifiers are applied to build the sentiment analysis model based on the extracted features. We conduct our experiments on the YouTube dataset and achieve an accuracy of 78\% which outperforms all previous systems. In term of processing speed, our method shows improvements of several orders of magnitude for feature extraction compared to CPU-based counterparts.

%%%%%%%%%%%%%%%%%%%%%%%%%%%%%%%%%%%%%%%%%%%%%%%%%%%%%%%%%%%%%%
%%%%%%%%%%%%%%%%%%%%%%%%%%%%%%%%%%%%%%%%%%%%%%%%%%%%%%%%%%%%%%
\section{Future Work}
\label{chap6:FutureWorks}

As an ongoing effort, we are exploring several extensions to the work in this dissertation. In the current systems, we take advantages of the GPU parallel processing power to perform most of the computational steps. 
The CPU is only used to execute small tasks and mainly to control the GPU work-flow. As a result, the computational capability of the CPU is under-utilized in our systems. For the future extension, we plan to investigate co-processing techniques that take into account both the computation resources. To do that, we need to find an efficient partitioning method to split the input data in order to balance the workload on both GPU and CPU. In addition, a GPU-CPU data transfer cost model is also required so that each operator in a query can utilize suitable processors - the CPU, the GPU, or both, for an optimized overall performance. For further enhance the parallel processing performance, multi-GPU algorithms must be taken into considerations.

The rich text information on the data storage layers of commonsense reasoning and query answering systems is also an open issue for our research. In the current setting, we use a dictionary to encode all string and URI values into numeric values (IDs). Thus, we only can examine the equality relation between two input values. In practice, most query engines allow users to input wild-card queries or accept regular expressions on text inputs. This is a potential research direction to improve our systems to support wider ranges of input queries. The crucial issue for this problem is building an efficient data layout to handle variable-sized data such as strings with unknown sizes. After that, a relevant parallel processing strategy is required to overcome the branch divergence problems.

Our GPU-based backward-chaining reasoning approach can only handle a small decidable fragment of description logics. For better reasoning and retrieving information on knowledge bases including both common and commonsense, we plan to extend our system to support higher levels of expressiveness and reasoning capabilities such as OWL 2 DL with existential and universal quantifications.

For multimodal sentiment analysis, gaze- and smile-based facial
expression features are usually found to be very useful for
sentiment classification. Our future research will aim to incorporate gaze and smile features, for facial-expression-based sentiment classification, in addition to focusing on the use of audio modality for the multimodal sentiment analysis task. Furthermore, we will explore the possibility of developing a culture- and language-independent multimodal sentiment classification framework. Finally, we will strive to improve the decision-level fusion process using a cognitively-inspired fusion engine. Subsequently, we will work on reducing the time complexities of our developed methods, in order to get closer to the ambitious goal of developing a real-time system for multimodal sentiment analysis. Hence, another aspect of our future work will be to effectively analyze and appropriately address the system's time complexity requirements in order to create an efficient and reliable multimodal sentiment analysis engine.
 % Conclusion & future work

%%%%%%%%%%%%%%%%%%%%%%%%%%%%%%%
% ~~~~~~~ Appendix ~~~~~~~~
%\addcontentsline{toc}{section}{\numberline{}\hspace{-.35in}{\bf Appendices}}

%\newpage
%\addcontentsline{toc}{section}{\numberline{}\hspace{-.35in}{\bf Appendix A}}
%\include{Appendices/AppendixA} 

%\newpage
%\addcontentsline{toc}{section}{\numberline{}\hspace{-.35in}{\bf Appendix B}}
%\include{Appendices/AppendixB} 

%\newpage
%\addcontentsline{toc}{section}{\numberline{}\hspace{-.35in}{\bf Appendix C}}
%\include{Appendices/AppendixC} 

%\newpage
%\addcontentsline{toc}{section}{\numberline{}\hspace{-.35in}{\bf Appendix D}}
%\include{Appendices/AppendixD} 

%\newpage
%\addcontentsline{toc}{section}{\numberline{}\hspace{-.35in}{\bf Appendix E}}
%\include{Appendices/AppendixE} 

%\newpage
%\addcontentsline{toc}{section}{\numberline{}\hspace{-.35in}{\bf Appendix F}}
%\include{Appendices/AppendixF} 

% ~~~~~~~ bibliography ~~~~~~~~
%\begin{spacing}{1.5}
%\include{append}
%\newpage
%\setlength{\baselineskip}{20pt plus 1pt minus 1pt}  % 1-1/2 spacing for whole document
%\addcontentsline{toc}{chapter}{References}

\bibliographystyle{IEEEtran}
\bibliography{reference}

%
%\end{spacing}

%%%%%%%%%%%%%%%%%%%%%%%%%%%%%%%
% ~~~~~~~ Author's Publications ~~~~~~~~
\newpage
\addcontentsline{toc}{section}{\numberline{}\hspace{-.35in}{\bf
Author's Publications}}
\renewcommand{\labelenumi}{[\arabic{enumi}]} 

\chapter*{Author's Publications}
\label{chap.publication} %\addcontentsline{toc}{chapter}{Publications}

\section*{Journal}
\begin{enumerate}
	
	\item \textbf{Tran H.N.}, Cambria E., Hussain A.,``Towards GPU-based Common-sense Reasoning: Using Fast Subgraph Matching", {\em Cognitive Computation 8 (6), 1074-1086}, 2016.

	\item \textbf{Tran H.N.}, Cambria E.,``A Survey of Graph Processing on Graphics Processing Units", {\em The Journal of Supercomputing} (in submission).
	
	\item \textbf{Tran H.N.}, Cambria E.,``Ensemble Application of ELM and GPU for Real-Time Multimodal Sentiment Analysis", {\em Memetic Computing} (in submission).
	
\end{enumerate}

\section*{Conference}
\begin{enumerate}
	
	\item \textbf{Tran H.N.}, Cambria E., ``GpSense: A GPU-friendly method for common-sense subgraph matching in massively parallel architectures", {\em International Conference on Intelligent Text Processing and Computational Linguistics (CICLing)}, 2016.
	
	\item \textbf{Tran, H.N.}, Kim J.J., He B., ``Fast subgraph matching on large graphs using
	graphics processors", {\em Renz M, Shahabi C, Zhou X, Cheema AM (eds) International Conference Database
	Systems for Advanced Applications (DASFAA)}, 2015.

	\item \textbf{Tran H.N.}, Do H.G., Cambria E.,``Fast Querying of Large-Scale RDF Data with Rule-Based Entailment Regimes using GPUs" (in submission).
	
\end{enumerate}

\end{document}